\documentclass[%
superscriptaddress,
nofootinbib,
 aps,
 prd,
 reprint
]{revtex4-2}

\usepackage{graphicx}
\usepackage{bm}% bold math
\usepackage[english]{babel}
\usepackage{csquotes}
\usepackage{amsmath}
\usepackage{amsthm}
\usepackage{latexsym}
\usepackage{amssymb}
\usepackage{mathrsfs}
\usepackage{epic}
\usepackage{epsfig}
\usepackage[dvipsnames]{xcolor}
\usepackage{dsfont}
\usepackage[colorlinks = true,breaklinks=true]{hyperref}
\usepackage{comment}
\usepackage[all]{xy}
\usepackage{enumitem}
\usepackage{eucal}
\usepackage{calc}
\usepackage[all]{xy}
\usepackage{epic}
\usepackage{tensor}
\usepackage{epsfig}
\usepackage[usenames,pdf]{pstricks}
\usepackage{pst-grad} % For gradients
\usepackage{pst-plot} % For axes
\usepackage[space]{grffile} % For spaces in paths
\usepackage{etoolbox} % For spaces in paths
\usepackage{subfigure}

\renewcommand{\Im}{ \mathrm{Im} }
\renewcommand{\Re}{ \mathrm{Re} }
\newcommand{\eps}{ \tilde{ \epsilon } } 

\newcommand{\lgrad}{\ell_{ \nabla R }}
\newcommand{\lw}{\ell_{ w } }
\newcommand{\lcurv}{\ell_{ R } }

\numberwithin{equation}{section}
\renewcommand\theequation{\arabic{section}.\arabic{equation}}

\begin{document}

\title{Spin Hall effects and the localization of massless spinning particles}

\author{Abraham I. Harte}
\email{abraham.harte@dcu.ie}
\address{Centre for Astrophysics and Relativity, School of Mathematical Sciences Dublin City University, Glasnevin, Dublin 9, Ireland}
	
\author{Marius A. Oancea}
\email{marius.oancea@univie.ac.at}
\address{Faculty of Physics, University of Vienna, Boltzmanngasse 5, 1090 Vienna, Austria}
\address{Erwin Schrödinger International Institute for Mathematics and Physics, University of Vienna, Boltzmanngasse 9, 1090 Vienna, Austria}
\address{Max Planck Institute for Gravitational Physics (Albert Einstein Institute), Am M\"uhlenberg 1, D-14476 Potsdam, Germany}

\begin{abstract} 
The spin Hall effects of light represent a diverse class of polarization-dependent physical phenomena involving the dynamics of electromagnetic wave packets. In a medium with an inhomogeneous refractive index, wave packets can be effectively described by massless spinning particles following polarization-dependent trajectories. Similarly, in curved spacetime the gravitational spin Hall effect of light is represented by polarization-dependent deviations from null geodesics. In this paper, we analyze the equations of motion describing the gravitational spin Hall effect of light. We show that these equations are a special case of the Mathisson-Papapetrou equations for spinning objects in general relativity. This allows us to use several known results for the Mathisson-Papapetrou equations, and apply them to the study of electromagnetic wave packets. We derive conservation laws, we discuss the limits of validity of the spin Hall equations, and we study how the energy centroids of wave packets, effectively described as massless spinning particles, depend on the external choice of a timelike vector field, representing a family of observers. In flat spacetime, the relativistic Hall effect and the Wigner(-Souriau) translations are recovered, while our equations also provide a generalization of these effects in arbitrary spacetimes. We construct a large class of wave packets that can be described by the spin Hall equations, but also find its limits by giving examples of wave packets which are more general and are not described by the spin Hall equations. Lastly, we examine the assumption that electromagnetic wave packets are massless. While this is approximately true in many contexts, it is not exact. We show that failing to carefully account for the limitations of the massless approximation results in the appearance of unphysical ``centroids'' which are nowhere near the wave packet itself.

\end{abstract}

\maketitle

\section{Introduction}

Many observations of the physical world---particularly in astrophysical contexts---involve measurements of electromagnetic and (more recently) gravitational radiation. Interpreting this radiation requires a theoretical model for its propagation. In the case of electromagnetic waves, one might begin with Maxwell's equations. In the case of gravitational waves, one might instead use the Einstein field equation. Regardless, exact solutions are rarely available and the geometric optics approximation\footnote{The geometric optics approximation, with or without higher-order corrections, is sometimes referred to as the high-frequency approximation, or as the Wentzel-Kramers-Brillouin (WKB) approximation.} is typically applied in order to make progress. This assumes that  wavelengths are small compared with all other relevant length scales, and forms the basis for most of the theory of gravitational lensing \cite{MTW, gravitational_lenses_book, Perlick2004, BHS7, Perlick2021}.

Mathematically, the geometric optics approximation allows the field equations, which are partial differential equations, to be approximated by a set of ordinary differential equations. The problem of solving partial differential equations is thereby reduced to the much simpler problem of solving ordinary differential equations. More specifically, this process shows that the amplitudes and polarization states of high-frequency electromagnetic and gravitational waves propagate along null geodesics. The resulting field acts, in this approximation, as though it were formed from a collection of noninteracting massless particles.

It is the purpose of this paper to investigate what happens beyond geometric optics, when wavelengths are small but not completely ignorable. More specifically, how do corrections to geometric optics affect propagation directions? While the equations which govern small corrections to geometric optics were derived long ago \cite{EhlersGeoOptics, Isaacson1} for electromagnetic and gravitational waves propagating through curved spacetimes, their consequences have not been thoroughly explored. It is nevertheless known that all reasonable definitions for the local ``propagation direction'' agree in geometric optics: The direction of the electromagnetic momentum density is identical for all observers, and that coincides with the direction of the local phase gradient, the direction along which ``information'' propagates, and the (necessarily degenerate) principal null direction of the electromagnetic field. Beyond geometric optics, different notions of propagation direction no longer agree: The direction of the 4-momentum density can be different for different observers, there can be two principal null directions, and phase gradients can depend on a choice of basis \cite{HarteOptics1}. Moreover, amplitude and polarization states no longer propagate independently along each ray. Instead, there is a transport of information between neighboring rays as well as along them. This means that beyond leading order, there is no well-defined direction which can be associated with ``information flow'' in a high-frequency field. 

This complexity requires that we be precise about what exactly it is whose propagation we would like to understand. In this paper, we focus on the ``bulk'' propagation of small\footnote{These wave packets must be large compared to their dominant wavelengths but small compared to all other length scales.} electromagnetic pulses in curved spacetimes. We choose a ``center'' for each pulse and ask how that center evolves in time. Pulses in geometric optics are simple: With reasonable assumptions, they travel along null geodesics \cite{sbierski2015characterisation}. Like their constituent rays, the centers of high-frequency pulses behave, at leading order, like massless monopolar particles. One order beyond geometric optics, the motion depends on a pulse's angular momentum. More subtly, it also depends on precisely which definition is used to describe the pulse's center. Regardless, there is a sense in which otherwise-identical wave packets with opposite circular polarizations can be deflected with respect to one another. This behavior may be summarized by stating that one order beyond geometric optics, the bulk motion is equivalent to that of a massless dipole.  

In the literature on flat-spacetime optics in nontrivial materials, spin-dependent corrections to the propagation of electromagnetic fields are sometimes described as spin Hall effects. There are in fact a number of different spin Hall effects which have been discussed theoretically, some of which have also been observed experimentally \cite{SHE_review, SOI_review, SHEL_review}. Some spin Hall effects are induced by, e.g., gradients in the refractive index \cite{OpticalMagnus, SHE-L_original, SHE_original, Bliokh2004, Bliokh2004_1, Duval2006, SHEL_experiment, Aiello2008, Bliokh2008, Bliokh2009}. Others arise even without any material inhomogeneities: The geometric spin Hall effect \cite{Aiello2009,Korger2011,Korger2014} and the related relativistic Hall effect \cite{Relativistic_Hall} and Wigner(-Souriau) translations \cite{Stone2015, Duval_chiral_fermions, DUVAL2015} all arise in vacuum and in flat spacetime. These three effects may be shown to be associated with differing definitions for the ``center'' of a given wave packet. More precisely, the relativistic Hall effect and the Wigner translations are related to differences between three-dimensional centroids which would naturally be associated with different observers. They are essentially the same as (unnamed) effects which have long been known for massive objects \cite{PryceCM, MollerLectures, DixonSR, Costa2015}. The geometric spin Hall effect is somewhat different, being instead concerned with differences between centroids which are defined on different two-dimensional cross sections. 

The spin-dependent propagation effects discussed in this paper arise in vacuum but in generic spacetimes, and are sometimes referred to as gravitational spin Hall effects \cite{GSHE_review, Oanceathesis}. Various approaches have been taken before to understand the motion of electromagnetic, and also gravitational, wave packets in this context. Some approaches have been based on classical high-frequency expansions in the spirit of geometric optics \cite{Frolov, Frolov2, covariantSpinoptics, Harte2018, spinorSpinoptics, spinorSpinoptics2, GSHE2020, shoom2020, Frolov2020, GSHE_GW, audretsch,rudiger}. Others have taken a semiclassical approach, using the  Bargmann-Wigner equations or Weyl equations \cite{SHE_QM1, SHE_Dirac, SHE_GW}. Still other approaches have not made any direct contact with an underlying field theory, but have instead claimed that the motion of a wave packet could be described using massless versions of the Mathisson-Papapetrou (MP) equations \cite{souriau1974modele, saturnini1976modele, Mashhoon1975, bailyn1977pole, bailyn1981pole, bini2006massless, semerak2015spinning, Duval2006}, equations which are known to describe classical spinning objects in curved spacetimes.

This paper focuses on the gravitational spin Hall effect of light, as described in Ref.~\cite{GSHE2020}. While the derivation there was based on a high-frequency approximation, it differs from other high-frequency approaches by being applicable in arbitrary spacetimes and by avoiding specific $3+1$ foliations. This paper endeavors to better understand the meaning, the domain of applicability, and the limitations of the gravitational spin Hall equations. It also unifies those equations with others which have appeared in different contexts: The gravitational spin Hall equations are shown to be a special case of the MP equations, and the flat-spacetime gradient-index spin Hall effect, the relativistic Hall effect, and the Wigner(-Souriau) translations are all shown to be special cases of the gravitational spin Hall equations. 

But before any equations of motion can be sensibly discussed, it is necessary to first explain what exactly those equations describe. This leads us to consider what can be meant by the centroid of an extended wave packet. One definition which has appeared in the literature is shown to be untenable. A large number of others remain, however, and we show the set of all such possible centroids is unbounded for massless---but not massive---objects. While this might at first appear to be a failure of the definitions, it is in fact a failure of masslessness. We show that wave packets with nonzero angular momentum cannot be massless, and this is essential to their localizability. While the massless \textit{approximation} can be useful for many purposes, ignoring its limitations can result in qualitatively incorrect conclusions.

The paper begins in Sec.~\ref{sec:GSHE_review} by reviewing the gravitational spin Hall effect of light, as presented in Ref.~\cite{GSHE2020}. For comparison with the spin Hall effect of light in flat-spacetime optics \cite{SOI_review}, we emphasize the role of the Berry phase and the Berry connection in describing polarization, as well as the role of the Berry curvature in the gravitational spin Hall equations. 

Section~\ref{Sec:MP} shows that with particular initial and spin supplementary conditions, the gravitational spin Hall equations emerge as a special case of the MP equations. This relation allows us to use the well-developed theory associated with the MP equations to clarify the meanings of the worldline and the momentum which arise in the spin Hall equations. It also allows us to write down conservation laws for those equations and to discuss their regimes of validity.

The spin Hall equations involve an arbitrary choice of timelike vector field, and we show in Sec.~\ref{sec:role_of_t} that this parametrizes different definitions for the centroid of an extended wave packet. Our main result regarding these centroids is that although massive spinning objects can be localized, massless ones cannot.

Section~\ref{sec:wavepackets} examines whether or not the initial conditions which reduce the MP equations to the gravitational spin Hall equations are in fact realized by reasonable wave packets. We use a high-frequency approximation to explicitly construct a large class of electromagnetic wave packets. For many members of this class, the appropriate conditions are indeed satisfied. However, we also find wave packets which do not have the expected properties. This implies that there are nontrivial assumptions on the nature of the wave packet which have been implicitly (and unknowingly) imposed in the prior literature. We also show that our approximate wave packets can fail to satisfy the dominant energy condition. This is an unphysical artifact of the high-frequency approximation, and is what leads to the apparent delocalization of spinning wave packets discussed in Sec.~\ref{sec:role_of_t}.

In Sec.~\ref{sec:shel}, we discuss an analogy between light propagation through an optical medium and light propagation through vacuum but in an effective optical metric. Using a standard optical metric, we recover the spin Hall effect of light in an inhomogeneous medium from the gravitational spin Hall equations. 

Finally, the Appendix demonstrates that at least in flat spacetime, the spin of any massless object which satisfies the dominant energy condition must vanish. It follows that, e.g., electromagnetic wave packets with nonzero spin cannot be exactly massless.  

\textit{Notation and conventions:} We work on an arbitrary smooth Lorentzian manifold $(M, g_{\alpha \beta})$, where the metric tensor $g_{\alpha \beta}$ has signature $(-\,+\,+\,+)$. Greek letters are used for spacetime indices and run from $0$ to $3$. We use bold symbols to denote $3$-vectors, and their components are labeled by Latin letters from the middle of the alphabet, $(i, j, k, \ldots)$, that run from $1$ to $3$. Units are used in which $G=c=1$, the Einstein summation convention is assumed, and we use the notation $a_\alpha b^\alpha = a \cdot b$, $a_\alpha a^\alpha = a \cdot a = a^2$. The Riemann tensor is defined such that $2 \nabla_{[\alpha} \nabla_{\beta]} \omega_\gamma = R_{\alpha\beta\gamma}{}^{\lambda} \omega_\lambda$ for any $\omega_\gamma$. When working with tensors $T$ defined at different spacetime points $x^\alpha, \tilde{x}^\alpha \in M$, we use the usual notation, $T\indices{_\alpha^\beta}(x)$, when the tensor is defined at $x^\alpha$, while we use primed indices, $T\indices{_{\alpha'}^{\beta'}}(\tilde{x})$, for tensors defined at $\tilde{x}^\alpha$.

\section{Gravitational spin Hall effect of light}
\label{sec:GSHE_review}

The equations of motion which describe the spin Hall effect for electromagnetic waves propagating through curved spacetimes were derived in Ref.~\cite{GSHE2020}. They were obtained by performing a covariant high-frequency analysis of the vacuum Maxwell equations. A similar approach was used in Ref.~\cite{GSHE_GW} to describe the spin Hall effect for gravitational waves propagating on curved backgrounds. For electromagnetic waves, the derivation starts with the WKB ansatz
\begin{equation}
    A_\alpha =  \Re \left[ \epsilon \left( \psi a_\alpha + \epsilon \psi^{(1)}_\alpha + \mathcal{O}(\epsilon^2) \right) e^{i u/\epsilon} \right] 
    \label{WKB0}
\end{equation}
for the electromagnetic potential, where $\epsilon$ is a small  parameter related to the wavelength, $u$ is a real phase function, $\psi$ is a real scalar amplitude, and $a_\alpha$ is a complex polarization vector normalized such that $a_\alpha \bar{a}^\alpha = 1$. Higher-order terms, such as $\psi^{(1)}_\alpha$, do not play any role in this section. The overall factor of $\epsilon$ is for convenience and ensures that the field strength $F_{\alpha \beta} = 2 \nabla_{[\alpha} A_{\beta]}$ is nontrivial and finite in the $\epsilon \to 0$ limit. If $u$ increases as with time, it is convenient to define the future-directed wave vector $k_\alpha = - \nabla_\alpha u$, and in terms of that, a timelike observer with $4$-velocity $t^\alpha$ will measure the wave frequency
\begin{equation}
    \omega = - t \cdot k/\epsilon.
    \label{omegaDef}
\end{equation}

The derivation of the spin Hall effect in Ref.~\cite{GSHE2020} relies on an analysis of the overall phase factor of the field, which consists of $u$ at the lowest order in $\epsilon$, together with a higher-order phase factor, referred to as the Berry phase, which comes from the polarization vector $a_\alpha$. Using the WKB ansatz above, together with the Maxwell equation
\begin{equation}
    \left( \nabla^\beta \nabla_\alpha - \delta^\beta_\alpha \nabla^\gamma \nabla_\gamma \right) A_\beta = 0,
\end{equation}
and the Lorenz gauge condition $\nabla_\alpha A^\alpha = 0$, the wave vector $k_\alpha$ must be be null and orthogonal to the polarization vector,
\begin{equation} \label{eq:orth}
    k \cdot k = k \cdot a = 0.
\end{equation}
Additionally, the scalar amplitude $\psi$ must satisfy the transport equation
\begin{equation}
    \nabla_\alpha \left( k^\alpha \psi^2 \right) = 0,
\end{equation}
and the polarization vector $a_\alpha$ must be parallel transported,
\begin{equation} \label{eq:transp}
    k^\beta \nabla_\beta a_\alpha = 0.
\end{equation}
These are the usual equations of geometric optics. The null geodesic rays of geometric optics are integral curves of $k^\alpha$.

It is convenient to expand the polarization vector in terms of a tetrad $\{ k_\alpha, t_\alpha, m_\alpha, \bar{m}_\alpha \}$, where the real covector $t_\alpha$ is timelike, $k_\alpha$, $m_\alpha$, and its complex conjugate $\bar{m}_\alpha$ are null, $t\cdot m = 0$, and $m \cdot \bar{m} = 1$. Given Eq. \eqref{eq:orth}, there must exist complex scalars $z_1$, $z_2$ and $z_3$ such that
\begin{equation} \label{eq:a_basis}
    a_\alpha = z_1 m_\alpha + z_2 \bar{m}_\alpha + z_3 k_\alpha.
\end{equation}
The complex covectors $m_\alpha$ and $\bar{m}_\alpha$ form a circular polarization basis, and the considered electromagnetic wave is circularly polarized when $z_1 = 0$ or $z_2 = 0$. The term proportional to $k_\alpha$ is pure gauge, not fixed by the Lorenz gauge condition, and it will not play any role in what follows.

The only element of the tetrad which is interpreted as being fixed by the field is $k_\alpha$. Supplementing that with $t_\alpha$ fixes the  $2$-plane spanned by $m_\alpha$ and $\bar{m}_\alpha$. And within that plane, there is still an additional freedom associated with the spin rotations $m_\alpha \mapsto e^{i \phi} m_\alpha$, where $\phi$ is any real scalar. Any change in $t_\alpha$ will result in a shift with the form $m_\alpha \mapsto m_\alpha + c k_\alpha$, which can only affect $z_3$ in \eqref{eq:a_basis}. Although that is interpreted as a  gauge transformation here, changes in $t_\alpha$ will act nontrivially and play an important role in the spin Hall equations\footnote{Changes of $t_\alpha$ are analogous to the Wigner translations discussed in Ref.~\cite{Stone2015}, which act as gauge transformations on plane waves, but act nontrivially on finite wave packets.}. Spin rotations instead affect the values of $z_1$ and $z_2$. Nevertheless, they do not affect the spin Hall equations.

We can now obtain a transport equation for $z_1$ and $z_2$ along the rays. Viewing $m_\alpha$ as a covector-valued field over the cotangent bundle, depending on both position and on $k_\alpha$, the parallel transport equation \eqref{eq:transp} implies that
\begin{equation} \label{eq:zdot}
    \frac{d }{d\tau } \begin{pmatrix} z_1 \\ z_2 \end{pmatrix}
    = i (k^\alpha B_\alpha) 
    \begin{pmatrix}
        1   &   0   \\
        0   &   -1
    \end{pmatrix}
    \begin{pmatrix} z_1 \\ z_2 \end{pmatrix},
\end{equation}
where
\begin{equation}
    B_\alpha = i \bar{m}^\beta \left( \nabla_\alpha  +  k_\gamma \Gamma^\gamma_{\alpha \lambda} \frac{\partial}{\partial k_\lambda} \right) m_\beta
\end{equation}
is the Berry connection. The operator in brackets in the Berry connection may be seen to be a horizontal covariant derivative on the cotangent bundle. Regardless, the transport equation \eqref{eq:zdot} can be integrated to yield
\begin{equation}
    z_1(\tau) =  e^{i \gamma} z_1(\tau_0), \qquad z_2(\tau) =  e^{-i \gamma} z_2(\tau_0),
\end{equation}
where 
\begin{equation}
    \gamma = \int_{\tau_0}^\tau d\tau' k^\alpha B_\alpha
\end{equation}
is the Berry phase. The Berry phase represents a higher-order correction to the overall phase of the WKB potential, and is generally responsible for the spin Hall effect of light \cite{Bliokh2004, Bliokh2008}. For circularly polarized electromagnetic waves, the leading-order field takes the form $A_\alpha =  \Re \left[ \epsilon \psi m_\alpha e^{i (u+\epsilon \gamma)/\epsilon} \right]$ or $A_\alpha =  \Re \left[ \epsilon \psi \bar{m}_\alpha e^{i (u-\epsilon \gamma)/\epsilon}\right]$, depending on the handedness of circular polarization state. The total phase at this order is therefore proportional to $u + \epsilon s \gamma$, where $s = \pm 1$. Note that both the Berry connection and the Berry phase depend on spin rotations $m_\alpha \mapsto e^{i \phi} m_\alpha$, transforming as $B_\alpha \mapsto B_\alpha - \nabla_\alpha \phi$ and $\gamma(\tau) \mapsto \gamma(\tau) - \phi(\tau) + \phi(\tau_0)$. 

In geometric optics, the dispersion relation $k \cdot k = 0$ may be viewed as a Hamilton-Jacobi equation for the phase function $u$. If that is solved using the method of characteristics, one recovers the null geodesic rays of geometric optics. In Ref.~\cite{GSHE2020}, the strategy was to generalize this procedure, deriving the spin Hall effect by looking for an effective dispersion relation involving the gradient of the corrected phase function $u + \epsilon s \gamma$. Letting $K_\alpha = - \nabla_\alpha (u + \epsilon s \gamma)$, we use $k \cdot k = 0$ and the definition of the Berry phase $\gamma$ to arrive at the following effective dispersion relation:
\begin{equation}
     K \cdot K -2 \epsilon s K \cdot B = \mathcal{O}(\epsilon^2).
\end{equation}
This can be viewed as a Hamilton-Jacobi equation for the total phase function $u + \epsilon s \gamma$. Using the method of characteristics, we can solve this Hamilton-Jacobi equation and obtain ray equations with polarization-dependent corrections to the geodesic equations of geometric optics. However, since these equations depend on the Berry connection $B_\alpha$, they are not invariant under spin rotations. This gauge dependence can be removed by switching to noncanonical coordinates\footnote{A similar approach is also used for the description of charged particles in an external electromagnetic field, where a coordinate transformation is used to rewrite the equations of motion in terms of the gauge-invariant Faraday tensor instead of the gauge-dependent vector potential.}, as described in Ref.~\cite[Sec.~IV.B.1]{GSHE2020} (see also Ref.~\cite{Littlejohn1991}).

The spin Hall equations which result from the use of these coordinates, which describe the polarization-dependent propagation of circularly polarized light, can be written as \cite{GSHE2020}
\begin{equation}
\label{spinHall}
\begin{split}
    \dot{x}^\alpha &= p^\alpha + \epsilon s p^\beta \Big[ \left( F_{p x} \right)\indices{_\beta^\alpha} + \Gamma^\gamma_{\lambda \beta} p_\gamma \left( F_{p p}\right)^{\lambda \alpha} \Big] ,   \\
    \dot{p}_\alpha &= \Gamma^\gamma_{\alpha \beta} p_\gamma p^\beta - \epsilon s  p^\beta \Big[  \left( F_{x x} \right)_{\alpha \beta} + \Gamma^\gamma_{\beta \lambda} p_\gamma  \left( F_{x p} \right)\indices{^\lambda_\alpha} \Big],
\end{split}
\end{equation}
where $x^\alpha$ denotes a position, $p_\alpha$ a momentum, the dot an ordinary (noncovariant) derivative $d/d\tau$, and $s = \pm 1$ depending on the handedness of the circular polarization state. The same equations, but with $s = \pm 2$ instead, describe the spin Hall effect for circularly polarized gravitational waves \cite{GSHE_GW}. They are understood to be valid up to terms of order $\epsilon^2$. The spin Hall equations are expressed in terms of the Berry curvature components
\begin{equation}
\begin{split}
    (F_{pp})^{\beta\alpha} &= 2 \Im\left( \frac{ \partial  m_\gamma }{ \partial p_{\alpha} } \frac{ \partial \bar{m}^\gamma }{ \partial p_{\beta} }  \right),
    \\
    (F_{xx})_{\beta\alpha} &= 2 \Im \left(\nabla_{\alpha} m_\gamma \nabla_{\beta} \bar{m}^{\gamma} +  m_\gamma \nabla_{[\alpha} \nabla_{\beta]} \bar{m}^\gamma \right),
    \\
    (F_{px})_{\alpha}{}^{\beta} &= - (F_{xp})^{\beta}{}_{\alpha} 
    = 2 \Im \left( \frac{ \partial m_\gamma }{ \partial p_{\beta}} \nabla_\alpha \bar{m}^\gamma \right) ,
\end{split}
\end{equation}
which are invariant with respect to spin rotations.

The procedure leading to the spin Hall equations has removed any dependence on spin rotations. However, it has introduced a physical dependence on the timelike covector $t_\alpha$. This can be made explicit by calculating the components of the Berry curvature using the properties of the tetrad\footnote{Here, $p_\alpha$ has replaced the $k_\alpha$ which appeared in the above discussion. $m_\alpha$ and $\bar{m}_\alpha$ are now viewed as functions of position and of $p_\alpha$.} $\{p_\alpha, t_\alpha, m_\alpha, \bar{m}_\alpha\}$, which results in \cite[Appendix C]{GSHE2020}
\begin{equation} \label{eq:Berry_curvature}
\begin{split}
    \left({F_{p p}}\right)^{\beta \alpha} &= \frac{ \Sigma^{\alpha \beta} }{(p \cdot t)^2} , 
    \\
    \left({F_{xx}}\right)_{\beta \alpha} &= \frac{\Sigma^{\gamma\lambda}}{2}  \bigg[ R_{\gamma  \lambda \alpha \beta} + \frac{2}{ (p \cdot t)^2}  p_\rho \Gamma^\rho_{\gamma [ \alpha }   \\ 
    & \qquad \qquad   ~ \times\bigg( \Gamma^\sigma_{\beta] \lambda} p_\sigma - 2 (p \cdot t)  \nabla_{\beta]} t_\lambda  \bigg) \bigg],
    \\
    \left({F_{x p}}\right)\indices{^\alpha_\beta} &= \frac{ \Sigma^{\alpha\gamma}  }{(p \cdot t)^2}  \left( p_\rho \Gamma^\rho_{\beta \gamma} -( p \cdot t ) \nabla_\beta t_\gamma \right).
\end{split}
\end{equation}
Each term here is linear in the real bivector 
\begin{equation}
    \Sigma^{\alpha \beta } = 2 i \bar{m}^{[\alpha} m^{\beta]} = \frac{ \varepsilon^{\alpha \beta \gamma \lambda} p_\gamma t_\lambda }{p \cdot t} ,
    \label{SigmaDef}
\end{equation}
which is invariant under spin rotations and is uniquely determined by $p_\alpha$ and $t_\alpha$. We shall see in Sec.~\ref{Sec:MP} that $\Sigma^{\alpha \beta}$ is proportional to the angular momentum tensor of the wave packet. Regardless, substituting Eq.~\eqref{eq:Berry_curvature} into \eqref{spinHall} shows that the spin Hall equations can be written in the more compact form 
\begin{subequations} \label{eq:gshe_eq}
    \begin{align}
    \dot{x}^\alpha &= p^\alpha + \frac{1}{p \cdot t} (\epsilon s \Sigma^{\alpha \beta}) p^\gamma \nabla_\gamma t_\beta, \label{eq:xdot}
    \\
    \frac{D p_\alpha }{ d \tau} &= -  \frac{1}{2} R_{\alpha \beta \gamma \lambda}  p^\beta (\epsilon s \Sigma^{\gamma \lambda}) , \label{eq:pdot}
    \end{align}
\end{subequations}
where $D/d\tau = \dot{x}^\alpha \nabla_\alpha$ denotes the covariant derivative along the worldline. It is now manifest that the only external choice relevant to these equations is the timelike vector field $t^\alpha$. We shall see below that that choice parametrizes the definition for $x^\alpha$.

Unlike the integral curves of $k^\alpha$, which are interpreted as rays within (say) a wave packet, the position $x^\alpha$ which appears in the spin Hall equations \eqref{eq:gshe_eq} is interpreted as describing the position of the wave packet as a whole: its ``centroid.'' Similarly, the momentum $p_\alpha$ is interpreted as the \textit{net} momentum of the wave packet, not as a momentum density within that wave packet. One consequence of the spin Hall equations is that the worldline is not necessary tangent to the momentum. Systems with this feature are sometimes referred to as having hidden momentum \cite{Shockley1967, Coleman1968, babson2009, Gralla2010, Costa2015} or anomalous velocity \cite{Berry_CM1, SHE_original, SHE_QM1, Stone2015(2), Stone2016}. Also note that inspection of Eq.~\eqref{eq:gshe_eq} shows that the affine parameter $\tau$ is dimensionless and that $\epsilon$ has units $\mbox{(length)}^2$. Examination of the spin Hall equations shows that $\tau$ has been chosen such that $\dot{x} \cdot t = p \cdot t$. It is straightforward to see from Eq.~\eqref{eq:pdot} that if $p_\alpha$ is initially null, it remains null for all time. Equation \eqref{eq:xdot} implies that when the momentum is null, so too is the worldline: $\dot{x} \cdot \dot{x} = \mathcal{O}(\epsilon^2)$. These equations are assumed to be used only for initial data in which $p_\alpha$, and therefore $\dot{x}^\alpha$, are indeed null.

There is not always a hidden momentum. As a particular case, suppose that $t^\alpha$ is parallel transported in the sense that
\begin{equation} \label{eq:t_parallel}
    \dot{x}^\beta \nabla_\beta t_\alpha = \mathcal{O}(\epsilon).
\end{equation}
With this choice, $\dot{x}^\alpha = p^\alpha + \mathcal{O}(\epsilon^2)$ and Eq.~\eqref{eq:gshe_eq} reduces to the polarization-dependent ray equations obtained by Frolov in Ref.~\cite[Eq.~110-112]{Frolov2020}. In this sense, the spin optics approximation in that paper describes a particular case of the gravitational spin Hall equations obtained in \cite{GSHE2020}. However, since one of our goals is to understand the role of $t^\alpha$ in Eq.~\eqref{eq:gshe_eq}, we do not assume any special choices for it in the remainder of this paper.

It was not clear in the derivation of the spin Hall equations precisely what $x^\alpha$, $p_\alpha$, or $t_\alpha$ are, what types of wave packets these equations describe, or what sorts of approximations are implicit in them (beyond the assumption of high-frequencies). These issues will be addressed below.

\section{Mathisson-Papapetrou equations and their implications for the spin Hall equations}
\label{Sec:MP}

The spin Hall equations \eqref{eq:gshe_eq} are interpreted as describing the motion of  circularly polarized electromagnetic wave packets. However, it is known from separate arguments that the motion of any sufficiently compact spinning object is governed by the Mathisson-Papapetrou (MP) equations\footnote{These are variously referred to as the Papapetrou, Mathisson-Papapetrou, and Mathisson-Papapetrou-Dixon equations. The same labels are commonly applied also to more general equations which involve the quadrupole and higher-order moments of the relevant object. As recounted in \cite{dixon2015new}, Mathisson \cite{Mathisson} appears to have been the first to obtain the pole-dipole equations \eqref{MP}. He did so before Papapetrou \cite{Papapetrou} and using a superior method. Mathisson also derived some of the quadrupole terms which are not included here. Dixon \cite{Dixon74} derived all quadrupole and higher-order terms, developing a full theory of multipole moments to all orders. This was later generalized to also allow for self-interaction \cite{HarteGrav, HarteReview}. Here we refer to the test body pole-dipole equations---without quadrupole or higher-order moments---as the MP equations.}
\begin{subequations}
\label{MP}
\begin{gather}
    \frac{Dp_\alpha }{d\tau}  = - \frac{1}{2}  R_{\alpha \beta \gamma \lambda } \dot{x}^\beta S^{\gamma \lambda}, 
    \label{MPp}
    \\
    \frac{DS^{\alpha \beta}  }{d\tau} = 2 p^{[\alpha} \dot{x}^{\beta]}.
    \label{MPS}
\end{gather}
\end{subequations}
These equations evolve an object's linear momentum $p_\alpha$ and its angular momentum $S^{\alpha \beta} = S^{[\alpha \beta]}$ along a specified worldline. They are very general: As long as the quadrupole and higher-order multipole moments of an object's stress-energy tensor can be ignored, the MP equations hold for all sufficiently compact objects with conserved stress-energy tensors \cite{Dixon74, HarteReview}. In particular, although much of the literature on these equations assumes that $p_\alpha$ is timelike, their derivation makes no use of that condition; null momenta are also admissible. Whether or not the momentum is null depends only on the nature of the underlying stress-energy tensor.

The generality of the MP equations can be understood, in part, from the fact that they are essentially kinematic. They arise as consequences of attempting to maintain Poincar\'{e} invariance as much as possible along the given worldline \cite{HarteSyms}. Indeed, they imply the presence of ten conserved quantities along that worldline, which correspond locally to the four translations, three rotations, and three boosts of a four-dimensional Minkowski spacetime (even when the actual spacetime is not Minkowski). The nontrivial physics which enters into this is that corrections to the MP equations---deviations due to the breakdown of Poincar\'{e} invariance---depend only an object's quadrupole and higher-order moments. It is expected from the equivalence principle that ``sufficiently compact'' objects should behave, at least locally, as though the spacetime is flat, and a calculation shows that the breakdown of the flat-spacetime conservation laws first occurs at quadrupolar order. 

What is relevant here is that electromagnetic wave packets are associated with conserved stress-energy tensors. Their bulk motion can therefore be described not only by the spin Hall equations, but also by the MP equations. We show in Sec.~\ref{Sec:MPtoSH} that there is a precise sense in which the spin Hall equations arise as a special case of the MP equations. Section~\ref{Sec:pDef} exploits this connection between the spin Hall equations and the MP equations to relate quantities in the spin Hall equations to an underlying stress-energy tensor. Section~\ref{Sec:consLaws} uses known results for the MP equations to write down previously unknown conservation laws associated with the spin Hall equations. Finally, Sec.~\ref{Sec:Approx} explores the approximations used in the spin Hall equations and discusses when those approximations hold.

\subsection{Spin Hall equations from MP equations}
\label{Sec:MPtoSH}

Our first task is to show that the spin Hall equations are a special case of the MP equations. We now show that the spin Hall equations arise after choosing appropriate initial data for the MP equations, fixing an appropriate definition for the centroid of an extended wave packet, and imposing a particular parameterization for the worldline of that centroid.

\textit{A priori}, it may appear that the spin Hall and MP equations do not even describe the same physical quantities. The spin Hall equations evolve $x^\alpha$ and $p_\alpha$ while $t^\alpha$ is specified independently. By contrast, the MP equations evolve $p_\alpha$ and $S^{\alpha \beta}$ while $x^\alpha$ is specified independently\footnote{The worldline which appears in the MP equations is to be interpreted as a choice of origin for a multipole expansion. It does not necessarily have any interpretation as a centroid. That interpretation arises only when  additional conditions are imposed on the worldline. However, except in maximally symmetric spacetimes, ignoring the quadrupole and higher-order moments cannot be justified unless there is some sense in which the worldline lies near an object's ``center.''}. This discrepancy is resolved by showing that in the present context, i) the angular momentum equation \eqref{MPS} can be trivially solved, and ii) the specification of $t^\alpha$ is equivalent to the specification of $x^\alpha$.

To summarize our result, given any future-directed timelike vector field $t^\alpha$ and any constant ``spin parameter'' $s \epsilon$, the MP equations reduce to the spin Hall equations, at least up to terms of order $\epsilon^2$, when:
\begin{enumerate}
	\item The worldline parameter $\tau$ is chosen such that
\begin{equation}
    \dot{x} \cdot t = p \cdot t
    \label{tauCond}
\end{equation}
	for all time.
	
	\label{cond1}
	
	\item The momentum $p_\alpha$ is at least initially null.
	
	\item The angular momentum satisfies
	\begin{equation}
		S^{\alpha \beta} p_\beta = 0	
		\label{Sp}
	\end{equation}
	at least initially, and
	\begin{equation}
	    S^{\alpha \beta} t_\beta = 0
    	\label{Sortho}
	\end{equation}
	for all time.
    \label{cond3} 
    
	\item The magnitude of the angular momentum is at least initially given by
	\begin{equation}
    	S^{\alpha \beta} S_{\alpha \beta} = 2 (s \epsilon)^2.
    \label{Smag}
\end{equation}

\label{cond4}
\end{enumerate}

The MP equations are reparametrization-invariant, so no generality is lost by imposing condition \ref{cond1} for all time, at least so long as the worldline is not orthogonal to $t^\alpha$. Equation~\eqref{tauCond} serves merely to use the object's energy to nondimensionalize the time parameter.

The interpretation of $S^{\alpha \beta} t_\beta = 0$, also known as the Corinaldesi-Papapetrou spin supplementary condition \cite{CP_ssc, Costa2015}, is more substantial. As discussed in more detail in Sec.~ \ref{Sect:St} below, it is an implicit definition for $x^\alpha$. The angular momentum of any object depends on the choice of origin\footnote{Recall that this is true even in Newtonian physics.}, and certain components of $S^{\alpha\beta}$ can always be eliminated by an appropriate choice of origin. Here, $x^\alpha$ is chosen to eliminate $S^{\alpha \beta} t_\beta$, which is proportional to the body's mass dipole moment with respect to an observer whose 4-velocity is tangent to $t^\alpha$. This definition allows $x^\alpha$ to be interpreted as a kind of centroid. However, that centroid clearly depends on $t^\alpha$. Different choices for $t^\alpha$ generically result in different centroids, and each of these is in principle observable. The different centroids represent slightly different notions of ``center'' for an extended object. Relations between them are discussed in Secs. \ref{Sect:dCentroids} and \ref{Sec:Loc} below.

A particular worldline can be fixed by choosing a particular $t^\alpha$. There should therefore exist an evolution equation for the tangent vector $\dot{x}^\alpha$ to that worldline. To derive that evolution equation, first combine \eqref{MPS}, \eqref{tauCond}, and \eqref{Sortho} to see that
\begin{align}
    0 &= \frac{D}{d\tau} \left( S^{\alpha \beta} t_\beta \right) 
    \nonumber \\
    &= (p \cdot t) (p^\alpha- \dot{x}^\alpha  ) + S^{\alpha \beta} \dot{x}^\gamma \nabla_\gamma t_\beta.
    \label{dSt}
\end{align}
Rearranging then results in the momentum-velocity relation
\begin{equation}
	\dot{x}^\alpha = p^\alpha + \frac{ 1 }{ p \cdot t } S^{\alpha \beta} \dot{x}^\gamma \nabla_\gamma t_\beta.
	\label{dotx}
\end{equation}
It follows that $\dot{x}^\alpha$ and $p^\alpha$ are not necessarily collinear. As long as the operator $\delta^\alpha_\beta - (p \cdot t)^{-1} S^{\alpha\gamma} \nabla_\beta t_\gamma$ can be inverted, \eqref{dotx} determines $\dot{x}^\alpha$ uniquely in terms of  $p^\alpha$, $S^{\alpha \beta}$, and $t^\alpha$. In the small-angular momentum context considered here, the invertibility requirement is trivially satisfied as long as $p \cdot t$ is not too small. In particular, $t^\alpha$ cannot be null and proportional to $p^\alpha$. This is discussed further in Sec.~\ref{Sect:St} below.

With the centroid fixed by \eqref{Sortho}, the superficially similar spin constraint \eqref{Sp} plays a very different role: It is interpreted as a genuine physical restriction on the types of systems which can be described by the spin Hall equations. Combining \eqref{Sp} and \eqref{Sortho} with \eqref{Smag} shows that the angular momentum is at least initially\footnote{The given constraints determine $S^{\alpha\beta}$ only up to an overall sign. Here we fix the sign in order for the MP and spin Hall equations to agree, which may also be viewed as fixing the definition for the sign of $s$.}
\begin{align}
    S^{\alpha \beta} = \epsilon s \Sigma^{\alpha\beta}
    = \frac{\epsilon s}{p \cdot t} \varepsilon^{\alpha \beta \gamma \lambda} p_\gamma t_\lambda.
    \label{S}
\end{align}
The spin constraint $S^{\alpha \beta} p_\beta = 0$ therefore amounts to a particular choice of initial condition for the angular momentum tensor. It implies that the spin is purely longitudinal. As explained in Sec.~\ref{Sect:PhGrad} below, this is consistent with a wide class of high-frequency electromagnetic wave packets. However, it is also shown there that there are reasonable high-frequency wave packets which are not consistent with \eqref{S}. It is a genuine physical restriction on the types of wave packets which can be described by the spin Hall equations.

Our next task is to show that if $S^{\alpha\beta}$ is initially given by \eqref{S}, it retains that form for all time. This can be demonstrated by showing that the constraints \eqref{Sp} and \eqref{Smag}, as well as the null character of $p_\alpha$, are preserved under time evolution. First consider the null character of $p_\alpha$. If we assume that $S^{\alpha\beta}= \mathcal{O}(\epsilon)$ for all time, \eqref{MPp} and \eqref{dotx} immediately imply that
\begin{equation}
	\frac{d}{d\tau} (p^\alpha p_\alpha) = \mathcal{O}(\epsilon^2).
\end{equation}
If $p_\alpha p^\alpha$ is initially zero, it can therefore grow to be at most of order $\epsilon^2$. Again applying the MP equations and the momentum-velocity relation,
\begin{align}
    \frac{D}{d\tau} ( S^{\alpha \beta} p_\beta ) = - \frac{1}{ p \cdot t } \left[ (S^{\gamma \beta} p_\beta) \dot{x}^\lambda \nabla_\lambda t_\gamma \right] p^\alpha + \mathcal{O}(\epsilon^2).
\end{align}
If $S^{\alpha \beta} p_\beta$ is initially zero, as is assumed in condition \ref{cond3} above, it therefore remains zero up to terms of order $\epsilon^2$. Equation~\eqref{Sp} is thus preserved under time evolution. Lastly, use of these results together with \eqref{MPS} shows that 
\begin{equation}
	\frac{d}{d\tau} (S^{\alpha \beta} S_{\alpha \beta}  ) = \mathcal{O}(\epsilon^3).
\end{equation}
This implies that the initial spin magnitude \eqref{Smag} is  preserved under time evolution. Combining these results shows that $S^{\alpha\beta}$ retains the form \eqref{S} for all time, at least up to terms of order $\epsilon^2$. 

The spin Hall equations \eqref{eq:gshe_eq} now follow, up to terms of order $\epsilon^2$, by substituting \eqref{S} into \eqref{MPp} and \eqref{dotx}. They may be viewed as the MP equations \eqref{MP} specialized to conditions \ref{cond1}-\ref{cond4} above. This result can also be established by directly showing that \eqref{S} is a solution to \eqref{MPS} and then using that to deduce the momentum-velocity relation \eqref{dotx} \cite[Sec.~2.4.4.]{Oanceathesis}. Regardless, the spin Hall equations of motion are equivalent to the equations of motion satisfied by a massless dipolar particle.

\subsection{The meaning of the momentum}
\label{Sec:pDef}

Now that we have established that the spin Hall equations follow from the MP equations, results known for the latter may be applied to the former. It is natural to ask what exactly is meant by the $p_\alpha$ and the $x^\alpha$ which appear in the spin Hall equations. The fundamental object in classical electromagnetism is the electromagnetic field, so there must be a relation between that field and (say) the momentum. Such a relation is not necessarily clear from the derivation of the spin Hall equations in Ref.~\cite{GSHE2020}. However, the MP equations can be derived by first defining $p_\alpha$ and $S^{\alpha\beta}$ as integrals over an object's stress-energy tensor and then using stress-energy conservation to deduce the evolution equations for those quantities \cite{Dixon70a, Dixon74, HarteReview}. Imposition of a centroid condition then provides a definition for $x^\alpha$ in terms of the underlying stress-energy tensor. To summarize, the field can be used to construct the stress-energy tensor, which can in turn  be used to construct the momenta and the centroid. 

As the spin Hall equations are special cases of the MP equations, we may identify momenta in the former with momenta in the latter. There are however subtleties. In particular, different definitions for the momenta may satisfy formally identical evolution equations. This is especially clear when the definitions differ by terms which are considered ``higher order.'' However, it can also occur in other cases. For example, if the triple $( x^\alpha, p_\alpha, S^{\alpha\beta} )$ satisfies the MP equations together with an appropriate centroid condition, so does $( x^\alpha, c p_\alpha, c S^{\alpha\beta} )$, where $c$ is any nonzero constant. It follows that at best, the momenta in the two frameworks can be identified only up to an overall constant. 

Despite this, we choose to interpret the momenta in the spin Hall equations to be exactly those which are typically used in derivations of the MP equations and their generalizations: If the object of interest has stress-energy tensor $T^{\alpha\beta}$, and if that object's worldtube is foliated by the 1-parameter family of hypersurfaces $\Sigma_\tau$, the linear and angular momenta at time $\tau$ are given by \cite[Eqs.~(5.1) and (5.2)]{Dixon70a}
\begin{subequations}
\label{pDef}
\begin{align}
    p_\alpha &= \int_{\Sigma_\tau} K_{\alpha}{}^{\alpha'} T^{\beta'}{}_{\alpha'} dS_{\beta'},
    \\
    S^{\alpha\beta} &= 2 \int_{\Sigma_\tau} \sigma^{[\beta}  H^{\alpha]\alpha'} T^{\beta'}{}_{\alpha' } dS_{\beta'}.
    \label{SDef}
\end{align}
\end{subequations}
Unprimed indices here are associated with $x^\alpha (\tau)$, which is assumed to lie in $\Sigma_\tau$. Primed indices are associated with the integration point $x'$. The bitensors $K_{\alpha}{}^{\alpha'} (x, x')$ and $\sigma^\beta (x, x' ) H^{\alpha \alpha'} (x, x')$ are Jacobi propagators; they can be used to form a basis for solutions to the geodesic deviation (or Jacobi) equation along the geodesic segment which connects $x$ to $x'$. The Jacobi propagators can be computed explicitly \cite{Dixon70a} using  derivatives of Synge's world function $\sigma(x, x')$, which is defined to be one half of the squared geodesic distance between its arguments \cite{Synge1960, Poisson2011}. In terms of this world function, $\sigma_\alpha = \nabla_\alpha \sigma$ and 
\begin{equation}
   H^{\alpha\alpha'} = [- \nabla_{\alpha'} \sigma_\alpha ]^{-1}, \qquad K_{\alpha}{}^{\alpha'} = H^{\beta \alpha'} \nabla_\alpha \sigma_\beta,
\end{equation}
where $[\ldots]^{-1}$ denotes an inverse operation. 

In flat spacetime and in inertial coordinates, the above propagators reduce to
\begin{equation}
    K_{\alpha}{}^{\alpha'} = \delta^{\alpha'}_\alpha, \qquad 
    \sigma^{\beta}  H^{\alpha\alpha'} = ( x^{\beta} - x'^{\beta} ) \eta^{\alpha\alpha'},
    \label{bitensorFlat}
\end{equation}
where $\eta^{\alpha\alpha'} = \mathrm{diag}(-1,1,1,1)$ is the usual Minkowski metric. Substituting these expressions into \eqref{pDef} recovers the standard special-relativistic definitions \cite{MTW, DixonSR} for the linear and angular momenta. Even in curved spacetimes, the special relativistic expressions for the momenta remain good approximations to the exact expressions \eqref{pDef} if the coordinates are taken to be Riemann normal coordinates with origin $x^\alpha$.

\subsection{Conservation laws}
\label{Sec:consLaws}

The specific choice of propagators appearing in the momenta \eqref{pDef} may appear to be obscure. They were chosen, in part, so that if $\kappa^\alpha$ is Killing, 
\begin{equation}
    \int_{\Sigma_\tau} T^{\beta'}{}_{\alpha'} \kappa^{\alpha'} dS_{\beta'} = p_\alpha \kappa^\alpha + \frac{1}{2} S^{\alpha \beta} \nabla_\alpha \kappa_\beta.
    \label{p_Int}
\end{equation}
This relation is exact. It is useful because the integral on the left-hand side---which is now identified as a linear combination of the linear and angular momenta---is conserved.

In fact, if $\kappa^\alpha$ is interpreted not as an ordinary Killing field, but as a \textit{generalized} Killing field\footnote{A generalized Killing field requires for its construction a choice of worldline and a foliation \cite{HarteSyms, HarteReview}. Each generalized Killing field is exactly Killing on the worldline in the sense that $\mathcal{L}_\kappa g_{ab} = \nabla_a \mathcal{L}_\kappa g_{bc} =0$ there. Away from the worldline, the generalized Killing fields are exact symmetries for  separation vectors (defined via the exponential map) rather than for the metric itself. Although the generalized Killing fields do not necessarily satisfy Killing's equation there, they do satisfy certain projections of it.}, the momentum definitions \eqref{pDef} ensure that \eqref{p_Int} remains valid in any spacetime; cf. Refs.~\cite{HarteSyms, HarteReview}. The space of generalized Killing fields is always ten dimensional in four spacetime dimensions. It also includes all ordinary Killing fields which may exist. If a generalized Killing field $\kappa^\alpha$ is not an ordinary Killing field, $p_\alpha \kappa^\alpha + \frac{1}{2} S^{\alpha \beta} \nabla_\alpha \kappa_\beta$ is not necessarily conserved, at least exactly. However, that quantity is \textit{approximately} conserved in the pole-dipole context in which the quadrupole and higher-order moments of a body are neglected. That is, the approximation in which the MP equations hold, and indeed, those equations are equivalent to the statement that
\begin{equation}
    p_\alpha \kappa^\alpha + \frac{1}{2} S^{\alpha\beta} \nabla_\alpha \kappa_\beta = \mathrm{const}.
    \label{consLaw}
\end{equation}
for all generalized Killing fields $\kappa^\alpha$. There are ten independent constants associated with the ten generalized Killing fields, and these completely determine the four components of $p_\alpha$ and the six components of $S^{\alpha \beta}$. The coupling of the linear and angular momenta in the MP equations \eqref{MP} is interpreted as a consequence of the fact that, e.g., a local rotation about one point on the worldline is equivalent to both a rotation and translation when viewed at another point on the worldline.

Regardless, \eqref{consLaw} is most useful when $\kappa^\alpha$ is an ordinary Killing field. In that case, the associated conservation law can be simplified in the spin Hall context. There, the angular momentum is given by Eq.~\eqref{S} so 
\begin{equation}
    p_\alpha \left[ \kappa^\alpha + \left( \frac{  \epsilon s }{ 2p \cdot t } \right) \varepsilon^{\alpha \beta\gamma\lambda} t_\beta \nabla_\gamma \kappa_\lambda \right] = \mathrm{const}. 
    \label{consLaw2}
\end{equation}
A particular component of the linear momentum is therefore conserved. Precisely which component is conserved generically depends on both the spin magnitude $\epsilon s$ and on the choice of $t^\alpha$.

There are, however, cases where the $s$-dependent terms vanish in Eq.~\eqref{consLaw2}. First, the Killing field may be covariantly constant. This occurs for all translations in Minkowski spacetime, and also for null translations along the direction of gravitational wave propagation in \textit{pp}-wave spacetimes. For different reasons, there can be no spin correction to the conservation of energy in static spacetimes, at least when $t^\alpha$ and $\kappa^\alpha$ are both identified with the static Killing field. The spin-dependent term in \eqref{consLaw2} will then be proportional to $t_{[\alpha} \nabla_{\beta} t_{\gamma]}$, which vanishes on account of the spacelike hypersurface which is orthogonal to $t^\alpha$ in that context. Statements of energy conservation in the Schwarzschild spacetime are therefore independent of spin in the spin Hall context. This is not the case in Kerr spacetimes with nonzero angular momentum, which are stationary but not static. 

We have only discussed conservation laws associated with Killing vector fields. Other conserved quantities, associated with the existence of Killing-Yano tensors, are known (at least approximately) for the massive MP equations coupled to appropriate centroid conditions \cite{rudiger1981conserved, rudiger1983conserved, Gibbons1993, MPD_conservation_2019, MPD_conservation_2020, MPD_conservation_2021}. However, we have not investigated whether or not these laws also hold for the massless case of interest here.

\subsection{Neglected terms}
\label{Sec:Approx}

The spin Hall equations \eqref{eq:gshe_eq} are expected to be valid only through first order in the ``small'' (although dimensionful) parameter $\epsilon$. Terms nonlinear in the spin have been ignored, as have any contributions from the quadrupole and higher-order moments of a wave packet's stress-energy tensor. Dixon has however found all multipolar corrections to the MP equations \cite{Dixon74}, and using his results, the neglected terms in the spin Hall equations can be estimated. We now discuss how to perform these estimates and under which conditions the spin Hall equations can be justified.

Dixon's laws of motion may be found in, e.g., \cite[Eqs.~(13.7) and (13.8)]{Dixon74}. See also \cite[Eqs.~(283), (284), and (290)]{dixon2015new} for a version of those laws which is truncated at quadrupolar order. Inspecting them shows that the MP evolution equation \eqref{MPp} for $p_\alpha$ is corrected by a force term proportional to  
\begin{equation}
    J^{\beta \gamma \lambda \rho} \nabla_\alpha R_{\beta \gamma \lambda \rho},
\end{equation}
where $J^{\beta \gamma \lambda \rho}$ denotes the quadrupole moment of the wave packet's stress-energy tensor. The evolution equation \eqref{MPS} for $S^{\alpha\beta}$ is corrected as well, acquiring a torque term proportional to 
\begin{equation}
    J^{\gamma \lambda \rho [\alpha} R^{\beta]}{}_{\gamma \lambda \rho}.
    \label{quadTorque}
\end{equation}
There are further forces and torques which couple a wave packet's octupole and higher-order moments to higher-order derivatives of the Riemann tensor, and together, these corrections provide a complete description for the evolution of the linear and angular momentum along a given worldline. If the worldline is fixed by adopting the centroid condition \eqref{Sortho}, there is, in addition, an evolution equation for $x^\alpha$ which generalizes the spin Hall equation \eqref{eq:xdot}. That generalization differs from its spin Hall counterpart due to the presence of terms involving the quadrupole and higher-order moments, as well as terms which are nonlinear in $S^{\alpha\beta}$.

To begin to estimate the consequences of neglecting all of these corrections to the spin Hall equations, it is first necessary to introduce a number of scales. To begin, the background spacetime is assumed to be characterized by a radius of curvature $\lcurv$ and a scale $\lgrad$ over which that radius varies,
\begin{equation}
    R_{\alpha \beta \gamma \lambda} \sim \frac{ 1 }{  \lcurv^2 }, \qquad \nabla_\rho R_{\alpha \beta \gamma \lambda} \sim \frac{ 1 }{ \ell_R^2 \lgrad}.
    \label{lRDef}
\end{equation}
These estimates, and all the similar ones below, are assumed to hold in a locally inertial frame which is instantaneously at rest with respect to $t^\alpha$, the vector field chosen to fix the centroid. Another important scale is $\ell_t$, which characterizes variations in $t^\alpha$,
\begin{equation}
    \frac{ \nabla_\alpha t_\beta }{ \sqrt{-t^2 } } \sim \frac{ 1 }{ \ell_t} .
\end{equation}
The length scales $\ell_R$, $\ell_{\nabla R}$, and $\ell_t$ characterize the external environment.

We now introduce several additional scales which characterize the wave packet itself. The first of these is the energy
\begin{equation}
    E \equiv - \frac{ p\cdot t }{ \sqrt{-t^2} }.
    \label{Energy}
\end{equation}
In terms of a wave packet's characteristic frequency $\omega$, it is suggested by \eqref{omegaDef} [and by \eqref{omegaDef2} below] that $E = \epsilon \omega$. Another relevant scale is provided by the angular momentum. Using \eqref{Smag}, this is given by
\begin{equation}
    S^{\alpha\beta} \sim \epsilon |s| = \frac{ |s| }{ \omega } E.
\end{equation}
While the derivation of the spin Hall equations in \cite{GSHE2020} suggested that  $s = \pm 1$ for circularly polarized electromagnetic wave packets, we shall see in Sec.~\ref{sec:wavepackets} below that this is true only when the wave packet has a relatively simple structure; it must have ``spin angular momentum'' but not ``orbital angular momentum.'' More generally, it follows from the integral expression \eqref{SDef} for $S^{\alpha\beta}$ that a rough bound is given by $|s|  < \omega \lw$, where $\lw$ denotes a characteristic width for the wave packet. Below, we assume that $|s|$ remains well below this bound in order not to violate the high-frequency approximation. But even so, it may still be that $|s| \gg 1$.

Our final estimate involves the quadrupole moment $J^{\alpha \beta \gamma\lambda}$. Unlike $p_\alpha$ and $S^{\alpha\beta}$, this rescales under changes in the worldline parameter. Using the dimensionless parameter $\tau$ which is associated with the normalization condition \eqref{tauCond}, it may be shown that the quadrupole moment has dimension $(\mbox{length})^4$. It is generically of order
\begin{equation}
    J^{\alpha \beta \gamma \lambda} \sim (E \lw)^2.
\end{equation}
Magnitudes of the higher-order moments are essentially the same except for the involvement of higher powers of $\lw$. Regardless, these scalings imply that a wave packet can be characterized by $s$, $\omega$, $E$, and $\lw$.

The approximations inherent in the spin Hall equations may now be summarized as
\begin{enumerate}
    \item Terms nonlinear in the spin can be neglected in the momentum-velocity relation: $\ell_t \gg |s|/\omega$.
    \label{assume1}
    
    \item The instantaneous quadrupole force can be neglected in comparison with the spin-curvature contribution to the linear momentum evolution: $\ell_w^2 \ll (|s|/\omega) \ell_{\nabla R}$.
    \label{assume2}
    
     \item The instantaneous quadrupole torque can be neglected in comparison with the $p^{[\alpha} \dot{x}^{\beta]}$ contribution to the angular momentum evolution: $\ell_w^2 \ll (|s|/\omega) \ell_R^2/\ell_t $.
    \label{assume3}
    
    \item The quadrupole torque negligibly affects the spin over the dimensionless integration timescale $\Delta \tau$: $E \Delta \tau \ll (|s|/\omega) (\ell_R/\ell_w)^2 $.
    \label{assume4}
    
\end{enumerate}
These constraints are all related to terms which are neglected when going from Dixon's laws of motion to the MP equations, and finally to the spin Hall equations. Separately, it is also necessary to assume that 
\begin{enumerate}
\setcounter{enumi}{4}
    \item The wave packet is large compared with its wavelength and it does not have nontrivial structure on very small scales, $\omega \lw \gg |s|$.
    \label{assume5}
\end{enumerate}
This is required for the approximate validity of geometric optics, which was used in the derivation of the spin Hall equations in \cite{GSHE2020}. Alternatively, if the MP equations are used as a starting point, geometric optics must be used to motivate the initial data considered here---for example the null character of $p_\alpha$. This viewpoint is discussed further in Sec.~\ref{sec:wavepackets} below. Regardless, condition 5 must be imposed in order for a pulse to maintain its structure. If it were violated, a wave packet would rapidly diffract away. 

Let us now examine the consequences of assumptions \ref{assume1}--\ref{assume5}. The first of these implies that $t^\alpha$ must vary sufficiently slowly that $\ell_t \gg |s| (\mbox{wavelengths})$. Assumptions \ref{assume1}, \ref{assume2}, \ref{assume3}, and \ref{assume5} imply that the wave packet must be small compared to the curvature scales,
\begin{equation}
    \lw \ll \min ( \ell_R, \ell_{\nabla R} ).
    \label{lCurv}
\end{equation}
This guarantees that, e.g., the octupole terms in the laws of motion are negligible compared with quadrupole terms. It is therefore unnecessary to impose that restriction separately from the ones above. 

Regardless, \eqref{lCurv} does not exhaust the content of the first three assumptions. When $|s| \sim 1$, for example, they imply much stricter bounds on $\ell_w$. Writing those bounds in dimensionless form while also incorporating assumption \ref{assume5},
\begin{equation}
    1 \ll \omega \lw \ll  \min \left[ \ell_R (\omega / \ell_t)^{ \frac{1}{2} }, (  \omega \ell_{\nabla R} )^{ \frac{1}{2} } \right].
    \label{omegaEll}
\end{equation}
In this sense, $\lw$ cannot be either too large or too small when compared with one wavelength.

Assumptions \ref{assume1}, \ref{assume2}, and \ref{assume3} arise from comparing the instantaneous magnitudes of different terms in the equations of motion. Assumption \ref{assume4} tells us how long those equations can be reliably integrated. If $\ell_w$ is approximately constant, it implies that the integrations remain valid over (dimensionful) timescales---or equivalently distances---of order
\begin{equation}
    \Delta t \equiv E \Delta \tau \ll \frac{ |s| }{ \omega } (\ell_R/\lw)^2 < \ell_R^2/\lw.
    \label{lBound1}
\end{equation}
Note that the allowable integration time here is much smaller when $|s| \sim 1$ than it is for a maximally spinning wave packet. This is because smaller spin effects are more easily overwhelmed by quadrupole corrections.

One subtlety in this discussion is that it is not necessarily justified to assume that $\lw$ remains constant over an integration timescale. Electromagnetic wave packets almost\footnote{There are nondiverging beams in flat spacetime, but these must be specially prepared and most do not decay rapidly enough to have well-defined momenta. In a curved spacetime, it is likely that except in very special circumstances, beam divergence will be even more rapid due to the defocusing of null geodesics.} invariably diffract and spread out as they propagate. This differs from the behavior of (some) solids and strongly self-gravitating fluids, which can---at least approximately---maintain their dimension over long timescales. We estimate the divergence of an electromagnetic wave packet by analogy with a Laguerre–Gauss beam in flat spacetime. If such a beam has minimum width $\ell^*_{w}$, its width at a distance $L \gg \ell^*_{w}$ away from where that minimum occurs is of order \cite{spotWidth}
\begin{equation}
    \lw \sim \frac{ L |s| }{ \omega \ell^*_{w} }.
    \label{lwDiverge}
\end{equation}
We assume that this relation holds not only for Laguerre–Gauss beams, but generically. Further assuming that the equations of motion are integrated beginning near the point where the beam has attained its minimum width, so $L \sim \Delta t$, substituting \eqref{lwDiverge} into \eqref{omegaEll} while also using \eqref{lBound1} shows that
\begin{equation}
     \Delta t \ll  \min \left[ \left( \frac{ \omega \ell_R^2 }{ \ell_t } \right)^{ \frac{1}{2} }  , \left( \frac{ |s|^{ \frac{1}{2} } \omega \ell_R^2 }{  \ell^*_{w} } \right)^{ \frac{1}{3} },   (\omega \ell_{\nabla R})^{ \frac{1}{2} }  \right] \frac{ \ell^*_{w} }{ |s|^{ \frac{1}{2} }  } .
     \label{LConstr2}
\end{equation}
Increasing $\omega$ at fixed $\ell^*_{w}$ therefore increases the upper bound on the integration time. This is consistent with what might have been expected from improving the high-frequency approximation. However, it is not possible to increase $\Delta t$  indefinitely: As shown by, e.g., \eqref{omegaEll} in the $|s| \sim 1$ case, increasing $\omega$ results in a decreasing upper bound on $\ell^*_{w}$. Also note that the maximum allowable integration time decreases when $|s| \gg 1$. 

One example which may be considered is that of a wave packet propagating at a distance $r$ from a static gravitating object of mass $M$. In this case, the curvature scales are $\lcurv \sim r (r/M)^{1/2}$ and $\lgrad \sim r$. Furthermore, if $t^\alpha$ is chosen to be parallel to the static Killing field, $\ell_t \sim r (r/M)$. This implies that $\ell_t \gg \lcurv \gg \lgrad$ in an approximately Newtonian regime. In that regime, it follows from assumptions \ref{assume2}, \ref{assume3}, and \ref{assume5} that the minimum beam width is bounded by
\begin{equation}
    1 \ll \omega \ell^*_{w} \ll (|s| \omega r)^{ \frac{1}{2} }.
\end{equation}
It also follows from \eqref{LConstr2} that if $r$ does not change too much over the integration time,
\begin{equation}
    \Delta t \ll  \min \left[ \left( \frac{ \omega r }{ |s| } \right)^{ \frac{1}{2} }, \left( \frac{ \omega  }{ |s| M \bar{\ell}_w } \right)^{ \frac{1}{3} } r \right] \ell^*_{w}.
\end{equation}
Both bounds together imply that $\Delta t \sim L \ll r$, which significantly limits the applicability of the spin Hall equations in astrophysical systems.

Except for assumption \ref{assume5} above, our discussion has focused only on neglected terms in the spin Hall equations of motion. However, there are separate errors incurred by using inaccurate initial data in those equations. As discussed in Sec.~\ref{Sec:MPtoSH}, it is assumed in the spin Hall context that $p_\alpha$ is null, $S^{\alpha\beta} S_{\alpha\beta} > 0$, and $S^{\alpha\beta} p_\beta = 0$. However, we show in the Appendix that in fact, there does not exist any exact wave packet with these properties. An electromagnetic field with nonzero angular momentum must have a timelike momentum, not a null one. Nevertheless, there are large classes of electromagnetic fields for which the spin Hall initial data is \textit{approximately} valid, and it is in that context that the spin Hall equations should be understood. We do not, however, attempt to estimate the errors incurred by this aspect of the approximation.

\section{The many centroids of extended objects}
\label{sec:role_of_t}

Whether an extended object is composed of ``ordinary'' matter, electromagnetic fields, or anything else, it is not possible to fully describe its location using only a single worldline. There are nevertheless situations in which it is useful to use a single worldline to describe the ``averaged'' location of an extended object. This is the role of a centroid. However, unlike in Newtonian mechanics, there are many centroids which might reasonably be associated with relativistic systems. One of these centroids might be more useful in one context, while another might be more useful in another context. The various centroids may be interpreted as a particular class of observables. 

The centroids considered here are associated with timelike vector fields. As stated in Sec.~\ref{Sec:MPtoSH} above, any such vector field may be associated with a centroid by requiring that $S^{\alpha\beta} t_\beta = 0$. This interpretation is verified in Sec.~\ref{Sect:St} below. Distinctions between the various centroids and the implications of those distinctions are discussed in Secs. \ref{Sect:dCentroids}--\ref{sec:CS_change_of_t}.

Many of the ideas described in this section were introduced long ago by, e.g., Pryce \cite{PryceCM} and M\o{}ller \cite{MollerLectures}. Some of those ideas have been rediscovered more recently by different communities, who have introduced different terminologies and interpretations. For example, displacements between different centroids have, in certain contexts, been described as relativistic Hall effects \cite{Relativistic_Hall} and also as Wigner \cite{Stone2015} or Wigner-Souriau \cite{Duval_chiral_fermions, DUVAL2015} translations. Regardless, properties of different centroids are well-understood for massive objects, where $p_\alpha$ is timelike.

What has not been so carefully explored in the literature is the massless case, where $p_\alpha$ is null. This section discusses both the massive and massless cases together. Our main new finding is concerned with the maximum possible separation between different centroids associated with the same physical object. For massive bodies, we recover the classical result \cite{MollerLectures, DixonSR, Herdeiro2012, Costa2015} that all possible centroids are confined to a disk with finite radius. The set of all centroids therefore localizes a massive object to a finite region, providing some reassurance that the centroid definition is a reasonable one. The massless case is different, however. We find that \textit{massless spinning objects cannot be localized} in this way; they possess centroids separated by arbitrarily large distances. This is potentially problematic, and resolving it involves examining certain subtleties of the approximations used to describe, e.g., electromagnetic wave packets. Our conclusion is that the delocalization of massless objects is not physically relevant because a wave packet cannot truly be massless.

The strategy taken in this section is to first discuss all issues in flat spacetime and in inertial coordinates. All arguments are then straightforward and all results are exact. There are no subtleties involving neglected higher-order terms in the laws of motion. Later, in Sec.~\ref{sec:CS_change_of_t}, we discuss how---with appropriate caveats---the same results carry over for sufficiently small objects in curved spacetime.

\subsection{Defining a centroid} \label{Sect:St}

Our first task is to show that, as claimed above, the choice of $t^\alpha$ is equivalent to a choice of worldline. For simplicity, we work in flat spacetime and use inertial coordinates.

To begin, recall that the definitions \eqref{pDef} for an object's linear and angular momentum supposed that a particular worldline had been fixed and that  $p_\alpha$ and $S^{\alpha\beta}$ depended only upon a parameter $\tau$ which had been associated with that worldline. Those definitions are easily generalized to avoid the introduction of any particular worldline. Instead, if the hypersurfaces $\Sigma_\tau$ are replaced by $\Sigma_x$, where $x^\alpha$ is now an arbitrary point (not yet associated with any particular worldline), $p_\alpha$ and $S^{\alpha\beta}$ may be viewed as functions of that point. With this redefinition in mind, as long as the $\Sigma_x$ foliate the support of the stress-energy tensor, stress-energy conservation implies that the left-hand side of \eqref{p_Int} must be independent of $x^\alpha$ for each Killing field $\kappa^\alpha$. The quantities $p_\alpha \kappa^\alpha + \frac{1}{2} S^{\alpha \beta} \nabla_\alpha \kappa_\beta$ are therefore conserved in the sense that they are independent of $x^\alpha$. Using this together with the fact that the flat spacetime Killing fields can be written as $\kappa^\alpha = \mathcal{T}^\alpha + \mathcal{B}^{\alpha\beta} x_\beta$, where the translation $\mathcal{T}^\alpha$ and the rotation or boost $\mathcal{B}^{\alpha\beta} = \mathcal{B}^{[\alpha\beta]}$ are arbitrary constants, the linear and angular momenta associated with two different points, $x$ and $\tilde{x}$, must be related via
\begin{subequations}
\begin{align}
    p_\alpha (\tilde{x}) &= p_\alpha(x), \\
    S^{\alpha\beta}(\tilde{x}) &= S^{\alpha\beta}(x) + 2 (x-\tilde{x})^{[\alpha} p^{\beta]}(x).
    \label{dS}
\end{align}
\end{subequations}
This describes how the linear and angular momenta transform under a shift of origin. In flat spacetime, these relations are exact. They have the same form as the Wigner-Souriau translations which arose in the study of chiral fermions in Refs.~\cite[Eq.~3.7]{Duval_chiral_fermions} and \cite[Eq.~2.7]{DUVAL2015}. They could also have been derived straightforwardly from the momentum definitions \eqref{pDef} as well as \eqref{bitensorFlat}.

Defining $\tilde{S}^{\alpha\beta} \equiv S^{\alpha\beta} (\tilde{x})$ together with the deviation vector $\xi^\alpha \equiv \tilde{x}^\alpha-x^\alpha$, it follows from \eqref{dS} that
\begin{equation}
    S^{\alpha\beta} t_\beta = \tilde{S}^{\alpha\beta} t_\beta - (-p \cdot t) \xi^\alpha - (\xi \cdot t) p^\alpha .
    \label{dx0}
\end{equation}
The claimed centroid condition \eqref{Sortho} now amounts to the vanishing of the left-hand side of this equation. And no matter how $\tilde{x}^\alpha$ has been chosen or what form $\tilde{S}^{\alpha\beta}$ may have, that can be arranged by choosing $x^\alpha$ such that
\begin{equation}
    x^\alpha = \tilde{x}^\alpha + T p^\alpha - \frac{ 1 }{ (- p \cdot t)} \tilde{S}^{\alpha\beta} t_\beta  , 
    \label{xCent}
\end{equation}
where $T$ is an arbitrary parameter. Varying over all possible values of this parameter recovers a worldline: what we call the centroid associated with $t^\alpha$. This is true regardless of whether $p_\alpha$ is timelike or null. It may also be seen that if $\nabla_\beta t^\alpha = 0$, the centroid is tangent to $p^\alpha$ and $T$ may be identified with the worldline parameter $\tau$ which is associated with the normalization condition \eqref{tauCond}. Both of these statements can fail when $t^\alpha$ is not constant.

Next, we verify that the centroid is deserving of its name. First, recall that stress-energy conservation implies that $S^{\alpha\beta}(x)$ does not depend on the hypersurface $\Sigma_x$, as long as all fields fall off sufficiently rapidly and all relevant hypersurfaces completely cut through the support of the stress-energy tensor. We may therefore choose $\Sigma_x$ to be the hyperplane which is orthogonal to $t^\alpha$ at $x^\alpha$. Doing so, while temporarily adopting inertial coordinates which are comoving with $t^\alpha$, use of \eqref{pDef} and \eqref{bitensorFlat} shows that $S^{\alpha\beta} t_\beta = 0$ holds only when
\begin{equation}
    x^i = \frac{1}{E} \int x'^{i} T^{0 0}(x') d^3 x' ,
    \label{xCentFlat}
\end{equation}
where 
\begin{equation}
E = \int T^{0 0}(x') d^3 x'
\end{equation}
is the energy \eqref{Energy}. This is the standard nonrelativistic center of mass definition, but with the nonrelativistic mass density replaced by the relativistic energy density $T_{\alpha\beta} t^\alpha t^\beta$. It follows that as long as $T_{\alpha\beta} t^\alpha t^\beta \geq 0$, the centroid must lie inside the convex hull of the spatial support of the stress-energy tensor. That the energy density should not be negative could be viewed as a consequence of, e.g., the dominant energy condition. That condition is satisfied by essentially all standard classical fields, including electromagnetic ones \cite{SandersEnergy}.

One subtlety which does not appear to have been recognized before is that although the dominant energy condition is satisfied by \textit{exact} electromagnetic field configurations, it is not necessarily satisfied by the \textit{approximate} fields which might be used to describe high-frequency wave packets. As discussed further in Sec.~\ref{Sect:PhGrad} below, there can exist timelike $t^\alpha$ for which the approximate energy density is positive in some regions and negative in others. This has a dramatic consequence: The centroid of an approximate wave packet can appear to lie arbitrarily far from the wave packet itself. Those centroids are of course spurious. They are a consequence of neglecting higher-order terms in the stress-energy tensor. See further discussion in  Secs. \ref{Sec:Loc} and \ref{Sect:PhGrad} below.

Another comment which can be made is concerned with the fact that it is common in the literature \cite{MollerLectures, Dixon74, EhlersRudolph, SchattnerCM, HarteReview} to use $S^{\alpha\beta} p_\beta = 0$ as a centroid condition instead of $S^{\alpha\beta} t_\beta = 0$, particularly---but not exclusively \cite{Mashhoon1975, Duval,  Duval2018, Duval2019}---for objects with timelike momenta. This has the apparent advantage that the results do not depend on extraneous choices such as that of $t^\alpha$. And in the massive case, there is nothing wrong with this; Eq.~\eqref{dotx} remains valid with $t^\alpha = p^\alpha$. But this fails for massless objects. In that case, \eqref{dx0} remains valid so replacing $t^\alpha$ there by $p^\alpha$ shows that $x$ must be a solution to
\begin{equation}
     \tilde{S}^{\alpha\beta} p_\beta = [(\tilde{x} - x) \cdot p] p^\alpha .
\end{equation}
If the left-hand side here is nonzero and not proportional to $p^\alpha$, no such solution exists. If the left-hand side is instead proportional to $p^\alpha$, any $x^\alpha$ which satisfies $(x- \tilde{x}) \cdot p = \mbox{const}$ will do. That restricts the centroid only to a three-dimensional null hypersurface, not a worldline. In either case, $S^{\alpha\beta} p_\beta = 0$ cannot be interpreted as a centroid condition for massless objects. While this has been noted before, details were scant \cite{Stone2015(2), Harte2018}.

In some of the literature which does attempt to use $S^{\alpha \beta} p_\beta = 0$ as a centroid condition for massless objects \cite{saturnini1976modele, Duval, Duval2018, Duval2019}, there is a relation derived between the momentum and the velocity which suggests that a centroid does indeed exist. However, that relation involves a ratio whose denominator (in a curved spacetime) is $R_{\alpha\beta\gamma\lambda} S^{\alpha\beta} S^{\gamma\lambda}$. The momentum-velocity relation therefore fails in the flat spacetime context of our present discussion. It also fails at least somewhere on many worldlines which might be considered in more general spacetimes. It does not appear to us to be viable to attempt to impose a condition which fails to be robust or to have reasonable limits. In particular, the lack of a viable flat-spacetime limit implies that even when $S^{\alpha \beta} p_\beta = 0$ does result in a unique worldline, it will not describe a centroid in the sense of \eqref{xCentFlat}.

\subsection{Displacements between different centroids}
\label{Sect:dCentroids}

As there are many different centroids which may be used to describe an extended object, it is natural to ask how these are related to one another. The answer has long been known for massive objects, as described in, e.g., Refs.~\cite{Moller, DixonSR, Herdeiro2012, Costa2015, vines2016canonical}. There, a canonical centroid was defined via $S^{\alpha\beta} p_\beta = 0$ and separations were derived between this centroid and others. As noted above, a canonical centroid cannot be defined in this way when considering massless objects. Nevertheless, only minor changes are needed to consider the differences between arbitrary reference centroids. We discuss both the massless and massive cases below. For simplicity, we also continue to work in flat spacetime and to use inertial coordinates.

Consider two future-directed timelike vector fields $t^\alpha$ and $\tilde{t}^\alpha$ and the corresponding centroid conditions
\begin{equation}
    S^{\alpha\beta} t_\beta = \tilde{S}^{\alpha\beta} \tilde{t}_\beta  = 0.
    \label{St2}
\end{equation}
These define two worldlines, the points on which may be denoted by $x^\alpha$ and $\tilde{x}^\alpha$. Finding a unique displacement $\xi^\alpha = \tilde{x}^\alpha - x^\alpha$ between them requires that  points on each worldline be identified in a particular way. It is convenient to do so by supposing that 
\begin{equation}
    \xi \cdot t = 0,
    \label{ptIdent}
\end{equation}
in which case \eqref{dx0} and \eqref{St2} imply that $\xi \cdot \tilde{t} = 0$ and
\begin{equation}
    \xi^\alpha = \frac{ S^{\alpha \beta} \tilde{t}_\beta }{ p \cdot \tilde{t} } = -  \frac{ \tilde{S}^{\alpha \beta} t_\beta }{ p \cdot t }.
    \label{xi}
\end{equation}
This displacement vector is exact in flat spacetime, is valid for both massive and massless objects, and there is no constraint on the nature of the angular momentum. One immediate consequence is that all centroids coincide for nonspinning objects.

Given Eq.~\eqref{St2}, it is always possible to introduce a spin vector $S^\alpha$ such that 
\begin{equation}
 S^{\alpha\beta} = \frac{ \varepsilon^{\alpha \beta\gamma \lambda} S_\gamma t_\lambda }{ \sqrt{-t^2}  },
\end{equation}
which is unique only up to arbitrary multiples of $t^\alpha$. In terms of any such spin vector, the displacement \eqref{xi} can be written as
\begin{equation}
    \xi^\alpha = \frac{ \varepsilon^{\alpha \beta\gamma \lambda} S_\beta t_\gamma \tilde{t}_\lambda }{ (p \cdot \tilde{t}) \sqrt{ - t^2} } .
    \label{xiGen}
\end{equation}
This is a spacelike vector orthogonal to $t^\alpha$, $\tilde{t}^\alpha$, and $S^\alpha$. It may be used to relate any two centroids to one another. It still does not make any assumptions regarding the nature of the spin or the object's mass.

If we now specialize to the spin Hall case where $S^{\alpha\beta}$ is given by Eq.~\eqref{S} and $p_\alpha$ is null, the spin vector may be  be identified with
\begin{equation}
    S_\alpha = \epsilon s \left( \frac{\sqrt{-t^2}}{p \cdot t} \right) p_\alpha.
    \label{Svect}
\end{equation}
As this is proportional to $p_\alpha$, it may be described as a ``longitudinal spin.'' The displacements in this case are given by
\begin{equation}
    \xi^\alpha = \epsilon s  \left[ \frac{ \varepsilon^{\alpha \beta \gamma \lambda}  p_\beta t_\gamma \tilde{t}_\lambda  }{(p \cdot t)  (p \cdot \tilde{t})} \right] ,
    \label{xi2}
\end{equation}
which are transverse to the momentum.

\subsection{Localization of extended objects}
\label{Sec:Loc}

As the choice of $t^\alpha$ is essentially arbitrary, one might hope that the centroids associated with different vector fields are not too different. In particular, it is natural to ask if they are all confined to a finite region---perhaps within the convex hull of the spacelike support of the object's stress-energy tensor. As noted above, this does indeed follow from \eqref{xCent} when $T_{\alpha\beta} t^\alpha t^\beta \geq 0$. It is also possible to show, without using $T_{\alpha\beta}$, that the set of all possible centroids is localized whenever $p_\alpha$ is timelike; cf. \cite[Sec.~6.3]{Moller} or \cite[Sec.~3.1b]{DixonSR}. We now discuss both the massive and the massless cases and show that in the latter context, some ``centroids'' can be arbitrarily distant from one another.

Assume that some future-directed timelike $t^\alpha$ has been fixed and measure all deviations as being with respect to the centroid for which $S^{\alpha\beta} t_\beta = 0$. If points on the centroids associated with $\tilde{t}^\alpha$ and $t^\alpha$ are identified using \eqref{ptIdent}, it follows from \eqref{xiGen} that the square of the proper distance between those points is
\begin{equation}
    \xi^2 = \left[ \frac{ (t \cdot \tilde{t})^2 - t^2 \tilde{t}^2 }{ (-t^2) ( p \cdot \tilde{t})^2 } \right] h_{\alpha \beta} S^{\alpha} S^{\beta}, 
    \label{xiLenGen}
\end{equation}
where
\begin{equation}
    h_{\alpha\beta} \equiv g_{\alpha\beta} + \frac{ t^2 \tilde{t}_\alpha \tilde{t}_\beta + \tilde{t}^2 t_\alpha t_\beta - 2 (t \cdot \tilde{t}) t_{(\alpha} \tilde{t}_{\beta)}  }{  (t \cdot \tilde{t})^2 - t^2 \tilde{t}^2 }
\end{equation}
projects vectors into the space orthogonal to both $t^\alpha$ and $\tilde{t}^\alpha$ (when those vectors are not parallel).

To discuss the implications of this in the massive case, it is convenient to now choose $t^\alpha = p^\alpha$ so all deviations are measured with respect to the centroid defined by $S^{\alpha\beta} p_\beta = 0$. Then \eqref{xiLenGen} implies that 
\begin{equation}
    \xi = \frac{ \mathcal{S} }{ m } ( V \sin \theta ),
    \label{xiMassive}
\end{equation}
where $\mathcal{S} \equiv ( \frac{1}{2} S^{\alpha\beta} S_{\alpha \beta} )^{1/2}$ characterizes the magnitude of the spin, $m \equiv ( - p^2)^{1/2}$ is the mass, 
\begin{equation}
   V \equiv [  1 - t^2 \tilde{t}^2 / (t \cdot \tilde{t})^2 ]^{1/2} < 1
\end{equation}
is the relative speed between $t^\alpha$ ($=p^\alpha$) and $\tilde{t}^\alpha$, and $\theta \in [0,\pi]$ is the angle between $\tilde{t}^\alpha$ and $S^{\alpha}$ which would be measured by an observer whose 4-velocity is tangent to $t^\alpha$. It is evident from \eqref{xiMassive} that the magnitude of the centroid displacement can be no larger than the M{\o}ller radius $\mathcal{S}/m$. All centroids are therefore confined to a disk with that radius. Unless energy conditions are violated, there is a sense in which the disk of centroids must be smaller than the object itself.

The massless case is more subtle. For simplicity, we do not discuss  the most general massless case, but only the spin Hall case in which the angular momentum is restricted via $S^{\alpha\beta}p_\beta = 0$. As it is not possible to choose $t^\alpha$ to be proportional to $p^\alpha$ in this context, we assume that $t^\alpha$ and its associated centroid have been fixed in some other way and that all other centroids are measured with respect to it. The distance between the centroid determined by $t^\alpha$ and the one determined by $\tilde{t}^\alpha$ is then found by substituting \eqref{Svect} into \eqref{xiLenGen}.  This yields
\begin{equation}
    \xi = \frac{ \mathcal{S} }{ E } \left( \frac{ V \sin \theta }{ 1- V \cos \theta} \right)
    \label{xiMassless}
\end{equation}
for a massless object, where $E$ is the energy \eqref{Energy}, and $\mathcal{S}$, $V$, and $\theta$ have the same meanings as in \eqref{xiMassive}. In this case, it also follows from \eqref{Smag} that $\mathcal{S} = \epsilon|s|$.

The prefactors are essentially the same in the massless displacement \eqref{xiMassless} and its massive counterpart \eqref{xiMassive}; the energy $E$ which appears in the massless case is simply replaced by the mass $m$, which is of course the energy in the zero-momentum frame. Up to this replacement, both displacements coincide when $V \ll 1$. Indeed, all centroids determined by ``nearly comoving'' observers satisfy $\xi \leq (\mathcal{S} / E) V$. In both the massless and the massive cases, they are contained within the ``generalized M{\o}ller radius'' $\mathcal{S}/E$.

If the magnitude of $V$ is not restricted, the massless and massive displacements still coincide when $\tilde{t}^\alpha$ is aligned, antialigned, or orthogonal to $S^\alpha$ in a frame comoving with $t^\alpha$. In the aligned and antialigned cases, there is no effect at all: $\xi = 0$. In the orthogonal case where $\theta = \pi/2$, we have instead that the proper distance between two centroids is $\xi= ( \mathcal{S} /E) V$. Since $V < 1$, this is again bounded by generalized M{\o}ller radius. In the massless case and for a high-frequency wave packet, \eqref{omegaDef2} below shows that this bound can be written as 
\begin{equation}
    \xi < \frac{ \mathcal{S}   }{E } =  \frac{ |s| }{ \omega } ,
\end{equation}
where $\omega$ denotes the angular frequency of the field in the frame comoving with $t^\alpha$. The displacement is therefore less than approximately $|s|$ wavelengths. It is in agreement with discussions of the relativistic Hall effect \cite{Relativistic_Hall} and the Wigner translations \cite{Stone2015}, where the energy centroid of a beam with nonzero angular momentum was shown to experience a similar shift after applying a boost orthogonal to the direction of propagation. 

What does not appear to have been noticed before is that the maximum displacement in the massless case does not occur at $\theta = \pi/2$ (except in the $V \ll 1$ limit). For fixed $V$, the angle which maximizes $\xi$ in \eqref{xiMassless} is instead 
\begin{equation}
    \theta = \cos^{-1} V.
    \label{thetaMax}
\end{equation}
Using that, the maximum displacement between massless centroids is found to be
\begin{equation}
    \xi = \frac{ \mathcal{S} }{ E } \left( \frac{ V }{ \sqrt{ 1-V^2 } } \right). 
    \label{xiMax}
\end{equation}
This diverges as $V \to 1$. Unlike in the massive case, \textit{the set of all possible centroids is not bounded for a massless spinning body}. Arbitrarily large displacements can occur between the centroids associated with $t^\alpha$ and $\tilde{t}^\alpha$ when those vectors differ by ultrarelativistic boosts which are almost---but not quite---parallel to the momentum. 

This presents an apparent problem for the formalism. One interpretation is simply that massless spinning objects, whatever those may be, cannot be localized. However, this is unacceptable if we interpret certain electromagnetic wave packets as examples of massless spinning objects. Physically realizable wave packets clearly can be localized, and any worldlines which fail to lie near the support of their stress-energy tensors are hardly deserving to be called ``centroids''. Therefore, either our centroid definition is inappropriate or there is something wrong with our interpretation of electromagnetic wave packets as spinning objects with null momenta. The first possibility can be discounted by recalling the discussion following \eqref{xCent}. 

The resolution is that the momentum of a spinning electromagnetic wave packet is not actually null. It must be timelike. We have been assuming above that the momentum is null, and given reasonable assumptions, this is \textit{approximately} true for a high-frequency wave packet. Indeed, it is true through leading and subleading orders in a high-frequency approximation, and that is all that the spin Hall equations can describe (as they omit terms of order $\epsilon^2$). However, it is demonstrated in Sec.~\ref{Sec:zeroPhase} below, using an explicit family of wave packets, that the momentum is always timelike when going to one higher order. The mass is found to be order $\epsilon/\ell_w$, where $\ell_w$ is again a characteristic width for the wave packet. Using Eq.~\eqref{xiMassive}, this implies that the maximum deviation between centroids is of order $\ell_w$. Although that is the intuitively expected result, establishing it requires that calculations be performed to a relatively high order. Truncating the approximation too early results in a conclusion which is not even qualitatively correct.

At lower orders in the high-frequency approximation, one can say only that the momentum is approximately null. Mathematically, we are considering 1-parameter families of wave packets in which, e.g.,  $\lim_{\epsilon \to 0} p^\alpha p_\alpha = 0$. However, what is physically interesting is an example of such a family at a particular (``small'') value of $\epsilon$. In that context, a vector can be ``approximately'' null only with respect to some restricted class of observers. If a vector is actually timelike, for example, there clearly exist some observers for whom it appears to be stationary and some observers for whom it appears to be ``nearly null.'' There is therefore a sense in which the high-frequency approximation implicitly selects a kind of rest frame. It is reliable only in frames which are not too highly boosted with respect to that rest frame.

\subsection{Centroids in curved spacetimes}
\label{sec:CS_change_of_t}

The main results obtained thus far in this section are that i) the displacements between different centroids are given by Eq.~\eqref{xiGen}, ii)  the maximum magnitude of the displacement is given by Eq.~\eqref{xiMassive} when $p_\alpha$ is timelike, and iii)  the displacement is unbounded when $p_\alpha$ is null. These results were derived in flat spacetime, and in that context, they are exact. There are no corrections due to higher-order spin effects, quadrupole moments, or anything else. We now discuss the sense in which our results remain at least approximately valid for sufficiently small objects in generic spacetimes. One would expect from the equivalence principle that everything remains at least approximately valid even in curved spacetimes, and indeed it does.

We begin by obtaining a curved spacetime form for the transformation law \eqref{dS} between the angular momentum evaluated about different points $x^\alpha$ and $\tilde{x}^\alpha$. It is useful to first use the exponential map to define a deviation vector $\xi^\alpha$ between those points, so
\begin{equation}
    \tilde{x}^\alpha = \exp_x \xi^\beta.
\end{equation}
In a Riemann normal coordinate system with origin $x^\alpha$, this takes the standard form $\xi^\alpha = \tilde{x}^\alpha - x^\alpha$. Regardless of the coordinate system, the displacement vector is the negative gradient of Synge's world function: $\xi^\alpha = - \sigma^\alpha (x,\tilde{x})$. We can use this to find covariant Taylor expansions in the style of, e.g., \cite[Sec.~6]{Poisson2011}. Letting primed indices be associated with $\tilde{x}^\alpha$ and unprimed ones with $x^\alpha$, the relevant expansion for the angular momentum tensor is
\begin{equation}
    \tilde{S}^{\alpha' \beta' } = g^{ \alpha' }{}_{\alpha} g^{\beta'}{}_{\beta} [ S^{\alpha \beta}  + \xi^\gamma \nabla_\gamma S^{\alpha\beta} + \mathcal{O}(\xi^2) ],
\end{equation}
where $g^{\alpha'}{}_{\alpha}$ denotes the bitensor which parallel propagates vectors from $x^\alpha$ to $\tilde{x}^\alpha$ along the geodesic segment which connects those points. A similar expansion may also be used to relate the linear momentum at $x^\alpha$ to the linear momentum at $\tilde{x}^\alpha$.

Regardless, continuing requires that we compute the gradients of $p_\alpha$ and $S^{\alpha\beta}$. While the argument can be generalized, consider for simplicity displacements $x^\alpha \mapsto \tilde{x}^\alpha$ which lie entirely within the same ``constant-time'' hypersurface, so $\Sigma_x = \Sigma_{\tilde{x}}$.  Then, associating double-primed indices with an integration point $x''$, it follows from \eqref{pDef} that
\begin{subequations}
\begin{align}
    \xi^\gamma \nabla_\gamma p_\alpha &= \xi^\gamma \int_{\Sigma_x} dS_{\beta''}T^{\beta''}{}_{\alpha''}  \nabla_\gamma K^{\alpha''}{}_{\alpha} ,
\\
    \xi^\gamma \nabla_\gamma S_{\alpha\beta} &= 2\xi^\gamma \int_{\Sigma_x} dS_{\beta''} T^{\beta''}{}_{\alpha''}  ( H^{\alpha''}{}_{[\alpha} \sigma_{\beta ] \gamma} 
    \nonumber
    \\
    & \qquad \qquad \qquad   ~  + \nabla_{\gamma} H^{\alpha''}{}_{[\alpha} \sigma_{\beta]}  )  .
\end{align}
\end{subequations}
All bitensors here are evaluated at $(x, x'')$. If the maximum distance, within $\Sigma_x$, between $x$ and any integration point where $T^{\beta''}{}_{\alpha''} \neq 0$ is of order $\ell_w$, standard coincidence limits for the world function \cite{Poisson2011} imply that $\sigma_{\alpha\beta} = g_{\alpha\beta} + \mathcal{O}(\ell_w^2/\ell_R^2 )$, $K^{\alpha'}{}_{\alpha} = H^{\alpha'}{}_{\alpha} + \mathcal{O}(\ell^2_w/ \ell_R^2 )$, and $\nabla_{\gamma} H^{\alpha'}{}_{\alpha}$ and $\nabla_{\gamma} K^{\alpha'}{}_{\alpha}$ are both of order $\ell_w/ \ell_R^2$, where $\ell_R$  denotes the curvature length scale introduced in Sec.~\ref{Sec:Approx}. Moreover, using the energy \eqref{Energy} to estimate the error terms, it follows that through the first order in $\xi^\alpha$,
\begin{subequations} \label{eq:delta_xpS}
\begin{align}
    \tilde{p}_{ \alpha' } &= g^{\alpha}{}_{ \alpha' } p_\alpha + \mathcal{O}(E \xi \ell_w/\ell_R^2), \\
    \tilde{S}^{ \alpha' \beta' } &= g^{ \alpha' }{}_{ \alpha } g\indices{^{ \beta' }_{\beta}} ( S^{\alpha \beta} + 2 p^{[\alpha} \xi^{\beta]} ) + \mathcal{O}(E \xi \ell_w^2/\ell_R^2). \label{eq:delta_S}
\end{align}
\end{subequations}
In flat spacetime, the error terms here are exactly zero; cf. \eqref{dS}.

We would now like to use \eqref{St2} to associate one centroid $\tilde{x}^\alpha$ with the timelike vector field $\tilde{t}^{ \alpha' }$ and another centroid $x^\alpha$ with the timelike vector field $t^\alpha$. Repeating the same steps as in Secs.~\ref{Sect:St} and \ref{Sect:dCentroids}, it is again convenient to identify points on both worldlines using $\xi \cdot t = 0$, which we now assume to be compatible with the assumption that $\Sigma_x = \Sigma_{\tilde{x}}$. Equation \eqref{eq:delta_xpS} can then be shown to imply that $g_{\alpha\alpha'} \xi^\alpha \tilde{t}^{\alpha'} = \mathcal{O}( \xi \ell_w^2 / \ell_R^2 )$. It also follows that \eqref{xi} generalizes to
\begin{equation}
    \xi^\alpha = \frac{ g_{\beta \beta'} S^{\alpha\beta}  \tilde{t}^{\beta'} }{ g^{\gamma}{}_{\gamma'}  p_\gamma \tilde{t}^{\gamma'} } + \mathcal{O}( \xi \ell_w^2/\ell_R^2).
    \label{xiGen1}
\end{equation}
This assumes that $\xi$ is not so large that terms of order $\xi^2$ become important in \eqref{eq:delta_xpS}. More generally, the basic special relativistic form for this expression remains valid when $\ell_w$ and $\xi$ are both much smaller than $\ell_R$. Most of the above special-relativistic results remain valid in this context. Technically, however, one can no longer conclude that there are massless centroids which are arbitrarily distant from one another, as then $\xi$ must be large. It is nevertheless clear from the flat spacetime limit that massless objects must still be problematic in curved spacetime.
\begin{figure}
\includegraphics[width=.47\textwidth]{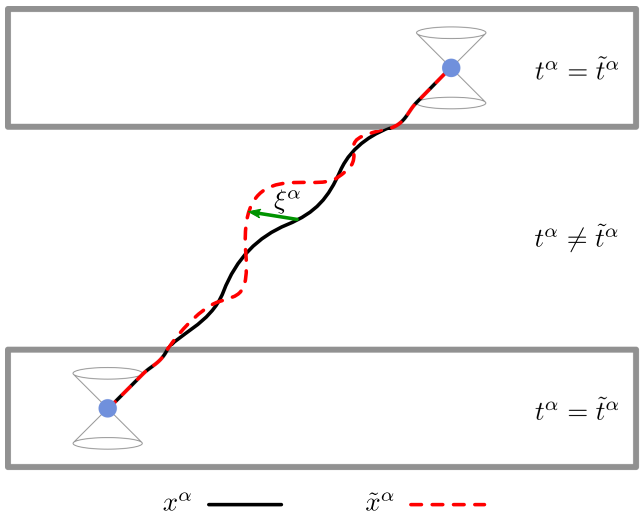}
\caption{A massless spinning object, as described by two different families of timelike observers, $t^\alpha$ and $\tilde{t}^\alpha$. The displacement between two points on the worldlines $x^\alpha$ and $\tilde{x}^\alpha$ is described by the shift vector $\xi^\alpha$. Since $t^\alpha = \tilde{t}^\alpha$ near the emitter and the receiver, we have $\xi^\alpha = 0$ in these regions, and the two worldlines coincide, up to relative error terms of order $(\ell_w/\ell_R)^2$. }  
\label{fig:worldlines}
\end{figure}

\subsection{The irrelevance of centroid conditions}

One important consequence of Eq.~\eqref{xiGen1} is that the centroid depends only (quasi)locally on the timelike vector field used to define it. If $t^\alpha$ and $\tilde{t}^\alpha$ coincide in, say, neighborhoods of emission and observation points, \textit{whatever they do in between the emitter and the observer is irrelevant}: The associated centroids will coincide at both the beginnings and ends of their journeys. This is illustrated schematically in Fig. \ref{fig:worldlines}. Physical meaning can be attributed to $t^\alpha$ only in the neighborhoods of the emission and the observation events. What it does elsewhere is essentially irrelevant.

As a consequence, the effects of different spin supplementary conditions can be understood without repeatedly solving the equations of motion with different conditions. Solving the Mathisson-Papapetrou equations---whether massive or massless---with the spin supplementary condition $S^{\alpha\beta} t_\beta = 0$  requires an apparently extraneous specification of the vector field $t^\alpha$. Indeed, that vector field must be specified not only at a point, but in a neighborhood of the entire trajectory. The observation of the previous paragraph shows that all that matters is the specification of the vector field near the emission and the observation points, at least if the displacement never gets too large. It may also be noted that it is only near the emitter and source that there necessarily exists a natural choice for $t^\alpha$: It may be identified near those points with the 4-velocities of the emitter and the observer.

The transformations \eqref{eq:delta_xpS} and \eqref{xiGen1} may also be interpreted as a way to generate new solutions to the equations of motion. Given a triple $(x^\alpha, p_\alpha, S^{\alpha\beta})$ which satisfies the MP equations with the centroid condition $S^{ab} t_b = 0$, the triple $(\exp_x \xi, \tilde{p}_{ \tilde{\alpha} }, \tilde{S}^{\tilde{\alpha} \tilde{\beta}} )$ satisfies those same equations but with the centroid condition $\tilde{S}^{ \tilde{\alpha} \tilde{\beta} } \tilde{t}_{ \tilde{\beta}} = 0$. This statement is exact in flat spacetime and approximate more generally.

\section{Momenta of circularly polarized wave packets}
\label{sec:wavepackets}

As discussed in Sec.~\ref{Sec:MP}, there are essentially only two conditions required for the validity of the spin Hall equations in the form considered here. First, the effect of the quadrupole and higher-order moments must be negligible. When this occurs can be estimated using the arguments in Sec.~\ref{Sec:Approx}. However, the spin Hall equations also require for their validity that $p_\alpha$ be null and that $S^{\alpha\beta}$ have the form \eqref{S}. These may be viewed as restrictions on the initial data for the MP equations. The claim has been that such conditions model an electromagnetic wave packet. However, this connection appears in the literature as an unsubstantiated (and usually unstated) hypothesis. It is implied by results in the Appendix that the spin Hall initial data cannot hold exactly, at least when $s \neq 0$. The purpose of this section is to understand if there is an appropriate \textit{approximate} sense in which the spin Hall initial data is actually associated with electromagnetic wave packets. 

We show that to the expected orders, there are indeed generic wave packets which are compatible with the spin Hall initial data. Nevertheless, we show that those data inevitably break down at one higher order; the momentum becomes timelike, for example. We also show that there are reasonable wave packets which are not even approximately described by the spin Hall initial data. For them, $S^{\alpha \beta} p_\beta$ is nonzero even at the lowest nontrivial order. Said differently, the spin is not purely longitudinal. The existence of these exceptions emphasizes that in applications, the connection between the ``microscopic'' (the electromagnetic field structure) and the ``macroscopic'' (the spin Hall equations or generalizations) is nontrivial and must be considered on a case-by-case basis.

All calculations in this section are performed in flat spacetime and in inertial coordinates. However, as we are concerned only with finding \textit{initial} linear and angular momenta for the MP equations, all calculations for sufficiently small wave packets are confined to small regions in spacetime. Flat spacetime calculations therefore remain excellent approximations for sufficiently compact wave packets even in curved spacetimes, as long as the inertial Minkowski coordinates are reinterpreted as an appropriate system of Riemann normal coordinates.

\subsection{A family of wave packets}

Our first task is to construct a sufficiently general class of approximate electromagnetic wave packets. We work in a high-frequency approximation and consider a family of vector potentials $A_\alpha$. These vector potentials are assumed to be given by the asymptotic series
\begin{equation}
    A_\alpha =  \Re \left[ \sum_{n=0}^\infty \eps^{n+1} \psi_\alpha^{(n)} e^{i u/\eps} \right],
    \label{WKB}
\end{equation}
where the amplitudes $\psi^{(n)}_\alpha$ and the eikonal $u$ are independent of the small parameter $\eps >0$. Note that the $\eps$ which appears here is related to, but generically distinct from, the $\epsilon$ which appears in the spin Hall equations [and in \eqref{WKB0}]. If $c$ is any constant, $A_\alpha/\eps$ is invariant, at leading order, under all transformations where $u \mapsto c u$ and $\eps \mapsto c \eps$. It is convenient for now to avail of this ambiguity by allowing $\eps$ to differ from $\epsilon$ by a convenient constant. Then $u$ is not restricted to have units $(\mbox{length})^2$.

Again defining the leading-order wave vector $k_\alpha \equiv - \nabla_\alpha u$, Maxwell's equations and the Lorenz gauge condition imply that $u$ must be a solution to the eikonal equation
\begin{equation}
    \nabla u \cdot \nabla u = k \cdot k = 0.
    \label{eikonal}
\end{equation}
Maxwell's equations and the gauge condition also imply that, for all $n \geq 0$, the amplitudes must satisfy the transport equations
\begin{equation}
    [2 (k \cdot \nabla) + (\nabla \cdot k) ] \psi^{(n)}_\alpha = - i \Box \psi^{(n-1)}_\alpha, 
    \label{xPort}
\end{equation}
and the constraint equations
\begin{equation}
        k^\alpha \psi_\alpha^{(n)} = - i \nabla^\alpha \psi_\alpha^{(n-1)},
        \label{constr}
\end{equation}
where $\psi^{(-1)}_\alpha \equiv 0$  \cite{Harte2018,GSHE2020}. These equations are hierarchical. A solution to the $n=0$ equation is required to solve the $n=1$ equation, an $n=1$ solution is required to solve the $n=2$ equation, etc.

We now specialize to fields which are, at least at the leading order, plane fronted\footnote{A plane fronted wave is not necessarily a plane wave. It is not necessarily uniform on each wavefront.} and traveling in the $+z$ direction in the inertial coordinate system $(t,x,y,z)$. This can be represented mathematically by choosing the eikonal 
\begin{equation}
    u = t-z.    
\end{equation}
It is then convenient to solve the transport and constraint equations by interpreting $u$ as a null coordinate and by defining
\begin{equation}
    v \equiv \frac{1}{2} (t+z), \qquad \zeta \equiv \frac{1}{\sqrt{2}} (x+i y).
\end{equation}
The four scalars $(u, v, \zeta, \bar{\zeta})$ form a null coordinate system in which the Minkowski line element reduces to
\begin{equation}
    ds^2 = 2 ( d\zeta d\bar{\zeta} - du dv).
\end{equation}
These coordinates can be associated with a complex null tetrad $( k_\alpha, n_\alpha, m_\alpha, \bar{m}_\alpha )$ via
\begin{equation}
    k_\alpha \equiv - \nabla_\alpha u, \quad n_\alpha \equiv - \nabla_\alpha v, \quad m_\alpha \equiv \nabla_\alpha \zeta.
    \label{tetrad}
\end{equation}
The only nonvanishing inner products among this tetrad are $m \cdot \bar{m} = 1$ and $ k \cdot n = -1$. In terms of it, the metric is $g_{\alpha \beta} = 2 [m_{(\alpha} \bar{m}_{\beta)} - k_{(\alpha} n_{\beta)}]$. Note that unlike in Sec.~\ref{sec:GSHE_review}, the $m_\alpha$ and $\bar{m}_\alpha$ here are ordinary fields on spacetime. They are not defined over the cotangent bundle.

We now restrict to waves which are not only plane fronted but also  circularly polarized at leading order. Mathematically, this is taken to mean that  $\psi^{(0)}_\alpha$ is assumed to be null (and nonzero). The constraint equation \eqref{constr} then implies that the leading-order amplitude must be proportional either to $m_\alpha + \chi k_\alpha$ or to $\bar{m}_\alpha + \chi k_\alpha$, where  $\chi$ is any scalar [see \eqref{eq:a_basis}]. Terms proportional to $k_\alpha$ are pure gauge at leading order, although not necessarily at higher orders\footnote{Terms in $\psi^{(0)}_\alpha$ which are proportional to $k_\alpha$ are physically equivalent to an ordinary (gauge-invariant) geometric optics field added to $\psi^{(1)}_\alpha$ \cite[Appendix B]{Harte2018}.}. Regardless, we set $\chi =0$ for simplicity. Circularly polarized fields are then described by 
\begin{equation}
    \psi^{(0)}_\alpha  = \psi m_\alpha 
    \label{A0}
\end{equation}
or $\psi^{(0)}_\alpha  = \psi \bar{m}_\alpha$, depending on the handedness of the field. Unlike in Sec.~\ref{sec:GSHE_review}, we do not assume that $\psi$ which appears here is necessarily real. Use of Eq.~\eqref{xPort} shows that it is independent of $v$ but otherwise arbitrary: $\psi = \psi (u, \zeta, \bar{\zeta})$. 

We may now substitute the leading-order amplitude \eqref{A0} into the transport equation \eqref{xPort} and the constraint equation \eqref{constr} in order to derive the subleading amplitude $\psi^{(1)}_\alpha$. One solution is
\begin{equation}
     \psi^{(1)}_\alpha = i \partial_{\bar{\zeta} } \left( \psi n_\alpha - v m_\alpha  \partial_\zeta \psi \right).
     \label{A1}
\end{equation}
Two orders beyond geometric optics, we find that
\begin{align}
    \psi^{(2)}_\alpha = \tfrac{1}{4} \partial_{ \bar{\zeta} } \Box \left[ v \left( 2 \psi  n_\alpha - v \partial_\zeta \psi m_\alpha \right) \right]
    .
    \label{A2}
\end{align}
Together, \eqref{WKB}, \eqref{A0}, \eqref{A1}, and \eqref{A2} describe an approximate family of circularly polarized electromagnetic fields in flat spacetime. That family is parametrized by the geometric-optics scalar amplitude $\psi$ and by the constant $\eps$, which is related to the inverse frequency of the field. Its properties are explored in the remainder of this section. If desired, fields with the opposite helicity can be considered by swapping $m_\alpha$ and $\bar{m}_\alpha$ in the amplitudes \eqref{A0}, \eqref{A1}, and \eqref{A2}.

Vector potentials are not directly measurable. More interesting are the field strengths $F_{\alpha\beta}$, and for any vector potential with the form \eqref{WKB}, a direct calculation shows that these are given by
\begin{align}
    F_{\alpha\beta} = 2 \Im \sum_{n=0}^\infty \eps^n \big( k_{[\alpha} \psi_{\beta]}^{(n)} + i \nabla_{[\alpha} \psi_{\beta]}^{(n-1)} \big) e^{i u/\eps} .
    \label{F}
\end{align}
For the amplitudes constructed above, this evaluates to 
\begin{align}
    F_{\alpha\beta}  =  2\Im \big\{  \big[  \left( \psi -\tfrac{1}{2} i \eps v^2 \Box ( v^{-1} \psi )
     - \tfrac{1}{8} \eps^2 v^2 \Box^2 \psi \right) 
    \nonumber
    \\
    {} ~ \times k_{[\alpha} m_{\beta]}  + \eps^2 \partial_{ \bar{\zeta}}^2 \psi n_{ [ \alpha } \bar{m}_{ \beta] } +i \eps \partial_{ \bar{\zeta} }  \left( \psi - \tfrac{1}{2} i\eps v \Box \psi \right) 
    \nonumber
    \\
     ~ \times \left(  k_{[\alpha} n_{\beta]} -  m_{[\alpha} \bar{m}_{\beta]} \right) \big] e^{i u/\eps} \big\} + \mathcal{O}(\eps^3) .
     \label{Ffull}
\end{align}
At leading (geometric optics) order, $F_{\alpha \beta}$ is a linear combination of $k_{[\alpha} m_{\beta]}$ and $k_{[\alpha} \bar{m}_{\beta]}$. At subleading order, its tensorial structure changes; it acquires terms proportional to $k_{[\alpha} n_{\beta]}$ and $ m_{[\alpha } \bar{m}_{\beta]}$. These corrections may be interpreted as modifying the apparent ``polarization state'' at higher orders. More broadly, the tensorial structure of the electromagnetic field may alternatively be understood by noting that the Newman-Penrose scalars associated with the tetrad $(k_\alpha, n_\alpha, m_\alpha, \bar{m}_\alpha)$ satisfy $\Phi_i = \mathcal{O}( \eps^{2-i} )$ for all $i=0,1,2$. At leading order, there is only $\Phi_2$. At subleading order, there is also $\Phi_1$. At two orders beyond geometric optics, there is also $\Phi_0$. This is a special case of the peeling result for high-frequency fields which was obtained in \cite{Harte2018}.

\subsection{Stress-energy tensors}

Our goal is to compute linear and angular momenta, which are determined by integrals of stress-energy tensors. The electromagnetic stress-energy tensor is
\begin{equation}
    T_{\alpha\beta} = \frac{1}{4\pi} \left( F_{\alpha \gamma} F_{\beta}{}^{\gamma} - \tfrac{1}{4} g_{\alpha\beta} F_{\gamma \lambda} F^{\gamma \lambda} \right),
    \label{Tem}
\end{equation}
and through leading and subleading orders, substitution of Eq.~\eqref{Ffull} into this expression results in
\begin{align}
    T_{\alpha\beta} = \frac{ 1 }{ 8 \pi } \big\{ \! \left[ |\psi |^2 + \eps v^2 \Im \left( \bar{\psi} \Box ( \psi /v ) \right) \right] k_\alpha k_\beta
    \nonumber
    \\
    ~{} - 4 \eps \Im \left( \bar{\psi} \partial_{ \bar{\zeta} } \psi k_{( \alpha} \bar{m}_{ \beta )} \right) \big\} + \mathcal{O}(\eps^2) .
\end{align}
In regions where $|\psi| \neq 0$, this can be written more suggestively as 
\begin{align}
    T_{\alpha\beta} = \frac{1}{ 8\pi }  \left[ | \psi |^2 + \eps v^2 \Im \left( \bar{ \psi } \Box ( \psi /v ) \right) \right]  
    \nonumber
    \\
     {} ~ \times \tilde{k}_\alpha \tilde{k}_\beta  + \mathcal{O}(\eps^2),
     \label{Tem2}
\end{align}
where the covector
\begin{equation}
    \tilde{k}_\alpha \equiv k_\alpha -2 \eps \Im \left[ ( \partial_{\bar{\zeta}} \ln \psi ) \bar{m}_\alpha  \right] 
    \label{Kdef}
\end{equation}
which appears here can be interpreted as a modified wave vector. Like the leading-order wave vector, this is null: $\tilde{k}_\alpha \tilde{k}^\alpha = \mathcal{O}(\eps^2)$. At leading order, $T_{\alpha\beta} \propto k_\alpha k_\beta + \mathcal{O}(\eps) $, and it is somewhat remarkable that at one order higher, this is modified only to $T_{\alpha\beta} \propto \tilde{k}_\alpha \tilde{k}_\beta +  \mathcal{O}(\eps^2)$. All observers therefore agree on the direction of momentum density, even at the subleading order.

Similar factorizations of the stress-energy tensor\footnote{Ref.~\cite{Harte2018} discusses these factorizations for the \textit{averaged} stress-energy tensor. However, averaging has no effect in the circularly polarized case considered here.} have been discussed in more general contexts in Ref.~\cite{Harte2018}. A result like the one found here was shown to arise whenever $k_\alpha$ is shear-free. If there is shear in the leading-order rays, different observers generically disagree on the direction of the subleading momentum density. Moreover, even in the shear-free case, observers  typically disagree on the direction of the momentum density once terms of order $\eps^2$ are included; stress-energy tensors at that order are generically more complicated.

\subsection{Linear and angular momenta}

The stress-energy tensor in Eq.~\eqref{Tem2} may be used to compute the net linear momentum on a $t = \mbox{const.}$ hypersurface. Using Eq.~\eqref{pDef}, and putting primes on integration variables and on the objects which depend on the integration variables,
\begin{align}
    p_\alpha = \frac{1}{8\pi} \int d^3 x' \left[ | \psi'|^2 + \eps v'^2 \Im \left( \bar{ \psi }' \Box' ( \psi' /v') \right) \right] 
    \nonumber
    \\
    ~{} \times \tilde{k}_\alpha   + \mathcal{O}(\eps^2).
    \label{pInt}
\end{align}
It follows from Eq.~\eqref{Kdef} that $p_\alpha p^\alpha = \mathcal{O}(\eps^2)$; the momentum is null through leading and subleading orders. However, it is not necessarily true that $p_\alpha$ is proportional to $k_\alpha$ beyond leading order. 

More interesting is the angular momentum $S^{\alpha\beta}$, which we compute about a point $x^\alpha = (t,x^i)$ which lies within the hypersurface of integration. There are two interesting sets of components. First, using Eqs.~\eqref{pDef} and \eqref{Tem2},
\begin{align}
    S^{i0} = \frac{1}{8\pi} \int d^3x'  \big[ |\psi'|^2 + \eps v'^2 \Im \left( \bar{\psi}' \Box' ( \psi' /v' ) \right) \big]
    \nonumber
    \\
    {} ~ \times (x'-x)^i  + \mathcal{O}(\eps^2).
     \label{S0i}
\end{align}
This can be interpreted as the dipole moment of the energy density with respect to a static observer at $x^\alpha$. The other relevant components of the angular momentum tensor are
\begin{align}
    S^{ij} = \frac{\eps}{2\pi} \Im \int d^3 x'  \left( \bar{ \psi }' \partial_{ \bar{\zeta}' } \psi' \right) \bar{m}^{[i }  (x'-x)^{j] } 
    \nonumber
    \\
    ~ - 2 k^{ [i} S^{j] 0}  + \mathcal{O}(\eps^2 ) . 
    \label{Sij}
\end{align}

\subsection{Vanishing phase gradients}
\label{Sec:zeroPhase}

The linear and angular momenta simplify considerably when the complex phase of $\psi$ is constant. Looking first at the linear momentum, if $\nabla_\alpha \arg \psi = 0$ and if $|\psi|$ decays to zero sufficiently rapidly at large transverse distances, Eq.~\eqref{pInt} reduces to
\begin{equation}
    p_\alpha = \frac{ 1 }{ 8\pi } \left( \int d^3 x |\psi|^2 \right) k_\alpha + \mathcal{O}(\eps^2). 
    \label{p2}
\end{equation}
The subleading contribution to the linear momentum vanishes and $p_\alpha$ is seen to be proportional to the constant leading-order wave vector $k_\alpha$. It is important to note, however, that the net momentum is in general distinct from the momentum \textit{density}. The former is proportional to $k_\alpha$ while the latter is proportional to $\tilde{k}_\alpha$. An observer with a high-resolution detector might therefore ascribe an apparent ``direction of propagation'' which differs, at $\mathcal{O}(\eps)$, from the direction of $p^\alpha$. More than this, the direction of the momentum density varies slightly across the wave packet. 

When the phase gradient vanishes, Eq.~\eqref{S0i} simplifies to
\begin{align}
    S^{i0} = \frac{1}{8\pi} \int d^3x' (x'-x)^i |\psi'|^2 
      + \mathcal{O}(\eps^2).
    \label{Si02}
\end{align}
Here too the subleading contribution vanishes. Furthermore, Eq.~\eqref{Sij} reduces to
\begin{equation}
    S^{ij} = 2 i \eps p^0 \bar{m}^{[i} m^{j]} - 2 k^{[i} S^{j] 0} + \mathcal{O}(\eps^2).
    \label{Sij2}
\end{equation}

One particularly simple choice for $x^i$ arises by enforcing the centroid condition \eqref{Sortho}. Temporarily assume that $t^\alpha = (1,0,0,0)$ so that condition reduces to $S^{i0} = 0$. It then follows from Eq.~\eqref{Si02} that this centroid condition implies that
\begin{equation}
    x^i = \frac{ 1 }{ 8\pi E } \int d^3x' |\psi'|^2 x'^i + \mathcal{O}(\eps^2),
    \label{xCentExample}
\end{equation}
where $E$ is again given by Eq.~\eqref{Energy}. In terms of the bivector $\Sigma^{\alpha \beta}$ which is defined by Eq.~\eqref{SigmaDef},
\begin{equation}
    S^{\alpha\beta} = \eps E \Sigma^{\alpha\beta} + \mathcal{O}(\eps^2).
    \label{S2}
\end{equation}
The angular momentum therefore satisfies $S^{\alpha \beta} p_\beta = \mathcal{O}(\eps^2)$. An angular momentum which differs from this only by a sign can be obtained by considering an otherwise-identical field with opposite helicity, which is accomplished by replacing the $m_\alpha$ which appears in Eq.~\eqref{A0} with $\bar{m}_\alpha$. 

Equation \eqref{S2} matches the form \eqref{S} for $S^{\alpha\beta}$, assuming that $s=1$ and 
\begin{equation}
    \epsilon =  \eps E . 
    \label{epseps}
\end{equation}
The two small parameters we have introduced are therefore proportional to one another. Physically, either one can be interpreted as related to the leading-order angular frequency which would be seen by an observer with 4-velocity $\tilde{t}^\alpha$:
\begin{equation}
    \omega = - (k \cdot \tilde{t} ) / \eps = - ( p \cdot \tilde{t}  )/\epsilon .
    \label{omegaDef2}
\end{equation}

To summarize, we have found sufficient conditions for the physical picture suggested in Sec.~\ref{Sec:MP}, namely that $p_\alpha$ is null, $s = \pm 1$, and $S^{\alpha \beta} p_\beta = 0$. This is valid, up to terms of order $\eps^2$, at least for all decaying, circularly polarized electromagnetic wave packets with planar wavefronts and vanishing phase gradients. It is shown in Sec.~\ref{Sect:PhGrad} below that this picture can change when $\psi$ has a nontrivial phase gradient.

\subsubsection{Violation of energy conditions}

It was shown in Sec.~\ref{sec:role_of_t} above that if $p_\alpha$ is null and $S^{\alpha\beta}$ has the form \eqref{S}, the set of all possible centroids determined by $S^{\alpha\beta} t_\beta = 0$ is not bounded in spacelike directions (when varying over all timelike $t^\alpha$). This suggests that the worldlines we refer to as centroids are perhaps poorly named; it may be that some of them are nowhere near the wave packet of interest. It is well known that this type of situation can occur for the Newtonian center of mass if the mass density switches sign\footnote{As an example, suppose that two masses, $m_+ > 0$ and $m_- = 2\delta - m_+$, are placed on a line at coordinates $\pm \ell_w$. Then the center of mass lies at $(m_+ /\delta -1)\ell_w$, which can be anywhere at all for appropriate choices of $\delta$.}. Relativistically, avoiding this kind of pathology involves requiring that the stress-energy tensor satisfy appropriate energy conditions. And in an exact context, electromagnetic stress-energy tensors with the form \eqref{Tem} are known to satisfy all standard energy conditions \cite{SandersEnergy}. However, it is not necessarily true that the \textit{approximate} electromagnetic stress-energy tensors of interest here also satisfy those energy conditions. We now show that they do not. It is this violation of energy conditions which is behind the peculiarly distant centroids associated with massless spinning objects.

Suppose that $\tilde{t}^\alpha$ has a form which maximizes the centroid displacements in Sec.~\ref{Sec:Loc}. Using our null tetrad \eqref{tetrad}, one possibility which is compatible with \eqref{thetaMax} is
\begin{equation}
    \tilde{t}^\alpha = \frac{\frac{1}{2} k^\alpha (1+V^2) + n^\alpha (1- V^2)  }{ \sqrt{ 1 - V^2 } }  + \frac{ V }{\sqrt{2}} (m^\alpha + \bar{m}^\alpha) ,
\end{equation}
where $V \in [0,1)$ denotes the relative speed between $\tilde{t}^\alpha$ and $(1,0,0,0)$. In the case of interest here, where there is no phase gradient, \eqref{Tem2} and \eqref{Kdef} imply that if $\mathcal{O}(\eps^2)$ terms are ignored,
\begin{align}
        T_{\alpha\beta} \tilde{t}^\alpha \tilde{t}^\beta &= \frac{ |\psi|^2 }{ 8\pi } ( \tilde{k} \cdot \tilde{t})^2 ,
        \nonumber
        \\
        &= \frac{ 1-V^2 }{8\pi} \left( |\psi|^2
         +  \frac{ \eps V \partial_y |\psi|^2 }{ \sqrt{1-V^2} } \right)  .
\end{align}
For any nontrivial bounded wave packet, there will be some regions in which $\partial_y |\psi|^2$ is negative and other regions in which it is positive. Furthermore, the first term here can always be made negligible compared to the second by choosing $V$ sufficiently close to 1. It follows that if $\mathcal{O}(\eps^2)$ terms are ignored, there are timelike vectors $\tilde{t}^\alpha$ for which $T_{\alpha\beta} \tilde{t}^\alpha \tilde{t}^\beta$ is negative in some parts of the wave packet and positive in others; the energy density switches sign. This amounts to a violation of the weak, strong, and dominant energy conditions. 

As noted above, this violation is an artifact of our approximation. All exact electromagnetic stress-energy tensors satisfy the weak, strong, and dominant energy conditions. That there is a problem with our approximation is not difficult to see in this context, as we are finding a  ``subleading'' term which dominates over the ``leading'' term. It is not particularly surprising that in such a scenario, terms of even higher order might not be negligible. It is less clear, however, that simply assuming that a wave packet is both massless and spinning is enough for it to be associated with spurious, arbitrarily distant centroids. That is, however, a consequence of the fact the high-frequency approximation breaks down for certain highly boosted observers.

\subsubsection{Momentum is timelike, not null}
\label{Sec:pTimelike}

It is shown in the Appendix that it is impossible for a truly massless wave packet to have nonzero spin. It is however clear that an electromagnetic wave packet can have nonzero spin. The conclusion is that spinning electromagnetic wave packets cannot truly be null. We now show that our wave packets are  timelike once we include terms two orders beyond those in geometric optics.

To establish this, first note that the electromagnetic field \eqref{Ffull} can be used to compute the stress-energy tensor to one higher order than shown in Eq.~\eqref{Tem2}. That may in turn be used to compute $p_\alpha$. The full stress-energy tensor is complicated, however. The calculation can be considerably simplified by noting that all we need to determine the causal character of $p_\alpha$ is the $\mathcal{O}(\eps^2)$ contribution to $k \cdot p$. That can in turn be computed by showing only that
\begin{equation}
    T_{\alpha \beta} k^\beta = - \frac{ \eps^2 }{16\pi}  |\nabla \psi|^2 k_\alpha + \mathcal{O}(\eps^3).
\end{equation}
This result and \eqref{p2} imply that
\begin{equation}
    p \cdot p = - \frac{\eps^2  E }{8\pi} \int d^3 x' |\nabla' \psi'|^2 + \mathcal{O}(\eps^3),
\end{equation}
where $E$ again denotes the energy seen by a stationary observer. Note that because $\partial_v \psi = 0$, this is always negative; every localized electromagnetic wave packet of the given form has the small nonzero rest mass
\begin{equation}
    m = \eps \left( \frac{ E }{ 8 \pi }  \int d^3 x' |\nabla' \psi'|^2  \right)^{1/2} + \mathcal{O}(\eps^2).
\end{equation}
If $\nabla \psi \sim \psi/\ell_w$ for some length scale $\ell_w$, this suggest that $m \sim \eps E/\ell_w = \epsilon/\ell_w$. Results on the localization of \textit{massive} objects which were reviewed in Sec.~\ref{Sec:Loc} therefore imply that all centroids associated with these wave packets are in fact confined to a disk whose radius is of order $\mathcal{S}/m \sim \ell_w$. This is the expected result. We emphasize, however, that it cannot be established---even qualitatively---in the massless approximation.

\subsection{Nonvanishing phase gradients and the limitations of the spin Hall framework}
\label{Sect:PhGrad}

If the phase gradient of $\psi$ does not vanish, the physical picture associated with the spin Hall equations might not hold. First, it is not necessarily true that $s = \pm 1$ for a circularly polarized electromagnetic wave packet. Additional contributions to this parameter can arise. This is referred to as orbital (as opposed to spin) angular momentum in the optics literature \cite{AM_Light2,AM_Light}. While the derivation of the spin Hall equations in \cite{GSHE2020} did not allow for the possibility of orbital angular momentum, their derivation as a special case of the MP equations makes it clear that orbital angular momentum requires no essential changes; $s$ merely takes on different integer values in the spin Hall equations\footnote{An analogous discussion for electromagnetic beams propagating in flat spacetime but in nontrivial materials may be found in \cite{Bliokh2006}.}. A more dramatic consequence of allowing nontrivial phase gradients is that it becomes possible to construct wave packets with spin vectors which are not longitudinal: $S^{\alpha \beta} p_\beta \neq 0$ even at the leading nontrivial order. In these cases, the form \eqref{S} for the angular momentum is incomplete and the spin Hall equations are no longer valid. Even then, however, the MP equations can still be applied.

We first consider the possibility of nonlongitudinal spin. If the centroid condition is imposed with $t^\alpha = (1,0,0,0)$, it follows from \eqref{Sij} that 
\begin{align}
    S^{\alpha\beta} p_\beta &= E S^{\alpha\beta} k_\beta + \mathcal{O}(\eps^2)
    \nonumber
    \\
    &= \frac{ \eps E }
    {4\pi} \Im \left[ \bar{m}^\alpha  \int d^3 x' (z'-z)  \bar{\psi}' \partial_{\bar{\zeta}' } \psi' \right] + \mathcal{O}(\eps^2) .
    \label{SpExample}
\end{align}
A nontrivial transverse angular momentum therefore requires that $\bar{\psi} \partial_{\bar{\zeta}} \psi$ have a nontrivial moment along the optical axis. That this is possible can be illustrated by examples. Suppose that
\begin{equation}
    \psi = |\psi| e^{ i x u / \ell_\perp^2},
    \label{psiEx1}
\end{equation}
where $\ell_\perp > 0$ is a parameter. Also assume that
\begin{equation}
    \int du' u' |\psi'|^2 = 0
    \label{upsiInt}
\end{equation}
so the $z$ component of the centroid lies at $z = t + \mathcal{O}(\eps)$. Substitution into \eqref{SpExample} then shows that
\begin{equation}
    S^{\alpha \beta} p_\beta = - \frac{ \eps E (m + \bar{m})^\alpha }{ 8 \sqrt{2} \pi \ell_\perp^2 }  \int d^3 x' |u' \psi'|^2 + \mathcal{O}(\eps^2),
\end{equation}
This is nonzero for any nontrivial $\psi$. It follows that for this class of wave packets, $S^{\alpha\beta}$ is not in the spin Hall form \eqref{S}. Equivalently, $S^\alpha$ cannot be parallel to $p^\alpha$; it must have a nonzero $y$ component. It is not possible to use the spin Hall equations to understand the motion of such a wave packet. However, there is no obstacle to using the MP equations in their more general form \eqref{MP}. The conclusion here is that $S^{\alpha\beta} p_\beta = 0$ is a physical restriction; it is not inevitable. The derivation of the spin Hall equations appears to have implicitly assumed that the wave packets have, e.g., vanishing phase gradients. 

Now consider a wave packet with no transverse angular momentum, but with potentially large amounts of longitudinal angular momentum. This can be produced by introducing polar coordinates $(r,\theta)$ in the $xy$ plane and then supposing that 
\begin{equation}
    \psi = |\psi| e^{i n \theta},
\end{equation}
where $n$ is an integer. Additionally, suppose for simplicity that $|\psi|$ depends only on $r$ and $u$ and that it satisfies \eqref{upsiInt}. It then follows from \eqref{Sij} that when evaluated at the centroid,
\begin{equation}
    S^{\alpha\beta} = (n+1) \eps E \Sigma^{\alpha\beta} + \mathcal{O}(\eps^2) ,
\end{equation}
where $\Sigma^{\alpha\beta}$ is again given by \eqref{SigmaDef}. If $\epsilon$ is again related to $\eps$ via \eqref{epseps} comparison with \eqref{S} shows that for these wave packets,
\begin{equation}
    s = n+1.
\end{equation}
If a wave packet with the opposite helicity had been considered, we would have found instead that $s = n-1$. Regardless, it is clear that $s$ is not necessarily equal to $\pm 1$. Much larger amounts  of angular momentum are possible than had been supposed in, e.g., \cite{GSHE2020}. Formally, $n$ can be arbitrarily large here. However, the high-frequency analysis breaks down when $\psi$ varies on the same scale as $e^{i u/\eps}$. If the spatial extent of the wave packet is of order $\ell_w$, this implies that our equations can be trusted only when $n \ll \ell_w/\eps = \omega \ell_w$.

\section{Spin Hall effect of light in an inhomogeneous medium}
\label{sec:shel}

As a final application, we now show how the ray equations describing the spin Hall effect of light in an inhomogeneous medium \cite{OpticalMagnus,SHE-L_original,SHE_original,Bliokh2004,Bliokh2004_1,Duval2006,Duval2007, Bliokh2008, Ruiz2015} can be recovered from the gravitational spin Hall equations \eqref{eq:gshe_eq}. The main tool used here is the well-known analogy between electromagnetic waves propagating inside a dielectric medium and electromagnetic waves propagating through vacuum but in an effective metric  \cite{Eddington, Gordon, Plebansky-Maxwell,  Birula_wavefunction1,  Birula_wavefunction2, cartographic_analog, covariant_dielectric}. More precisely, consider a background metric $\tilde{g}_{\alpha \beta}$ and a dielectric medium with a varying refractive index $n$ and a 4-velocity $u^\alpha$. It has then been shown in Ref.~\cite{Gordon, Synge1960} (see also Ref.~\cite{covariant_dielectric}) that the combined effect of the background spacetime and the dielectric medium on light rays can be studied by considering vacuum propagation in the optical metric
\begin{equation} \label{eq:optical_metric}
    g_{\alpha \beta} = \tilde{g}_{\alpha \beta} + (1 - n^{-2}) u_\alpha u_\beta,
\end{equation}
where the indices on the 4-velocities here have been lowered using $\tilde{g}_{\alpha\beta}$.

To describe the spin Hall effect of light in an inhomogeneous medium, we take the background metric to be the Minkowski one in inertial coordinates $(t,x,y,z)$ so $\tilde{g}_{\alpha\beta} = \eta_{\alpha\beta}$. We also suppose that the medium is stationary in these coordinates so $\partial_t n =0$ and $u^\alpha = (1,0,0,0)$.  The spin Hall equations \eqref{eq:gshe_eq} additionally require the choice of a timelike vector field $t^\alpha$ to fix the centroid definition, and this may be identified here with $u^\alpha$. A calculation then shows that with the effective metric $g_{\alpha\beta}$, $\Sigma^{\alpha\beta} R_{\alpha\beta\gamma}{}^{\lambda}  = 0$. The spin Hall equations in this metric therefore reduce to
\begin{subequations} \label{eq:gshe_minkowski}
\begin{align}
    \dot{x}^\alpha &= g^{\alpha\beta} \left( p_\beta + \frac{1}{p \cdot t} S_{\beta \rho} g^{\gamma\lambda} p_\gamma \nabla_\lambda t^\rho \right),
    \\
    \dot{p}_\alpha &= \Gamma^\beta_{\alpha \gamma} g^{\gamma \lambda} p_\beta p_\lambda  . 
\end{align}
\end{subequations}
As $u^\alpha$ is Killing and the effective metric is static, the spin-dependent terms in the conservation law \eqref{consLaw2} vanish so $E = - p_\alpha u^\alpha = \mbox{const}$. Introducing a $3$-vector notation, the momentum must be null with respect to $g_{\alpha\beta}$, meaning that
\begin{equation}
    p_i = \bm{p}, \qquad
    p_0 = - \frac{\sqrt{\bm{p} \cdot \bm{p}}}{n} = - \frac{p}{n}.
\end{equation}
Energy conservation in the effective metric therefore implies that $E = p/n$ is constant. It is only the direction of $\bm{p}$ which must be determined from the equations of motion.

A calculation shows that the nontrivial components of \eqref{eq:gshe_minkowski} reduce to
\begin{subequations}
\begin{align}
    \frac{ dt }{ d\tau} &= n p = n^2 E ,  \\
     \frac{ d\bm{x}}{ d\tau } &= \bm{p} + \frac{ \epsilon s }{ p^3} \left( \frac{d \bm{p}}{d\tau} \times  \bm{p}\right),  \\
    \frac{d\bm{p}}{d\tau} &= \frac{p^2}{n} \boldsymbol{\nabla} n = \frac{1}{2} \bm{\nabla} (n E)^2 ,
\end{align}
\end{subequations}
To obtain the same form of the ray equations as in the optics literature, we can reparametrize everything in terms of $t$ instead of $\tau$. Doing so,
\begin{subequations}\label{eq:she_optics}
\begin{align} 
    \frac{d \bm{x}}{d t} &= \frac{\bm{p}}{n p} + \frac{  \epsilon s }{p^3} \left( \frac{d\bm{p}}{d t} \times  \bm{p} \right),  \\
    \frac{d \bm{p}}{d t} &= \frac{p}{n^2} \boldsymbol{\nabla} n,
\end{align}
\end{subequations}
These are the ray equations describing the spin Hall effect of light in an inhomogeneous medium, as obtained in Refs.~\cite{SHE_original, Onoda2006,Ruiz2015}. They can be rewritten in the form presented in Refs.~\cite{Bliokh2004,Bliokh2004_1,Bliokh2008,Bliokh2009} by rescaling the momentum and time, as mentioned in Ref.~\cite{Ruiz2015} (see also Ref.~\cite{Duval2006}). 

Deriving the spin Hall effect of light in an inhomogeneous medium from \eqref{eq:gshe_eq} is important for several reasons. First, it establishes that the gravitational spin Hall equations really are related to the spin Hall effects described in flat-spacetime optics; the gravitational spin Hall equations thus have been given an appropriate name. Second, the ray equations usually used in the optical literature implicitly fix $t^\alpha$ at the outset and do not allow it to vary. Beginning instead with the gravitational spin Hall equations, where $t^\alpha$ is arbitrary, instead allows a unified description of the spin Hall effect of light, determined by the gradient of $n$, and the relativistic Hall effect \cite{Relativistic_Hall, Stone2015}, determined by  changes in $t^\alpha$. Lastly, the spin Hall effect of light, as described by Eqs.~\eqref{eq:she_optics}, has been confirmed experimentally in Refs.~\cite{SHEL_experiment,Bliokh2008}. The present connection between Eqs.~\eqref{eq:gshe_eq} and Eqs.~\eqref{eq:she_optics} gives some level of confidence in the theoretical predictions of Eqs.~\eqref{eq:gshe_eq} and in the existence of a genuinely gravitational spin Hall effect of light. 

As another application of the type of analysis presented in this section, one might consider the propagation of light in a plasma which is in a curved spacetime, perhaps near a black hole. In some regimes, the plasma can be expected to have an effective refractive index \cite{plasma_lensing1,plasma_lensing2,plasma_lensing3,plasma_lensing4,plasma_lensing5}. The spin Hall equations \eqref{eq:gshe_eq} together with the optical metric \eqref{eq:optical_metric} could then be used to derive polarization-dependent corrections to the propagation of electromagnetic pulses in the presence of an astrophysical plasma.

\section{Conclusions}
\label{sec:conclusions}

This paper has investigated the implications, properties, and limitations of the gravitational spin Hall equations derived in Refs.~\cite{GSHE2020,GSHE_GW}. In the electromagnetic case of interest here, these equations describe the motion of high-frequency circularly polarized electromagnetic pulses which propagate in vacuum but in arbitrary background spacetimes. In this context, the spin Hall effect refers to the transverse deflection of a pulse due to its spin. 

Our first class of results concern the meanings of the position and the momentum which appear in the gravitational spin Hall equations. In general, a spinning wave packet must be extended and the adoption of any equation of motion must be associated with a particular choice of centroid. We have found that the position appearing in the gravitational spin Hall equations is a centroid whose definition is parametrized by the timelike vector field $t^\alpha$ which appears in those equations. That position may be interpreted as a spin supplementary condition, chosen to ensure that the angular momentum satisfies $S^{\alpha\beta} t_\beta = 0$. More physically, the centroid is the center of energy of the wave packet in a frame which is instantaneously at rest with respect to $t^\alpha$. 

Different choices for $t^\alpha$ are in general associated with different centroids, and we have computed the shifts between those centroids. For massive objects with timelike momentum, we recover the known result that these shifts are always bounded: At any fixed time, all centroids lie within a finite disk. However, we have shown that this is no longer true in the (massless) null case. Massless spinning objects have arbitrarily distant centroids. In this sense, they cannot be localized.

Although this might appear to be problematic, we show that there is no spinning electromagnetic field configuration which is in fact massless. More generally, no massless object of any composition can have spin unless it violates the dominant energy condition. Nevertheless, there is a sense in which high-frequency electromagnetic wave packets can be \textit{approximately} null. Truncating the high-frequency approximation at subleading order results in an approximate stress-energy tensor which violates the dominant energy condition. There is a large amount of both positive and negative energy density in certain highly boosted frames, and it is this negative energy density which makes it appear as though there are centroids far outside of the wave packet itself. These energy densities---and the associated distant centroids---are not real. They are unphysical artifacts of the high-frequency approximation. If higher-order terms are included, a spinning electromagnetic wave packet would be seen to satisfy all standard energy conditions and to have a positive rest mass. This rest mass guarantees that all centroids remain in a finite region; real electromagnetic wave packets can be localized.

When working in the approximately massless approximation associated with the spin Hall equations, one must be careful about the limitations of that approximation. The concept of something being ``approximately null'' can make sense only in a class of frames, and the high-frequency approximation breaks down in very different frames. In particular, some weak restrictions must be placed on the $t^\alpha$ appearing in the gravitational spin Hall equations in order to avoid regimes where those equations are no longer valid.

We have also addressed other aspects of the approximations inherent in the gravitational spin Hall equations. First, there is the question whether or not the initial data assumed in the gravitational spin Hall equations does indeed describe reasonable high-frequency wave packets. We argue that it does, to the expected degree of accuracy, at least when there are negligible phase gradients across the wavefronts. Some cases of nontrivial phase gradients can still be described by the spin Hall equations, just with larger amounts of angular momentum. In other cases with significant phase gradients, $S^{\alpha \beta} p_\beta \neq 0$ so the angular momentum is no longer longitudinal and the spin Hall equations cannot be applied. Nevertheless, those cases can still be described by the Mathisson-Papapetrou equations, which are more general than the spin Hall equations. 

Besides the approximations involved in the initial conditions used in the equations of motion, there are also neglected terms in the equations of motion themselves. For example, the quadrupole moment of the wave packet is neglected. We have provide a detailed discussion of when such terms can be neglected and when they cannot. This is subtler than for the  nearly-rigid massive objects whose quadrupole moments are more commonly considered, as electromagnetic fields do not hold themselves together as they propagate. Electromagnetic wave packets generically spread out over time, increasing the quadrupole and higher-order moments and eventually invalidating the equations of motion. 

Another theme in this paper has been to relate the gravitational spin Hall equations to other equations which have also been proposed in the literature to  describe the motion of spinning electromagnetic wave packets---sometimes in quite different contexts. First, we have shown that the gravitational spin Hall equations are special cases of the Mathisson-Papapetrou equations, which govern the motion of generic (not necessarily electromagnetic) spinning objects in curved spacetimes. The gravitational spin Hall equations arise from the MP equations with a particular choice of spin supplementary (or centroid) condition, a particular type of initial data, and a particular worldline parameterization. Second, we have shown that the spin Hall effect of light in an inhomogeneous medium can be obtained from the gravitational spin Hall equations with the use of an effective optical metric. This provides a connection between the gravitational spin Hall and the MP frameworks, and an effect which has been experimentally observed \cite{SHEL_experiment,Bliokh2008}. 

Lastly, we have shown that the observer dependence of the gravitational spin Hall equations is directly related to the relativistic Hall effect \cite{Relativistic_Hall} and the Wigner(-Souriau) translations \cite{Stone2015,  Duval_chiral_fermions, DUVAL2015}. While these effects are exactly recovered (as previously discussed) in Minkowski spacetime, the discussion here generalizes them to arbitrary curved spacetimes. We have also pointed out that this effect has long been known  in the relativistic theory of motion as applied to massive objects \cite{PryceCM, MollerLectures, DixonSR, Costa2015}, and the approximately massless electromagnetic case is not significantly different (except in the aforementioned unboundedness of the set of all massless centroids). 

Our analysis of different centroids and their properties has been purely classical. In a quantum mechanical context, there are various results which state that massless particles cannot be localized when their spins are greater than $1/2$ \cite{NewtonWigner, localizability1967, localizability1969, Hegerfeldt1974, Hegerfeldt1980, bacry, bacry1988, SKAGERSTAM, SHE_QM2, KOSINSKI2018, finster2020}. While this appears to be at least qualitatively related to our result that massless classical objects with finite spin cannot be localized, the meanings of ``particle'' and ``localization'' are different in both contexts. It would nevertheless be interesting to better understand the connections between these results.

\section*{Acknowledgments}

We would like to thank Lars Andersson, Peter Horv\'{a}thy, J\'er\'emie Joudioux and Michael Stone for helpful discussions. M.A.O. is grateful for financial support from the Erwin Schrödinger International Institute for Mathematics and Physics.

\appendix

\section*{Appendix: Massless spinning objects violate energy conditions}
\label{appendix}

\renewcommand{\theequation}{A.\arabic{equation}}

The purpose of this Appendix is to show that if a massless object has finite momentum and satisfies the dominant energy condition, its spin must vanish. For simplicity, we work in flat spacetime. We also assume that the hypersurface used to compute the linear and angular momenta is spacelike and that  the angular momentum is computed with respect to a centroid defined using \eqref{Sortho}. It follows from our results that at least within classical physics---where the dominant energy condition is usually expected to hold---``massless spinning particles'' are unphysical except as approximations. This applies, in particular, to all classical electromagnetic wave packets. 

We work here with linear and angular momenta defined via the integrals \eqref{pDef}. Using inertial coordinates together with \eqref{bitensorFlat}, first note that the massless condition $p^\alpha p_\alpha = 0$ can be written as 
 \begin{equation}
 	\int_{\Sigma_\tau} \! dS_\alpha \int_{\Sigma_\tau} \! dS'_\beta T^{\alpha}{}_{\gamma}(x) T^{\beta \gamma}(x') = 0.
 	\label{ppCond}
 \end{equation}
As $\Sigma_\tau$ is spacelike by assumption, $dS_\alpha$ must be past-directed timelike. This and the dominant energy condition, which requires that $T_{\alpha\beta} v^\alpha w^\beta \geq 0$ for any co-oriented timelike vectors $v^\alpha$ and $w^\alpha$, imply that $T^{\alpha \beta} dS_\beta$ must be future-directed causal. The inner product between any two future-directed causal vectors can never be positive, so the integrand in \eqref{ppCond} is nonpositive. If the integrand were anywhere negative, the integral would not vanish. The massless condition therefore requires that for all $x$ and $x'$ in $\Sigma_\tau$, 
\begin{equation}
    T^{\alpha}{}_\gamma (x) T^{\beta \gamma}(x') dS_\alpha dS'_\beta = 0 .
\end{equation}
Setting $x=x'$ here implies that $T^{\alpha \beta} dS_\beta$ must be null. Allowing $x$ and $x'$ to differ shows that in addition, the direction of this vector field must be constant. More precisely,
\begin{equation}
	T^{\alpha \beta} (x) dS_\beta \propto p^\alpha.
	\label{Tk}
\end{equation}

It now follows from \eqref{pDef}, \eqref{bitensorFlat}, and \eqref{Tk} that the angular momentum must have the form $S^{\alpha \beta} = p^{[\alpha} s^{\beta]}$, for some vector $s^\beta$. This makes no assumptions regarding the origin used to compute the angular momentum. However, as explained in Sec.~\ref{sec:role_of_t}, given any timelike vector $t^\alpha$, it is always possible to adjust the origin in order to ensure that $S^{\alpha\beta} t_\beta = 0$. Doing so in this case makes \textit{all} of the angular momentum  vanish: $S^{\alpha \beta} =0$.

\newpage

\bibliography{references}

\begin{thebibliography}{121}%
\makeatletter
\providecommand \@ifxundefined [1]{%
 \@ifx{#1\undefined}
}%
\providecommand \@ifnum [1]{%
 \ifnum #1\expandafter \@firstoftwo
 \else \expandafter \@secondoftwo
 \fi
}%
\providecommand \@ifx [1]{%
 \ifx #1\expandafter \@firstoftwo
 \else \expandafter \@secondoftwo
 \fi
}%
\providecommand \natexlab [1]{#1}%
\providecommand \enquote  [1]{``#1''}%
\providecommand \bibnamefont  [1]{#1}%
\providecommand \bibfnamefont [1]{#1}%
\providecommand \citenamefont [1]{#1}%
\providecommand \href@noop [0]{\@secondoftwo}%
\providecommand \href [0]{\begingroup \@sanitize@url \@href}%
\providecommand \@href[1]{\@@startlink{#1}\@@href}%
\providecommand \@@href[1]{\endgroup#1\@@endlink}%
\providecommand \@sanitize@url [0]{\catcode `\\12\catcode `\$12\catcode
  `\&12\catcode `\#12\catcode `\^12\catcode `\_12\catcode `\%12\relax}%
\providecommand \@@startlink[1]{}%
\providecommand \@@endlink[0]{}%
\providecommand \url  [0]{\begingroup\@sanitize@url \@url }%
\providecommand \@url [1]{\endgroup\@href {#1}{\urlprefix }}%
\providecommand \urlprefix  [0]{URL }%
\providecommand \Eprint [0]{\href }%
\providecommand \doibase [0]{https://doi.org/}%
\providecommand \selectlanguage [0]{\@gobble}%
\providecommand \bibinfo  [0]{\@secondoftwo}%
\providecommand \bibfield  [0]{\@secondoftwo}%
\providecommand \translation [1]{[#1]}%
\providecommand \BibitemOpen [0]{}%
\providecommand \bibitemStop [0]{}%
\providecommand \bibitemNoStop [0]{.\EOS\space}%
\providecommand \EOS [0]{\spacefactor3000\relax}%
\providecommand \BibitemShut  [1]{\csname bibitem#1\endcsname}%
\let\auto@bib@innerbib\@empty
%</preamble>
\bibitem [{\citenamefont {Misner}\ \emph {et~al.}(1973)\citenamefont {Misner},
  \citenamefont {Thorne},\ and\ \citenamefont {Wheeler}}]{MTW}%
  \BibitemOpen
  \bibfield  {author} {\bibinfo {author} {\bibfnamefont {C.~W.}\ \bibnamefont
  {Misner}}, \bibinfo {author} {\bibfnamefont {K.~S.}\ \bibnamefont {Thorne}},\
  and\ \bibinfo {author} {\bibfnamefont {J.~A.}\ \bibnamefont {Wheeler}},\
  }\href@noop {} {\emph {\bibinfo {title} {Gravitation}}}\ (\bibinfo
  {publisher} {W. H. Freeman San Francisco},\ \bibinfo {year}
  {1973})\BibitemShut {NoStop}%
\bibitem [{\citenamefont {{Schneider}}\ \emph {et~al.}(1992)\citenamefont
  {{Schneider}}, \citenamefont {{Ehlers}},\ and\ \citenamefont
  {{Falco}}}]{gravitational_lenses_book}%
  \BibitemOpen
  \bibfield  {author} {\bibinfo {author} {\bibfnamefont {P.}~\bibnamefont
  {{Schneider}}}, \bibinfo {author} {\bibfnamefont {J.}~\bibnamefont
  {{Ehlers}}},\ and\ \bibinfo {author} {\bibfnamefont {E.~E.}\ \bibnamefont
  {{Falco}}},\ }\href {https://doi.org/10.1007/978-3-662-03758-4} {\emph
  {\bibinfo {title} {Gravitational Lenses}}}\ (\bibinfo  {publisher}
  {Springer-Verlag, Berlin, Heidelberg, New York},\ \bibinfo {year}
  {1992})\BibitemShut {NoStop}%
\bibitem [{\citenamefont {Perlick}(2004)}]{Perlick2004}%
  \BibitemOpen
  \bibfield  {author} {\bibinfo {author} {\bibfnamefont {V.}~\bibnamefont
  {Perlick}},\ }\bibfield  {title} {\bibinfo {title} {{Gravitational lensing
  from a spacetime perspective}},\ }\href {https://doi.org/10.12942/lrr-2004-9}
  {\bibfield  {journal} {\bibinfo  {journal} {Living Reviews in Relativity}\
  }\textbf {\bibinfo {volume} {7}},\ \bibinfo {pages} {9} (\bibinfo {year}
  {2004})}\BibitemShut {NoStop}%
\bibitem [{\citenamefont {Cunha}\ and\ \citenamefont {Herdeiro}(2018)}]{BHS7}%
  \BibitemOpen
  \bibfield  {author} {\bibinfo {author} {\bibfnamefont {P.~V.~P.}\
  \bibnamefont {Cunha}}\ and\ \bibinfo {author} {\bibfnamefont {C.~A.~R.}\
  \bibnamefont {Herdeiro}},\ }\bibfield  {title} {\bibinfo {title} {{Shadows
  and strong gravitational lensing: A brief review}},\ }\href
  {https://doi.org/10.1007/s10714-018-2361-9} {\bibfield  {journal} {\bibinfo
  {journal} {General Relativity and Gravitation}\ }\textbf {\bibinfo {volume}
  {50}},\ \bibinfo {pages} {42} (\bibinfo {year} {2018})}\BibitemShut {NoStop}%
\bibitem [{\citenamefont {Perlick}\ and\ \citenamefont
  {Tsupko}(2022)}]{Perlick2021}%
  \BibitemOpen
  \bibfield  {author} {\bibinfo {author} {\bibfnamefont {V.}~\bibnamefont
  {Perlick}}\ and\ \bibinfo {author} {\bibfnamefont {O.~Y.}\ \bibnamefont
  {Tsupko}},\ }\bibfield  {title} {\bibinfo {title} {{Calculating black hole
  shadows: Review of analytical studies}},\ }\href
  {https://doi.org/https://doi.org/10.1016/j.physrep.2021.10.004} {\bibfield
  {journal} {\bibinfo  {journal} {Physics Reports}\ }\textbf {\bibinfo {volume}
  {947}},\ \bibinfo {pages} {1} (\bibinfo {year} {2022})}\BibitemShut {NoStop}%
\bibitem [{\citenamefont {Ehlers}(1967)}]{EhlersGeoOptics}%
  \BibitemOpen
  \bibfield  {author} {\bibinfo {author} {\bibfnamefont {J.}~\bibnamefont
  {Ehlers}},\ }\bibfield  {title} {\bibinfo {title} {Zum Übergang von der
  wellenoptik zur geometrischen optik in der allgemeinen
  relativitätstheorie},\ }\href {https://doi.org/10.1515/zna-1967-0906}
  {\bibfield  {journal} {\bibinfo  {journal} {Zeitschrift f\"{u}r
  Naturforschung A}\ }\textbf {\bibinfo {volume} {22}},\ \bibinfo {pages}
  {1328} (\bibinfo {year} {1967})}\BibitemShut {NoStop}%
\bibitem [{\citenamefont {Isaacson}(1968)}]{Isaacson1}%
  \BibitemOpen
  \bibfield  {author} {\bibinfo {author} {\bibfnamefont {R.~A.}\ \bibnamefont
  {Isaacson}},\ }\bibfield  {title} {\bibinfo {title} {Gravitational radiation
  in the limit of high frequency. i. the linear approximation and geometrical
  optics},\ }\href {https://doi.org/10.1103/PhysRev.166.1263} {\bibfield
  {journal} {\bibinfo  {journal} {Physical Review}\ }\textbf {\bibinfo {volume}
  {166}},\ \bibinfo {pages} {1263} (\bibinfo {year} {1968})}\BibitemShut
  {NoStop}%
\bibitem [{\citenamefont {Harte}(2019{\natexlab{a}})}]{HarteOptics1}%
  \BibitemOpen
  \bibfield  {author} {\bibinfo {author} {\bibfnamefont {A.~I.}\ \bibnamefont
  {Harte}},\ }\bibfield  {title} {\bibinfo {title} {{Gravitational lensing
  beyond geometric optics: I. Formalism and observables}},\ }\href
  {https://doi.org/10.1007/s10714-018-2494-x} {\bibfield  {journal} {\bibinfo
  {journal} {General Relativity and Gravitation}\ }\textbf {\bibinfo {volume}
  {51}},\ \bibinfo {pages} {14} (\bibinfo {year}
  {2019}{\natexlab{a}})}\BibitemShut {NoStop}%
\bibitem [{\citenamefont {Sbierski}(2015)}]{sbierski2015characterisation}%
  \BibitemOpen
  \bibfield  {author} {\bibinfo {author} {\bibfnamefont {J.}~\bibnamefont
  {Sbierski}},\ }\bibfield  {title} {\bibinfo {title} {{Characterisation of the
  energy of Gaussian beams on Lorentzian manifolds: With applications to black
  hole spacetimes}},\ }\href {https://doi.org/10.2140/apde.2015.8.1379}
  {\bibfield  {journal} {\bibinfo  {journal} {Analysis \& PDE}\ }\textbf
  {\bibinfo {volume} {8}},\ \bibinfo {pages} {1379} (\bibinfo {year}
  {2015})}\BibitemShut {NoStop}%
\bibitem [{\citenamefont {Sinova}\ \emph {et~al.}(2015)\citenamefont {Sinova},
  \citenamefont {Valenzuela}, \citenamefont {Wunderlich}, \citenamefont
  {Back},\ and\ \citenamefont {Jungwirth}}]{SHE_review}%
  \BibitemOpen
  \bibfield  {author} {\bibinfo {author} {\bibfnamefont {J.}~\bibnamefont
  {Sinova}}, \bibinfo {author} {\bibfnamefont {S.~O.}\ \bibnamefont
  {Valenzuela}}, \bibinfo {author} {\bibfnamefont {J.}~\bibnamefont
  {Wunderlich}}, \bibinfo {author} {\bibfnamefont {C.~H.}\ \bibnamefont
  {Back}},\ and\ \bibinfo {author} {\bibfnamefont {T.}~\bibnamefont
  {Jungwirth}},\ }\bibfield  {title} {\bibinfo {title} {{Spin Hall effects}},\
  }\href {https://doi.org/10.1103/RevModPhys.87.1213} {\bibfield  {journal}
  {\bibinfo  {journal} {Reviews of Modern Physics}\ }\textbf {\bibinfo {volume}
  {87}},\ \bibinfo {pages} {1213} (\bibinfo {year} {2015})}\BibitemShut
  {NoStop}%
\bibitem [{\citenamefont {Bliokh}\ \emph {et~al.}(2015)\citenamefont {Bliokh},
  \citenamefont {Rodr\'{i}guez-Fortu\~{n}o}, \citenamefont {Nori},\ and\
  \citenamefont {Zayats}}]{SOI_review}%
  \BibitemOpen
  \bibfield  {author} {\bibinfo {author} {\bibfnamefont {K.~Y.}\ \bibnamefont
  {Bliokh}}, \bibinfo {author} {\bibfnamefont {F.~J.}\ \bibnamefont
  {Rodr\'{i}guez-Fortu\~{n}o}}, \bibinfo {author} {\bibfnamefont
  {F.}~\bibnamefont {Nori}},\ and\ \bibinfo {author} {\bibfnamefont {A.~V.}\
  \bibnamefont {Zayats}},\ }\bibfield  {title} {\bibinfo {title} {{Spin-orbit
  interactions of light}},\ }\href {https://doi.org/10.1038/nphoton.2015.201}
  {\bibfield  {journal} {\bibinfo  {journal} {Nature Photonics}\ }\textbf
  {\bibinfo {volume} {9}},\ \bibinfo {pages} {796} (\bibinfo {year}
  {2015})}\BibitemShut {NoStop}%
\bibitem [{\citenamefont {Ling}\ \emph {et~al.}(2017)\citenamefont {Ling},
  \citenamefont {Zhou}, \citenamefont {Huang}, \citenamefont {Liu},
  \citenamefont {Qiu}, \citenamefont {Luo},\ and\ \citenamefont
  {Wen}}]{SHEL_review}%
  \BibitemOpen
  \bibfield  {author} {\bibinfo {author} {\bibfnamefont {X.}~\bibnamefont
  {Ling}}, \bibinfo {author} {\bibfnamefont {X.}~\bibnamefont {Zhou}}, \bibinfo
  {author} {\bibfnamefont {K.}~\bibnamefont {Huang}}, \bibinfo {author}
  {\bibfnamefont {Y.}~\bibnamefont {Liu}}, \bibinfo {author} {\bibfnamefont
  {C.-W.}\ \bibnamefont {Qiu}}, \bibinfo {author} {\bibfnamefont
  {H.}~\bibnamefont {Luo}},\ and\ \bibinfo {author} {\bibfnamefont
  {S.}~\bibnamefont {Wen}},\ }\bibfield  {title} {\bibinfo {title} {{Recent
  advances in the spin Hall effect of light}},\ }\href
  {http://stacks.iop.org/0034-4885/80/i=6/a=066401} {\bibfield  {journal}
  {\bibinfo  {journal} {Reports on Progress in Physics}\ }\textbf {\bibinfo
  {volume} {80}},\ \bibinfo {pages} {066401} (\bibinfo {year}
  {2017})}\BibitemShut {NoStop}%
\bibitem [{\citenamefont {Dooghin}\ \emph {et~al.}(1992)\citenamefont
  {Dooghin}, \citenamefont {Kundikova}, \citenamefont {Liberman},\ and\
  \citenamefont {Zel'dovich}}]{OpticalMagnus}%
  \BibitemOpen
  \bibfield  {author} {\bibinfo {author} {\bibfnamefont {A.~V.}\ \bibnamefont
  {Dooghin}}, \bibinfo {author} {\bibfnamefont {N.~D.}\ \bibnamefont
  {Kundikova}}, \bibinfo {author} {\bibfnamefont {V.~S.}\ \bibnamefont
  {Liberman}},\ and\ \bibinfo {author} {\bibfnamefont {B.~Y.}\ \bibnamefont
  {Zel'dovich}},\ }\bibfield  {title} {\bibinfo {title} {{Optical Magnus
  effect}},\ }\href {https://doi.org/10.1103/PhysRevA.45.8204} {\bibfield
  {journal} {\bibinfo  {journal} {Physical Review A}\ }\textbf {\bibinfo
  {volume} {45}},\ \bibinfo {pages} {8204} (\bibinfo {year}
  {1992})}\BibitemShut {NoStop}%
\bibitem [{\citenamefont {Liberman}\ and\ \citenamefont
  {Zel'dovich}(1992)}]{SHE-L_original}%
  \BibitemOpen
  \bibfield  {author} {\bibinfo {author} {\bibfnamefont {V.~S.}\ \bibnamefont
  {Liberman}}\ and\ \bibinfo {author} {\bibfnamefont {B.~Y.}\ \bibnamefont
  {Zel'dovich}},\ }\bibfield  {title} {\bibinfo {title} {{Spin-orbit
  interaction of a photon in an inhomogeneous medium}},\ }\href
  {https://doi.org/10.1103/PhysRevA.46.5199} {\bibfield  {journal} {\bibinfo
  {journal} {Physical Review A}\ }\textbf {\bibinfo {volume} {46}},\ \bibinfo
  {pages} {5199} (\bibinfo {year} {1992})}\BibitemShut {NoStop}%
\bibitem [{\citenamefont {Onoda}\ \emph {et~al.}(2004)\citenamefont {Onoda},
  \citenamefont {Murakami},\ and\ \citenamefont {Nagaosa}}]{SHE_original}%
  \BibitemOpen
  \bibfield  {author} {\bibinfo {author} {\bibfnamefont {M.}~\bibnamefont
  {Onoda}}, \bibinfo {author} {\bibfnamefont {S.}~\bibnamefont {Murakami}},\
  and\ \bibinfo {author} {\bibfnamefont {N.}~\bibnamefont {Nagaosa}},\
  }\bibfield  {title} {\bibinfo {title} {Hall effect of light},\ }\href
  {https://doi.org/10.1103/PhysRevLett.93.083901} {\bibfield  {journal}
  {\bibinfo  {journal} {Physical Review Letters}\ }\textbf {\bibinfo {volume}
  {93}},\ \bibinfo {pages} {083901} (\bibinfo {year} {2004})}\BibitemShut
  {NoStop}%
\bibitem [{\citenamefont {Bliokh}\ and\ \citenamefont
  {Bliokh}(2004{\natexlab{a}})}]{Bliokh2004}%
  \BibitemOpen
  \bibfield  {author} {\bibinfo {author} {\bibfnamefont {K.~Y.}\ \bibnamefont
  {Bliokh}}\ and\ \bibinfo {author} {\bibfnamefont {Y.~P.}\ \bibnamefont
  {Bliokh}},\ }\bibfield  {title} {\bibinfo {title} {{Modified geometrical
  optics of a smoothly inhomogeneous isotropic medium: The anisotropy, Berry
  phase, and the optical Magnus effect}},\ }\href
  {https://doi.org/10.1103/PhysRevE.70.026605} {\bibfield  {journal} {\bibinfo
  {journal} {Physical Review E}\ }\textbf {\bibinfo {volume} {70}},\ \bibinfo
  {pages} {026605} (\bibinfo {year} {2004}{\natexlab{a}})}\BibitemShut
  {NoStop}%
\bibitem [{\citenamefont {Bliokh}\ and\ \citenamefont
  {Bliokh}(2004{\natexlab{b}})}]{Bliokh2004_1}%
  \BibitemOpen
  \bibfield  {author} {\bibinfo {author} {\bibfnamefont {K.~Y.}\ \bibnamefont
  {Bliokh}}\ and\ \bibinfo {author} {\bibfnamefont {Y.~P.}\ \bibnamefont
  {Bliokh}},\ }\bibfield  {title} {\bibinfo {title} {{Topological spin
  transport of photons: The optical Magnus effect and Berry phase}},\ }\href
  {https://doi.org/https://doi.org/10.1016/j.physleta.2004.10.035} {\bibfield
  {journal} {\bibinfo  {journal} {Physics Letters A}\ }\textbf {\bibinfo
  {volume} {333}},\ \bibinfo {pages} {181} (\bibinfo {year}
  {2004}{\natexlab{b}})}\BibitemShut {NoStop}%
\bibitem [{\citenamefont {Duval}\ \emph {et~al.}(2006)\citenamefont {Duval},
  \citenamefont {Horv\'ath},\ and\ \citenamefont {Horv\'athy}}]{Duval2006}%
  \BibitemOpen
  \bibfield  {author} {\bibinfo {author} {\bibfnamefont {C.}~\bibnamefont
  {Duval}}, \bibinfo {author} {\bibfnamefont {Z.}~\bibnamefont {Horv\'ath}},\
  and\ \bibinfo {author} {\bibfnamefont {P.~A.}\ \bibnamefont {Horv\'athy}},\
  }\bibfield  {title} {\bibinfo {title} {Fermat principle for spinning light},\
  }\href {https://doi.org/10.1103/PhysRevD.74.021701} {\bibfield  {journal}
  {\bibinfo  {journal} {Physical Review D}\ }\textbf {\bibinfo {volume} {74}},\
  \bibinfo {pages} {021701} (\bibinfo {year} {2006})}\BibitemShut {NoStop}%
\bibitem [{\citenamefont {Hosten}\ and\ \citenamefont
  {Kwiat}(2008)}]{SHEL_experiment}%
  \BibitemOpen
  \bibfield  {author} {\bibinfo {author} {\bibfnamefont {O.}~\bibnamefont
  {Hosten}}\ and\ \bibinfo {author} {\bibfnamefont {P.}~\bibnamefont {Kwiat}},\
  }\bibfield  {title} {\bibinfo {title} {{Observation of the spin Hall effect
  of light via weak measurements}},\ }\href
  {https://doi.org/10.1126/science.1152697} {\bibfield  {journal} {\bibinfo
  {journal} {Science}\ }\textbf {\bibinfo {volume} {319}},\ \bibinfo {pages}
  {787} (\bibinfo {year} {2008})}\BibitemShut {NoStop}%
\bibitem [{\citenamefont {Aiello}\ and\ \citenamefont
  {Woerdman}(2008)}]{Aiello2008}%
  \BibitemOpen
  \bibfield  {author} {\bibinfo {author} {\bibfnamefont {A.}~\bibnamefont
  {Aiello}}\ and\ \bibinfo {author} {\bibfnamefont {J.~P.}\ \bibnamefont
  {Woerdman}},\ }\bibfield  {title} {\bibinfo {title} {{Role of beam
  propagation in Goos--H\"{a}nchen and Imbert--Fedorov shifts}},\ }\href
  {https://doi.org/10.1364/OL.33.001437} {\bibfield  {journal} {\bibinfo
  {journal} {Optics Letters}\ }\textbf {\bibinfo {volume} {33}},\ \bibinfo
  {pages} {1437} (\bibinfo {year} {2008})}\BibitemShut {NoStop}%
\bibitem [{\citenamefont {Bliokh}\ \emph {et~al.}(2008)\citenamefont {Bliokh},
  \citenamefont {Niv}, \citenamefont {Kleiner},\ and\ \citenamefont
  {Hasman}}]{Bliokh2008}%
  \BibitemOpen
  \bibfield  {author} {\bibinfo {author} {\bibfnamefont {K.~Y.}\ \bibnamefont
  {Bliokh}}, \bibinfo {author} {\bibfnamefont {A.}~\bibnamefont {Niv}},
  \bibinfo {author} {\bibfnamefont {V.}~\bibnamefont {Kleiner}},\ and\ \bibinfo
  {author} {\bibfnamefont {E.}~\bibnamefont {Hasman}},\ }\bibfield  {title}
  {\bibinfo {title} {{Geometrodynamics of spinning light}},\ }\href
  {http://dx.doi.org/10.1038/nphoton.2008.229} {\bibfield  {journal} {\bibinfo
  {journal} {Nature Photonics}\ }\textbf {\bibinfo {volume} {2}},\ \bibinfo
  {pages} {748} (\bibinfo {year} {2008})}\BibitemShut {NoStop}%
\bibitem [{\citenamefont {Bliokh}(2009)}]{Bliokh2009}%
  \BibitemOpen
  \bibfield  {author} {\bibinfo {author} {\bibfnamefont {K.~Y.}\ \bibnamefont
  {Bliokh}},\ }\bibfield  {title} {\bibinfo {title} {{Geometrodynamics of
  polarized light: Berry phase and spin Hall effect in a gradient-index
  medium}},\ }\href {https://doi.org/10.1088/1464-4258/11/9/094009} {\bibfield
  {journal} {\bibinfo  {journal} {Journal of Optics A: Pure and Applied
  Optics}\ }\textbf {\bibinfo {volume} {11}},\ \bibinfo {pages} {094009}
  (\bibinfo {year} {2009})}\BibitemShut {NoStop}%
\bibitem [{\citenamefont {Aiello}\ \emph {et~al.}(2009)\citenamefont {Aiello},
  \citenamefont {Lindlein}, \citenamefont {Marquardt},\ and\ \citenamefont
  {Leuchs}}]{Aiello2009}%
  \BibitemOpen
  \bibfield  {author} {\bibinfo {author} {\bibfnamefont {A.}~\bibnamefont
  {Aiello}}, \bibinfo {author} {\bibfnamefont {N.}~\bibnamefont {Lindlein}},
  \bibinfo {author} {\bibfnamefont {C.}~\bibnamefont {Marquardt}},\ and\
  \bibinfo {author} {\bibfnamefont {G.}~\bibnamefont {Leuchs}},\ }\bibfield
  {title} {\bibinfo {title} {{Transverse Angular Momentum and Geometric Spin
  Hall Effect of Light}},\ }\href
  {https://doi.org/10.1103/PhysRevLett.103.100401} {\bibfield  {journal}
  {\bibinfo  {journal} {Physical Review Letters}\ }\textbf {\bibinfo {volume}
  {103}},\ \bibinfo {pages} {100401} (\bibinfo {year} {2009})}\BibitemShut
  {NoStop}%
\bibitem [{\citenamefont {Korger}\ \emph {et~al.}(2011)\citenamefont {Korger},
  \citenamefont {Aiello}, \citenamefont {Gabriel}, \citenamefont {Banzer},
  \citenamefont {Kolb}, \citenamefont {Marquardt},\ and\ \citenamefont
  {Leuchs}}]{Korger2011}%
  \BibitemOpen
  \bibfield  {author} {\bibinfo {author} {\bibfnamefont {J.}~\bibnamefont
  {Korger}}, \bibinfo {author} {\bibfnamefont {A.}~\bibnamefont {Aiello}},
  \bibinfo {author} {\bibfnamefont {C.}~\bibnamefont {Gabriel}}, \bibinfo
  {author} {\bibfnamefont {P.}~\bibnamefont {Banzer}}, \bibinfo {author}
  {\bibfnamefont {T.}~\bibnamefont {Kolb}}, \bibinfo {author} {\bibfnamefont
  {C.}~\bibnamefont {Marquardt}},\ and\ \bibinfo {author} {\bibfnamefont
  {G.}~\bibnamefont {Leuchs}},\ }\bibfield  {title} {\bibinfo {title}
  {{Geometric Spin Hall Effect of Light at polarizing interfaces}},\ }\href
  {https://doi.org/10.1007/s00340-011-4400-z} {\bibfield  {journal} {\bibinfo
  {journal} {Applied Physics B}\ }\textbf {\bibinfo {volume} {102}},\ \bibinfo
  {pages} {427} (\bibinfo {year} {2011})}\BibitemShut {NoStop}%
\bibitem [{\citenamefont {Korger}\ \emph {et~al.}(2014)\citenamefont {Korger},
  \citenamefont {Aiello}, \citenamefont {Chille}, \citenamefont {Banzer},
  \citenamefont {Wittmann}, \citenamefont {Lindlein}, \citenamefont
  {Marquardt},\ and\ \citenamefont {Leuchs}}]{Korger2014}%
  \BibitemOpen
  \bibfield  {author} {\bibinfo {author} {\bibfnamefont {J.}~\bibnamefont
  {Korger}}, \bibinfo {author} {\bibfnamefont {A.}~\bibnamefont {Aiello}},
  \bibinfo {author} {\bibfnamefont {V.}~\bibnamefont {Chille}}, \bibinfo
  {author} {\bibfnamefont {P.}~\bibnamefont {Banzer}}, \bibinfo {author}
  {\bibfnamefont {C.}~\bibnamefont {Wittmann}}, \bibinfo {author}
  {\bibfnamefont {N.}~\bibnamefont {Lindlein}}, \bibinfo {author}
  {\bibfnamefont {C.}~\bibnamefont {Marquardt}},\ and\ \bibinfo {author}
  {\bibfnamefont {G.}~\bibnamefont {Leuchs}},\ }\bibfield  {title} {\bibinfo
  {title} {{Observation of the Geometric Spin Hall Effect of Light}},\ }\href
  {https://doi.org/10.1103/PhysRevLett.112.113902} {\bibfield  {journal}
  {\bibinfo  {journal} {Physical Review Letters}\ }\textbf {\bibinfo {volume}
  {112}},\ \bibinfo {pages} {113902} (\bibinfo {year} {2014})}\BibitemShut
  {NoStop}%
\bibitem [{\citenamefont {Bliokh}\ and\ \citenamefont
  {Nori}(2012)}]{Relativistic_Hall}%
  \BibitemOpen
  \bibfield  {author} {\bibinfo {author} {\bibfnamefont {K.~Y.}\ \bibnamefont
  {Bliokh}}\ and\ \bibinfo {author} {\bibfnamefont {F.}~\bibnamefont {Nori}},\
  }\bibfield  {title} {\bibinfo {title} {{Relativistic Hall Effect}},\ }\href
  {https://doi.org/10.1103/PhysRevLett.108.120403} {\bibfield  {journal}
  {\bibinfo  {journal} {Physical Review Letters}\ }\textbf {\bibinfo {volume}
  {108}},\ \bibinfo {pages} {120403} (\bibinfo {year} {2012})}\BibitemShut
  {NoStop}%
\bibitem [{\citenamefont {Stone}\ \emph
  {et~al.}(2015{\natexlab{a}})\citenamefont {Stone}, \citenamefont {Dwivedi},\
  and\ \citenamefont {Zhou}}]{Stone2015}%
  \BibitemOpen
  \bibfield  {author} {\bibinfo {author} {\bibfnamefont {M.}~\bibnamefont
  {Stone}}, \bibinfo {author} {\bibfnamefont {V.}~\bibnamefont {Dwivedi}},\
  and\ \bibinfo {author} {\bibfnamefont {T.}~\bibnamefont {Zhou}},\ }\bibfield
  {title} {\bibinfo {title} {{Wigner Translations and the Observer Dependence
  of the Position of Massless Spinning Particles}},\ }\href
  {https://doi.org/10.1103/PhysRevLett.114.210402} {\bibfield  {journal}
  {\bibinfo  {journal} {Physical Review Letters}\ }\textbf {\bibinfo {volume}
  {114}},\ \bibinfo {pages} {210402} (\bibinfo {year}
  {2015}{\natexlab{a}})}\BibitemShut {NoStop}%
\bibitem [{\citenamefont {Duval}\ and\ \citenamefont
  {Horv\'athy}(2015)}]{Duval_chiral_fermions}%
  \BibitemOpen
  \bibfield  {author} {\bibinfo {author} {\bibfnamefont {C.}~\bibnamefont
  {Duval}}\ and\ \bibinfo {author} {\bibfnamefont {P.~A.}\ \bibnamefont
  {Horv\'athy}},\ }\bibfield  {title} {\bibinfo {title} {Chiral fermions as
  classical massless spinning particles},\ }\href
  {https://doi.org/10.1103/PhysRevD.91.045013} {\bibfield  {journal} {\bibinfo
  {journal} {Physical Review D}\ }\textbf {\bibinfo {volume} {91}},\ \bibinfo
  {pages} {045013} (\bibinfo {year} {2015})}\BibitemShut {NoStop}%
\bibitem [{\citenamefont {Duval}\ \emph {et~al.}(2015)\citenamefont {Duval},
  \citenamefont {Elbistan}, \citenamefont {Horváthy},\ and\ \citenamefont
  {Zhang}}]{DUVAL2015}%
  \BibitemOpen
  \bibfield  {author} {\bibinfo {author} {\bibfnamefont {C.}~\bibnamefont
  {Duval}}, \bibinfo {author} {\bibfnamefont {M.}~\bibnamefont {Elbistan}},
  \bibinfo {author} {\bibfnamefont {P.}~\bibnamefont {Horváthy}},\ and\
  \bibinfo {author} {\bibfnamefont {P.-M.}\ \bibnamefont {Zhang}},\ }\bibfield
  {title} {\bibinfo {title} {{Wigner–Souriau translations and Lorentz
  symmetry of chiral fermions}},\ }\href
  {https://doi.org/https://doi.org/10.1016/j.physletb.2015.01.048} {\bibfield
  {journal} {\bibinfo  {journal} {Physics Letters B}\ }\textbf {\bibinfo
  {volume} {742}},\ \bibinfo {pages} {322 } (\bibinfo {year}
  {2015})}\BibitemShut {NoStop}%
\bibitem [{\citenamefont {Pryce}(1948)}]{PryceCM}%
  \BibitemOpen
  \bibfield  {author} {\bibinfo {author} {\bibfnamefont {M.~H.~L.}\
  \bibnamefont {Pryce}},\ }\bibfield  {title} {\bibinfo {title} {{The
  mass-centre in the restricted theory of relativity and its connexion with the
  quantum theory of elementary particles}},\ }\href
  {https://doi.org/10.1098/rspa.1948.0103} {\bibfield  {journal} {\bibinfo
  {journal} {Proceedings of the Royal Society A: Mathematical, Physical and
  Engineering Sciences}\ }\textbf {\bibinfo {volume} {195}},\ \bibinfo {pages}
  {62} (\bibinfo {year} {1948})}\BibitemShut {NoStop}%
\bibitem [{\citenamefont {M\o{}ller}(1949)}]{MollerLectures}%
  \BibitemOpen
  \bibfield  {author} {\bibinfo {author} {\bibfnamefont {C.}~\bibnamefont
  {M\o{}ller}},\ }\bibfield  {title} {\bibinfo {title} {On the definition of
  the centre of gravity of an arbitrary closed system in the theory of
  relativity},\ }\href {https://philarchive.org/archive/MLLOTDv1} {\bibfield
  {journal} {\bibinfo  {journal} {Communications of the Dublin Institute for
  Advanced Studies, Series A}\ }\textbf {\bibinfo {volume} {5}},\ \bibinfo
  {pages} {3} (\bibinfo {year} {1949})}\BibitemShut {NoStop}%
\bibitem [{\citenamefont {Dixon}(1982)}]{DixonSR}%
  \BibitemOpen
  \bibfield  {author} {\bibinfo {author} {\bibfnamefont {W.~G.}\ \bibnamefont
  {Dixon}},\ }\href@noop {} {\emph {\bibinfo {title} {{Special Relativity: The
  Foundations of Macroscopic Physics}}}}\ (\bibinfo  {publisher} {Cambridge
  University Press, Cambridge},\ \bibinfo {year} {1982})\BibitemShut {NoStop}%
\bibitem [{\citenamefont {Costa}\ and\ \citenamefont
  {Nat{\'a}rio}(2015)}]{Costa2015}%
  \BibitemOpen
  \bibfield  {author} {\bibinfo {author} {\bibfnamefont {L.~F.~O.}\
  \bibnamefont {Costa}}\ and\ \bibinfo {author} {\bibfnamefont
  {J.}~\bibnamefont {Nat{\'a}rio}},\ }\bibinfo {title} {Center of mass, spin
  supplementary conditions, and the momentum of spinning particles},\ in\ \href
  {https://doi.org/10.1007/978-3-319-18335-0_6} {\emph {\bibinfo {booktitle}
  {Equations of Motion in Relativistic Gravity}}},\ \bibinfo {editor} {edited
  by\ \bibinfo {editor} {\bibfnamefont {D.}~\bibnamefont {Puetzfeld}}, \bibinfo
  {editor} {\bibfnamefont {C.}~\bibnamefont {L{\"a}mmerzahl}},\ and\ \bibinfo
  {editor} {\bibfnamefont {B.}~\bibnamefont {Schutz}}}\ (\bibinfo  {publisher}
  {Springer},\ \bibinfo {address} {Cham},\ \bibinfo {year} {2015})\ pp.\
  \bibinfo {pages} {215--258}\BibitemShut {NoStop}%
\bibitem [{\citenamefont {Oancea}\ \emph {et~al.}(2019)\citenamefont {Oancea},
  \citenamefont {Paganini}, \citenamefont {Joudioux},\ and\ \citenamefont
  {Andersson}}]{GSHE_review}%
  \BibitemOpen
  \bibfield  {author} {\bibinfo {author} {\bibfnamefont {M.~A.}\ \bibnamefont
  {Oancea}}, \bibinfo {author} {\bibfnamefont {C.~F.}\ \bibnamefont
  {Paganini}}, \bibinfo {author} {\bibfnamefont {J.}~\bibnamefont {Joudioux}},\
  and\ \bibinfo {author} {\bibfnamefont {L.}~\bibnamefont {Andersson}},\
  }\bibfield  {title} {\bibinfo {title} {{An overview of the gravitational spin
  Hall effect}},\ }\href {https://arxiv.org/abs/1904.09963} {\bibfield
  {journal} {\bibinfo  {journal} {arXiv:1904.09963}\ } (\bibinfo {year}
  {2019})}\BibitemShut {NoStop}%
\bibitem [{\citenamefont {Oancea}(2021)}]{Oanceathesis}%
  \BibitemOpen
  \bibfield  {author} {\bibinfo {author} {\bibfnamefont {M.~A.}\ \bibnamefont
  {Oancea}},\ }\emph {\bibinfo {title} {{Spin Hall effects in General
  Relativity}}},\ \href {https://doi.org/10.25932/publishup-50229} {\bibinfo
  {type} {{PhD thesis}}},\ \bibinfo  {school} {University of Potsdam} (\bibinfo
  {year} {2021})\BibitemShut {NoStop}%
\bibitem [{\citenamefont {Frolov}\ and\ \citenamefont {Shoom}(2011)}]{Frolov}%
  \BibitemOpen
  \bibfield  {author} {\bibinfo {author} {\bibfnamefont {V.~P.}\ \bibnamefont
  {Frolov}}\ and\ \bibinfo {author} {\bibfnamefont {A.~A.}\ \bibnamefont
  {Shoom}},\ }\bibfield  {title} {\bibinfo {title} {Spinoptics in a stationary
  spacetime},\ }\href {https://doi.org/10.1103/PhysRevD.84.044026} {\bibfield
  {journal} {\bibinfo  {journal} {Physical Review D}\ }\textbf {\bibinfo
  {volume} {84}},\ \bibinfo {pages} {044026} (\bibinfo {year}
  {2011})}\BibitemShut {NoStop}%
\bibitem [{\citenamefont {Frolov}\ and\ \citenamefont {Shoom}(2012)}]{Frolov2}%
  \BibitemOpen
  \bibfield  {author} {\bibinfo {author} {\bibfnamefont {V.~P.}\ \bibnamefont
  {Frolov}}\ and\ \bibinfo {author} {\bibfnamefont {A.~A.}\ \bibnamefont
  {Shoom}},\ }\bibfield  {title} {\bibinfo {title} {Scattering of circularly
  polarized light by a rotating black hole},\ }\href
  {https://doi.org/10.1103/PhysRevD.86.024010} {\bibfield  {journal} {\bibinfo
  {journal} {Physical Review D}\ }\textbf {\bibinfo {volume} {86}},\ \bibinfo
  {pages} {024010} (\bibinfo {year} {2012})}\BibitemShut {NoStop}%
\bibitem [{\citenamefont {Yoo}(2012)}]{covariantSpinoptics}%
  \BibitemOpen
  \bibfield  {author} {\bibinfo {author} {\bibfnamefont {C.-M.}\ \bibnamefont
  {Yoo}},\ }\bibfield  {title} {\bibinfo {title} {Notes on spinoptics in a
  stationary spacetime},\ }\href {https://doi.org/10.1103/PhysRevD.86.084005}
  {\bibfield  {journal} {\bibinfo  {journal} {Physical Review D}\ }\textbf
  {\bibinfo {volume} {86}},\ \bibinfo {pages} {084005} (\bibinfo {year}
  {2012})}\BibitemShut {NoStop}%
\bibitem [{\citenamefont {Harte}(2019{\natexlab{b}})}]{Harte2018}%
  \BibitemOpen
  \bibfield  {author} {\bibinfo {author} {\bibfnamefont {A.~I.}\ \bibnamefont
  {Harte}},\ }\bibfield  {title} {\bibinfo {title} {{Gravitational lensing
  beyond geometric optics: I. Formalism and observables}},\ }\href
  {https://doi.org/10.1007/s10714-018-2494-x} {\bibfield  {journal} {\bibinfo
  {journal} {General Relativity and Gravitation}\ }\textbf {\bibinfo {volume}
  {51}},\ \bibinfo {pages} {14} (\bibinfo {year}
  {2019}{\natexlab{b}})}\BibitemShut {NoStop}%
\bibitem [{\citenamefont {Dolan}(2018{\natexlab{a}})}]{spinorSpinoptics}%
  \BibitemOpen
  \bibfield  {author} {\bibinfo {author} {\bibfnamefont {S.~R.}\ \bibnamefont
  {Dolan}},\ }\bibfield  {title} {\bibinfo {title} {{Higher-order geometrical
  optics for circularly-polarized electromagnetic waves}},\ }\href
  {https://arxiv.org/abs/1801.02273} {\bibfield  {journal} {\bibinfo  {journal}
  {arXiv:1801.02273}\ } (\bibinfo {year} {2018}{\natexlab{a}})}\BibitemShut
  {NoStop}%
\bibitem [{\citenamefont {Dolan}(2018{\natexlab{b}})}]{spinorSpinoptics2}%
  \BibitemOpen
  \bibfield  {author} {\bibinfo {author} {\bibfnamefont {S.~R.}\ \bibnamefont
  {Dolan}},\ }\bibfield  {title} {\bibinfo {title} {{Geometrical optics for
  scalar, electromagnetic and gravitational waves on curved spacetime}},\
  }\href {https://doi.org/10.1142/S0218271818430101} {\bibfield  {journal}
  {\bibinfo  {journal} {International Journal of Modern Physics D}\ }\textbf
  {\bibinfo {volume} {27}},\ \bibinfo {pages} {1843010} (\bibinfo {year}
  {2018}{\natexlab{b}})}\BibitemShut {NoStop}%
\bibitem [{\citenamefont {Oancea}\ \emph {et~al.}(2020)\citenamefont {Oancea},
  \citenamefont {Joudioux}, \citenamefont {Dodin}, \citenamefont {Ruiz},
  \citenamefont {Paganini},\ and\ \citenamefont {Andersson}}]{GSHE2020}%
  \BibitemOpen
  \bibfield  {author} {\bibinfo {author} {\bibfnamefont {M.~A.}\ \bibnamefont
  {Oancea}}, \bibinfo {author} {\bibfnamefont {J.}~\bibnamefont {Joudioux}},
  \bibinfo {author} {\bibfnamefont {I.~Y.}\ \bibnamefont {Dodin}}, \bibinfo
  {author} {\bibfnamefont {D.~E.}\ \bibnamefont {Ruiz}}, \bibinfo {author}
  {\bibfnamefont {C.~F.}\ \bibnamefont {Paganini}},\ and\ \bibinfo {author}
  {\bibfnamefont {L.}~\bibnamefont {Andersson}},\ }\bibfield  {title} {\bibinfo
  {title} {{Gravitational spin Hall effect of light}},\ }\href
  {https://doi.org/10.1103/PhysRevD.102.024075} {\bibfield  {journal} {\bibinfo
   {journal} {Physical Review D}\ }\textbf {\bibinfo {volume} {102}},\ \bibinfo
  {pages} {024075} (\bibinfo {year} {2020})}\BibitemShut {NoStop}%
\bibitem [{\citenamefont {Shoom}(2021)}]{shoom2020}%
  \BibitemOpen
  \bibfield  {author} {\bibinfo {author} {\bibfnamefont {A.~A.}\ \bibnamefont
  {Shoom}},\ }\bibfield  {title} {\bibinfo {title} {{Gravitational Faraday and
  spin-Hall effects of light}},\ }\href
  {https://doi.org/10.1103/PhysRevD.104.084007} {\bibfield  {journal} {\bibinfo
   {journal} {Physical Review D}\ }\textbf {\bibinfo {volume} {104}},\ \bibinfo
  {pages} {084007} (\bibinfo {year} {2021})}\BibitemShut {NoStop}%
\bibitem [{\citenamefont {Frolov}(2020)}]{Frolov2020}%
  \BibitemOpen
  \bibfield  {author} {\bibinfo {author} {\bibfnamefont {V.~P.}\ \bibnamefont
  {Frolov}},\ }\bibfield  {title} {\bibinfo {title} {Maxwell equations in a
  curved spacetime: Spin optics approximation},\ }\href
  {https://doi.org/10.1103/PhysRevD.102.084013} {\bibfield  {journal} {\bibinfo
   {journal} {Physical Review D}\ }\textbf {\bibinfo {volume} {102}},\ \bibinfo
  {pages} {084013} (\bibinfo {year} {2020})}\BibitemShut {NoStop}%
\bibitem [{\citenamefont {Andersson}\ \emph {et~al.}(2021)\citenamefont
  {Andersson}, \citenamefont {Joudioux}, \citenamefont {Oancea},\ and\
  \citenamefont {Raj}}]{GSHE_GW}%
  \BibitemOpen
  \bibfield  {author} {\bibinfo {author} {\bibfnamefont {L.}~\bibnamefont
  {Andersson}}, \bibinfo {author} {\bibfnamefont {J.}~\bibnamefont {Joudioux}},
  \bibinfo {author} {\bibfnamefont {M.~A.}\ \bibnamefont {Oancea}},\ and\
  \bibinfo {author} {\bibfnamefont {A.}~\bibnamefont {Raj}},\ }\bibfield
  {title} {\bibinfo {title} {Propagation of polarized gravitational waves},\
  }\href {https://doi.org/10.1103/PhysRevD.103.044053} {\bibfield  {journal}
  {\bibinfo  {journal} {Physical Review D}\ }\textbf {\bibinfo {volume}
  {103}},\ \bibinfo {pages} {044053} (\bibinfo {year} {2021})}\BibitemShut
  {NoStop}%
\bibitem [{\citenamefont {Audretsch}(1981)}]{audretsch}%
  \BibitemOpen
  \bibfield  {author} {\bibinfo {author} {\bibfnamefont {J.}~\bibnamefont
  {Audretsch}},\ }\bibfield  {title} {\bibinfo {title} {{Trajectories and spin
  motion of massive spin-$\frac{1}{2}$ particles in gravitational fields}},\
  }\href {https://doi.org/https://doi.org/10.1088/0305-4470/14/2/017}
  {\bibfield  {journal} {\bibinfo  {journal} {Journal of Physics A:
  Mathematical and General}\ }\textbf {\bibinfo {volume} {14}},\ \bibinfo
  {pages} {411} (\bibinfo {year} {1981})}\BibitemShut {NoStop}%
\bibitem [{\citenamefont {R{\"u}diger}(1981{\natexlab{a}})}]{rudiger}%
  \BibitemOpen
  \bibfield  {author} {\bibinfo {author} {\bibfnamefont {R.}~\bibnamefont
  {R{\"u}diger}},\ }\bibfield  {title} {\bibinfo {title} {{The Dirac equation
  and spinning particles in general relativity}},\ }\href
  {https://doi.org/https://doi.org/10.1098/rspa.1981.0132} {\bibfield
  {journal} {\bibinfo  {journal} {Proceedings of the Royal Society A:
  Mathematical, Physical and Engineering Sciences}\ }\textbf {\bibinfo {volume}
  {377}},\ \bibinfo {pages} {417} (\bibinfo {year}
  {1981}{\natexlab{a}})}\BibitemShut {NoStop}%
\bibitem [{\citenamefont {Gosselin}\ \emph
  {et~al.}(2007{\natexlab{a}})\citenamefont {Gosselin}, \citenamefont
  {B{\'e}rard},\ and\ \citenamefont {Mohrbach}}]{SHE_QM1}%
  \BibitemOpen
  \bibfield  {author} {\bibinfo {author} {\bibfnamefont {P.}~\bibnamefont
  {Gosselin}}, \bibinfo {author} {\bibfnamefont {A.}~\bibnamefont
  {B{\'e}rard}},\ and\ \bibinfo {author} {\bibfnamefont {H.}~\bibnamefont
  {Mohrbach}},\ }\bibfield  {title} {\bibinfo {title} {{Spin Hall effect of
  photons in a static gravitational field}},\ }\href
  {https://doi.org/10.1103/PhysRevD.75.084035} {\bibfield  {journal} {\bibinfo
  {journal} {Physical Review D}\ }\textbf {\bibinfo {volume} {75}},\ \bibinfo
  {pages} {084035} (\bibinfo {year} {2007}{\natexlab{a}})}\BibitemShut
  {NoStop}%
\bibitem [{\citenamefont {Gosselin}\ \emph
  {et~al.}(2007{\natexlab{b}})\citenamefont {Gosselin}, \citenamefont
  {B{\'e}rard},\ and\ \citenamefont {Mohrbach}}]{SHE_Dirac}%
  \BibitemOpen
  \bibfield  {author} {\bibinfo {author} {\bibfnamefont {P.}~\bibnamefont
  {Gosselin}}, \bibinfo {author} {\bibfnamefont {A.}~\bibnamefont
  {B{\'e}rard}},\ and\ \bibinfo {author} {\bibfnamefont {H.}~\bibnamefont
  {Mohrbach}},\ }\bibfield  {title} {\bibinfo {title} {{Semiclassical dynamics
  of Dirac particles interacting with a static gravitational field}},\ }\href
  {https://doi.org/https://doi.org/10.1016/j.physleta.2007.04.022} {\bibfield
  {journal} {\bibinfo  {journal} {Physics Letters A}\ }\textbf {\bibinfo
  {volume} {368}},\ \bibinfo {pages} {356} (\bibinfo {year}
  {2007}{\natexlab{b}})}\BibitemShut {NoStop}%
\bibitem [{\citenamefont {Yamamoto}(2018)}]{SHE_GW}%
  \BibitemOpen
  \bibfield  {author} {\bibinfo {author} {\bibfnamefont {N.}~\bibnamefont
  {Yamamoto}},\ }\bibfield  {title} {\bibinfo {title} {{Spin Hall effect of
  gravitational waves}},\ }\href {https://doi.org/10.1103/PhysRevD.98.061701}
  {\bibfield  {journal} {\bibinfo  {journal} {Physical Review D}\ }\textbf
  {\bibinfo {volume} {98}},\ \bibinfo {pages} {061701} (\bibinfo {year}
  {2018})}\BibitemShut {NoStop}%
\bibitem [{\citenamefont {Souriau}(1974)}]{souriau1974modele}%
  \BibitemOpen
  \bibfield  {author} {\bibinfo {author} {\bibfnamefont {J.-M.}\ \bibnamefont
  {Souriau}},\ }\bibfield  {title} {\bibinfo {title} {Modèle de particule à
  spin dans le champ électromagnétique et gravitationnel},\ }\href
  {http://www.numdam.org/item/AIHPA_1974__20_4_315_0/} {\bibfield  {journal}
  {\bibinfo  {journal} {Annales de l'Institut Henri Poincar{\'e} A}\ }\textbf
  {\bibinfo {volume} {20}},\ \bibinfo {pages} {315} (\bibinfo {year}
  {1974})}\BibitemShut {NoStop}%
\bibitem [{\citenamefont {Saturnini}(1976)}]{saturnini1976modele}%
  \BibitemOpen
  \bibfield  {author} {\bibinfo {author} {\bibfnamefont {P.}~\bibnamefont
  {Saturnini}},\ }\emph {\bibinfo {title} {{Un mod{\`e}le de particule {\`a}
  spin de masse nulle dans le champ de gravitation}}},\ \href
  {https://hal.archives-ouvertes.fr/tel-01344863} {\bibinfo {type} {{PhD
  thesis}}},\ \bibinfo  {school} {{Universit{\'e} de Provence}} (\bibinfo
  {year} {1976})\BibitemShut {NoStop}%
\bibitem [{\citenamefont {Mashhoon}(1975)}]{Mashhoon1975}%
  \BibitemOpen
  \bibfield  {author} {\bibinfo {author} {\bibfnamefont {B.}~\bibnamefont
  {Mashhoon}},\ }\bibfield  {title} {\bibinfo {title} {Massless spinning test
  particles in a gravitational field},\ }\href
  {https://doi.org/10.1016/0003-4916(75)90304-8} {\bibfield  {journal}
  {\bibinfo  {journal} {Annals of Physics}\ }\textbf {\bibinfo {volume} {89}},\
  \bibinfo {pages} {254} (\bibinfo {year} {1975})}\BibitemShut {NoStop}%
\bibitem [{\citenamefont {Bailyn}\ and\ \citenamefont
  {Ragusa}(1977)}]{bailyn1977pole}%
  \BibitemOpen
  \bibfield  {author} {\bibinfo {author} {\bibfnamefont {M.}~\bibnamefont
  {Bailyn}}\ and\ \bibinfo {author} {\bibfnamefont {S.}~\bibnamefont
  {Ragusa}},\ }\bibfield  {title} {\bibinfo {title} {Pole-dipole model of
  massless particles},\ }\href
  {https://doi.org/https://doi.org/10.1103/PhysRevD.15.3543} {\bibfield
  {journal} {\bibinfo  {journal} {Physical Review D}\ }\textbf {\bibinfo
  {volume} {15}},\ \bibinfo {pages} {3543} (\bibinfo {year}
  {1977})}\BibitemShut {NoStop}%
\bibitem [{\citenamefont {Bailyn}\ and\ \citenamefont
  {Ragusa}(1981)}]{bailyn1981pole}%
  \BibitemOpen
  \bibfield  {author} {\bibinfo {author} {\bibfnamefont {M.}~\bibnamefont
  {Bailyn}}\ and\ \bibinfo {author} {\bibfnamefont {S.}~\bibnamefont
  {Ragusa}},\ }\bibfield  {title} {\bibinfo {title} {{Pole-dipole model of
  massless particles. II}},\ }\href
  {https://doi.org/https://doi.org/10.1103/PhysRevD.23.1258} {\bibfield
  {journal} {\bibinfo  {journal} {Physical Review D}\ }\textbf {\bibinfo
  {volume} {23}},\ \bibinfo {pages} {1258} (\bibinfo {year}
  {1981})}\BibitemShut {NoStop}%
\bibitem [{\citenamefont {Bini}\ \emph {et~al.}(2006)\citenamefont {Bini},
  \citenamefont {Cherubini}, \citenamefont {Geralico},\ and\ \citenamefont
  {Jantzen}}]{bini2006massless}%
  \BibitemOpen
  \bibfield  {author} {\bibinfo {author} {\bibfnamefont {D.}~\bibnamefont
  {Bini}}, \bibinfo {author} {\bibfnamefont {C.}~\bibnamefont {Cherubini}},
  \bibinfo {author} {\bibfnamefont {A.}~\bibnamefont {Geralico}},\ and\
  \bibinfo {author} {\bibfnamefont {R.~T.}\ \bibnamefont {Jantzen}},\
  }\bibfield  {title} {\bibinfo {title} {Massless spinning test particles in
  algebraically special vacuum space--times},\ }\href
  {https://doi.org/10.1142/S0218271806008498} {\bibfield  {journal} {\bibinfo
  {journal} {International Journal of Modern Physics D}\ }\textbf {\bibinfo
  {volume} {15}},\ \bibinfo {pages} {737} (\bibinfo {year} {2006})}\BibitemShut
  {NoStop}%
\bibitem [{\citenamefont {Semer{\'a}k}(2015)}]{semerak2015spinning}%
  \BibitemOpen
  \bibfield  {author} {\bibinfo {author} {\bibfnamefont {O.}~\bibnamefont
  {Semer{\'a}k}},\ }\bibfield  {title} {\bibinfo {title} {{Spinning particles
  in vacuum spacetimes of different curvature types: Natural reference tetrads
  and massless particles}},\ }\href
  {https://doi.org/https://doi.org/10.1103/PhysRevD.92.124036} {\bibfield
  {journal} {\bibinfo  {journal} {Physical Review D}\ }\textbf {\bibinfo
  {volume} {92}},\ \bibinfo {pages} {124036} (\bibinfo {year}
  {2015})}\BibitemShut {NoStop}%
\bibitem [{\citenamefont {Littlejohn}\ and\ \citenamefont
  {Flynn}(1991)}]{Littlejohn1991}%
  \BibitemOpen
  \bibfield  {author} {\bibinfo {author} {\bibfnamefont {R.~G.}\ \bibnamefont
  {Littlejohn}}\ and\ \bibinfo {author} {\bibfnamefont {W.~G.}\ \bibnamefont
  {Flynn}},\ }\bibfield  {title} {\bibinfo {title} {Geometric phases in the
  asymptotic theory of coupled wave equations},\ }\href
  {https://doi.org/10.1103/PhysRevA.44.5239} {\bibfield  {journal} {\bibinfo
  {journal} {Physical Review A}\ }\textbf {\bibinfo {volume} {44}},\ \bibinfo
  {pages} {5239} (\bibinfo {year} {1991})}\BibitemShut {NoStop}%
\bibitem [{\citenamefont {Shockley}\ and\ \citenamefont
  {James}(1967)}]{Shockley1967}%
  \BibitemOpen
  \bibfield  {author} {\bibinfo {author} {\bibfnamefont {W.}~\bibnamefont
  {Shockley}}\ and\ \bibinfo {author} {\bibfnamefont {R.~P.}\ \bibnamefont
  {James}},\ }\bibfield  {title} {\bibinfo {title} {{"Try Simplest Cases"
  Discovery of "Hidden Momentum" Forces on "Magnetic Currents"}},\ }\href
  {https://doi.org/10.1103/PhysRevLett.18.876} {\bibfield  {journal} {\bibinfo
  {journal} {Physical Review Letters}\ }\textbf {\bibinfo {volume} {18}},\
  \bibinfo {pages} {876} (\bibinfo {year} {1967})}\BibitemShut {NoStop}%
\bibitem [{\citenamefont {Coleman}\ and\ \citenamefont
  {Van~Vleck}(1968)}]{Coleman1968}%
  \BibitemOpen
  \bibfield  {author} {\bibinfo {author} {\bibfnamefont {S.}~\bibnamefont
  {Coleman}}\ and\ \bibinfo {author} {\bibfnamefont {J.~H.}\ \bibnamefont
  {Van~Vleck}},\ }\bibfield  {title} {\bibinfo {title} {Origin of "hidden
  momentum forces" on magnets},\ }\href
  {https://doi.org/10.1103/PhysRev.171.1370} {\bibfield  {journal} {\bibinfo
  {journal} {Physical Review}\ }\textbf {\bibinfo {volume} {171}},\ \bibinfo
  {pages} {1370} (\bibinfo {year} {1968})}\BibitemShut {NoStop}%
\bibitem [{\citenamefont {Babson}\ \emph {et~al.}(2009)\citenamefont {Babson},
  \citenamefont {Reynolds}, \citenamefont {Bjorkquist},\ and\ \citenamefont
  {Griffiths}}]{babson2009}%
  \BibitemOpen
  \bibfield  {author} {\bibinfo {author} {\bibfnamefont {D.}~\bibnamefont
  {Babson}}, \bibinfo {author} {\bibfnamefont {S.~P.}\ \bibnamefont
  {Reynolds}}, \bibinfo {author} {\bibfnamefont {R.}~\bibnamefont
  {Bjorkquist}},\ and\ \bibinfo {author} {\bibfnamefont {D.~J.}\ \bibnamefont
  {Griffiths}},\ }\bibfield  {title} {\bibinfo {title} {Hidden momentum, field
  momentum, and electromagnetic impulse},\ }\href
  {https://doi.org/10.1119/1.3152712} {\bibfield  {journal} {\bibinfo
  {journal} {American Journal of Physics}\ }\textbf {\bibinfo {volume} {77}},\
  \bibinfo {pages} {826} (\bibinfo {year} {2009})}\BibitemShut {NoStop}%
\bibitem [{\citenamefont {Gralla}\ \emph {et~al.}(2010)\citenamefont {Gralla},
  \citenamefont {Harte},\ and\ \citenamefont {Wald}}]{Gralla2010}%
  \BibitemOpen
  \bibfield  {author} {\bibinfo {author} {\bibfnamefont {S.~E.}\ \bibnamefont
  {Gralla}}, \bibinfo {author} {\bibfnamefont {A.~I.}\ \bibnamefont {Harte}},\
  and\ \bibinfo {author} {\bibfnamefont {R.~M.}\ \bibnamefont {Wald}},\
  }\bibfield  {title} {\bibinfo {title} {Bobbing and kicks in electromagnetism
  and gravity},\ }\href {https://doi.org/10.1103/PhysRevD.81.104012} {\bibfield
   {journal} {\bibinfo  {journal} {Physical Review D}\ }\textbf {\bibinfo
  {volume} {81}},\ \bibinfo {pages} {104012} (\bibinfo {year}
  {2010})}\BibitemShut {NoStop}%
\bibitem [{\citenamefont {Sundaram}\ and\ \citenamefont
  {Niu}(1999)}]{Berry_CM1}%
  \BibitemOpen
  \bibfield  {author} {\bibinfo {author} {\bibfnamefont {G.}~\bibnamefont
  {Sundaram}}\ and\ \bibinfo {author} {\bibfnamefont {Q.}~\bibnamefont {Niu}},\
  }\bibfield  {title} {\bibinfo {title} {{Wave-packet dynamics in slowly
  perturbed crystals: Gradient corrections and Berry-phase effects}},\ }\href
  {https://doi.org/10.1103/PhysRevB.59.14915} {\bibfield  {journal} {\bibinfo
  {journal} {Physical Review B}\ }\textbf {\bibinfo {volume} {59}},\ \bibinfo
  {pages} {14915} (\bibinfo {year} {1999})}\BibitemShut {NoStop}%
\bibitem [{\citenamefont {Stone}\ \emph
  {et~al.}(2015{\natexlab{b}})\citenamefont {Stone}, \citenamefont {Dwivedi},\
  and\ \citenamefont {Zhou}}]{Stone2015(2)}%
  \BibitemOpen
  \bibfield  {author} {\bibinfo {author} {\bibfnamefont {M.}~\bibnamefont
  {Stone}}, \bibinfo {author} {\bibfnamefont {V.}~\bibnamefont {Dwivedi}},\
  and\ \bibinfo {author} {\bibfnamefont {T.}~\bibnamefont {Zhou}},\ }\bibfield
  {title} {\bibinfo {title} {{Berry phase, Lorentz covariance, and anomalous
  velocity for Dirac and Weyl particles}},\ }\href
  {https://doi.org/10.1103/PhysRevD.91.025004} {\bibfield  {journal} {\bibinfo
  {journal} {Physical Review D}\ }\textbf {\bibinfo {volume} {91}},\ \bibinfo
  {pages} {025004} (\bibinfo {year} {2015}{\natexlab{b}})}\BibitemShut
  {NoStop}%
\bibitem [{\citenamefont {Stone}(2016)}]{Stone2016}%
  \BibitemOpen
  \bibfield  {author} {\bibinfo {author} {\bibfnamefont {M.}~\bibnamefont
  {Stone}},\ }\bibfield  {title} {\bibinfo {title} {{Berry phase and anomalous
  velocity of Weyl fermions and Maxwell photons}},\ }\href
  {https://doi.org/10.1142/S0217979215502495} {\bibfield  {journal} {\bibinfo
  {journal} {International Journal of Modern Physics B}\ }\textbf {\bibinfo
  {volume} {30}},\ \bibinfo {pages} {1550249} (\bibinfo {year}
  {2016})}\BibitemShut {NoStop}%
\bibitem [{\citenamefont {Dixon}(2015)}]{dixon2015new}%
  \BibitemOpen
  \bibfield  {author} {\bibinfo {author} {\bibfnamefont {W.~G.}\ \bibnamefont
  {Dixon}},\ }\bibfield  {title} {\bibinfo {title} {{The new mechanics of Myron
  Mathisson and its subsequent development}},\ }in\ \href
  {https://doi.org/https://doi.org/10.1007/978-3-319-18335-0_1} {\emph
  {\bibinfo {booktitle} {Equations of Motion in Relativistic Gravity}}}\
  (\bibinfo  {publisher} {Springer},\ \bibinfo {year} {2015})\ pp.\ \bibinfo
  {pages} {1--66}\BibitemShut {NoStop}%
\bibitem [{\citenamefont {Mathisson}(2010)}]{Mathisson}%
  \BibitemOpen
  \bibfield  {author} {\bibinfo {author} {\bibfnamefont {M.}~\bibnamefont
  {Mathisson}},\ }\bibfield  {title} {\bibinfo {title} {Republication of: New
  mechanics of material systems},\ }\href
  {https://doi.org/10.1007/s10714-010-0939-y} {\bibfield  {journal} {\bibinfo
  {journal} {General Relativity and Gravitation}\ }\textbf {\bibinfo {volume}
  {42}},\ \bibinfo {pages} {1011} (\bibinfo {year} {2010})}\BibitemShut
  {NoStop}%
\bibitem [{\citenamefont {Papapetrou}(1951)}]{Papapetrou}%
  \BibitemOpen
  \bibfield  {author} {\bibinfo {author} {\bibfnamefont {A.}~\bibnamefont
  {Papapetrou}},\ }\bibfield  {title} {\bibinfo {title} {{Spinning
  test-particles in general relativity. I}},\ }\href
  {https://doi.org/10.1098/rspa.1951.0200} {\bibfield  {journal} {\bibinfo
  {journal} {Proceedings of the Royal Society A: Mathematical, Physical and
  Engineering Sciences}\ }\textbf {\bibinfo {volume} {209}},\ \bibinfo {pages}
  {248} (\bibinfo {year} {1951})}\BibitemShut {NoStop}%
\bibitem [{\citenamefont {{Dixon}}(1974)}]{Dixon74}%
  \BibitemOpen
  \bibfield  {author} {\bibinfo {author} {\bibfnamefont {W.~G.}\ \bibnamefont
  {{Dixon}}},\ }\bibfield  {title} {\bibinfo {title} {{Dynamics of Extended
  Bodies in General Relativity. III. Equations of Motion}},\ }\href
  {https://doi.org/10.1098/rsta.1974.0046} {\bibfield  {journal} {\bibinfo
  {journal} {Philosophical Transactions of the Royal Society A: Mathematical,
  Physical and Engineering Sciences}\ }\textbf {\bibinfo {volume} {277}},\
  \bibinfo {pages} {59} (\bibinfo {year} {1974})}\BibitemShut {NoStop}%
\bibitem [{\citenamefont {Harte}(2012)}]{HarteGrav}%
  \BibitemOpen
  \bibfield  {author} {\bibinfo {author} {\bibfnamefont {A.~I.}\ \bibnamefont
  {Harte}},\ }\bibfield  {title} {\bibinfo {title} {Mechanics of extended
  masses in general relativity},\ }\href
  {https://doi.org/10.1088/0264-9381/29/5/055012} {\bibfield  {journal}
  {\bibinfo  {journal} {Classical and Quantum Gravity}\ }\textbf {\bibinfo
  {volume} {29}},\ \bibinfo {pages} {055012} (\bibinfo {year}
  {2012})}\BibitemShut {NoStop}%
\bibitem [{\citenamefont {Harte}(2015)}]{HarteReview}%
  \BibitemOpen
  \bibfield  {author} {\bibinfo {author} {\bibfnamefont {A.~I.}\ \bibnamefont
  {Harte}},\ }\bibfield  {title} {\bibinfo {title} {Motion in classical field
  theories and the foundations of the self-force problem},\ }in\ \href
  {https://doi.org/https://doi.org/10.1007/978-3-319-18335-0_12} {\emph
  {\bibinfo {booktitle} {Equations of motion in relativistic gravity}}},\
  \bibinfo {series} {Fundamental Theories of Physics}, Vol.\ \bibinfo {volume}
  {179},\ \bibinfo {editor} {edited by\ \bibinfo {editor} {\bibfnamefont
  {D.}~\bibnamefont {Puetzfeld}}, \bibinfo {editor} {\bibfnamefont
  {C.}~\bibnamefont {L\"{a}mmerzahl}},\ and\ \bibinfo {editor} {\bibfnamefont
  {B.}~\bibnamefont {Schutz}}}\ (\bibinfo  {publisher} {Springer, Cham},\
  \bibinfo {year} {2015})\ pp.\ \bibinfo {pages} {327--398}\BibitemShut
  {NoStop}%
\bibitem [{\citenamefont {Harte}(2008)}]{HarteSyms}%
  \BibitemOpen
  \bibfield  {author} {\bibinfo {author} {\bibfnamefont {A.~I.}\ \bibnamefont
  {Harte}},\ }\bibfield  {title} {\bibinfo {title} {Approximate spacetime
  symmetries and conservation laws},\ }\href
  {https://doi.org/10.1088/0264-9381/25/20/205008} {\bibfield  {journal}
  {\bibinfo  {journal} {Classical and Quantum Gravity}\ }\textbf {\bibinfo
  {volume} {25}},\ \bibinfo {pages} {205008} (\bibinfo {year}
  {2008})}\BibitemShut {NoStop}%
\bibitem [{\citenamefont {Corinaldesi}\ and\ \citenamefont
  {Papapetrou}(1951)}]{CP_ssc}%
  \BibitemOpen
  \bibfield  {author} {\bibinfo {author} {\bibfnamefont {E.}~\bibnamefont
  {Corinaldesi}}\ and\ \bibinfo {author} {\bibfnamefont {A.}~\bibnamefont
  {Papapetrou}},\ }\bibfield  {title} {\bibinfo {title} {{Spinning
  test-particles in general relativity. II}},\ }\href
  {https://doi.org/10.1098/rspa.1951.0201} {\bibfield  {journal} {\bibinfo
  {journal} {Proceedings of the Royal Society A: Mathematical, Physical and
  Engineering Sciences}\ }\textbf {\bibinfo {volume} {209}},\ \bibinfo {pages}
  {259} (\bibinfo {year} {1951})}\BibitemShut {NoStop}%
\bibitem [{\citenamefont {{Dixon}}(1970)}]{Dixon70a}%
  \BibitemOpen
  \bibfield  {author} {\bibinfo {author} {\bibfnamefont {W.~G.}\ \bibnamefont
  {{Dixon}}},\ }\bibfield  {title} {\bibinfo {title} {{Dynamics of Extended
  Bodies in General Relativity. I. Momentum and Angular Momentum}},\ }\href
  {https://doi.org/10.1098/rspa.1970.0020} {\bibfield  {journal} {\bibinfo
  {journal} {Proceedings of the Royal Society A: Mathematical, Physical and
  Engineering Sciences}\ }\textbf {\bibinfo {volume} {314}},\ \bibinfo {pages}
  {499} (\bibinfo {year} {1970})}\BibitemShut {NoStop}%
\bibitem [{\citenamefont {Synge}(1960)}]{Synge1960}%
  \BibitemOpen
  \bibfield  {author} {\bibinfo {author} {\bibfnamefont {J.~L.}\ \bibnamefont
  {Synge}},\ }\href@noop {} {\emph {\bibinfo {title} {Relativity: The General
  Theory}}}\ (\bibinfo  {publisher} {North-Holland Publishing Company,
  Amsterdam},\ \bibinfo {year} {1960})\BibitemShut {NoStop}%
\bibitem [{\citenamefont {Poisson}\ \emph {et~al.}(2011)\citenamefont
  {Poisson}, \citenamefont {Pound},\ and\ \citenamefont {Vega}}]{Poisson2011}%
  \BibitemOpen
  \bibfield  {author} {\bibinfo {author} {\bibfnamefont {E.}~\bibnamefont
  {Poisson}}, \bibinfo {author} {\bibfnamefont {A.}~\bibnamefont {Pound}},\
  and\ \bibinfo {author} {\bibfnamefont {I.}~\bibnamefont {Vega}},\ }\bibfield
  {title} {\bibinfo {title} {The motion of point particles in curved
  spacetime},\ }\href {https://doi.org/10.12942/lrr-2011-7} {\bibfield
  {journal} {\bibinfo  {journal} {Living Reviews in Relativity}\ }\textbf
  {\bibinfo {volume} {14}},\ \bibinfo {pages} {7} (\bibinfo {year}
  {2011})}\BibitemShut {NoStop}%
\bibitem [{\citenamefont
  {R{\"u}diger}(1981{\natexlab{b}})}]{rudiger1981conserved}%
  \BibitemOpen
  \bibfield  {author} {\bibinfo {author} {\bibfnamefont {R.}~\bibnamefont
  {R{\"u}diger}},\ }\bibfield  {title} {\bibinfo {title} {{Conserved quantities
  of spinning test particles in general relativity. I}},\ }\href
  {https://doi.org/https://doi.org/10.1098/rspa.1981.0046} {\bibfield
  {journal} {\bibinfo  {journal} {Proceedings of the Royal Society A:
  Mathematical, Physical and Engineering Sciences}\ }\textbf {\bibinfo {volume}
  {375}},\ \bibinfo {pages} {185} (\bibinfo {year}
  {1981}{\natexlab{b}})}\BibitemShut {NoStop}%
\bibitem [{\citenamefont {R{\"u}diger}(1983)}]{rudiger1983conserved}%
  \BibitemOpen
  \bibfield  {author} {\bibinfo {author} {\bibfnamefont {R.}~\bibnamefont
  {R{\"u}diger}},\ }\bibfield  {title} {\bibinfo {title} {{Conserved quantities
  of spinning test particles in general relativity. II}},\ }\href
  {https://doi.org/https://doi.org/10.1098/rspa.1983.0012} {\bibfield
  {journal} {\bibinfo  {journal} {Proceedings of the Royal Society A:
  Mathematical, Physical and Engineering Sciences}\ }\textbf {\bibinfo {volume}
  {385}},\ \bibinfo {pages} {229} (\bibinfo {year} {1983})}\BibitemShut
  {NoStop}%
\bibitem [{\citenamefont {Gibbons}\ \emph {et~al.}(1993)\citenamefont
  {Gibbons}, \citenamefont {Rietdijk},\ and\ \citenamefont {{van
  Holten}}}]{Gibbons1993}%
  \BibitemOpen
  \bibfield  {author} {\bibinfo {author} {\bibfnamefont {G.~W.}\ \bibnamefont
  {Gibbons}}, \bibinfo {author} {\bibfnamefont {R.~H.}\ \bibnamefont
  {Rietdijk}},\ and\ \bibinfo {author} {\bibfnamefont {J.~W.}\ \bibnamefont
  {{van Holten}}},\ }\bibfield  {title} {\bibinfo {title} {{SUSY in the sky}},\
  }\href {https://doi.org/https://doi.org/10.1016/0550-3213(93)90472-2}
  {\bibfield  {journal} {\bibinfo  {journal} {Nuclear Physics B}\ }\textbf
  {\bibinfo {volume} {404}},\ \bibinfo {pages} {42} (\bibinfo {year}
  {1993})}\BibitemShut {NoStop}%
\bibitem [{\citenamefont {Witzany}(2019)}]{MPD_conservation_2019}%
  \BibitemOpen
  \bibfield  {author} {\bibinfo {author} {\bibfnamefont {V.}~\bibnamefont
  {Witzany}},\ }\bibfield  {title} {\bibinfo {title} {{Hamilton-Jacobi equation
  for spinning particles near black holes}},\ }\href
  {https://doi.org/10.1103/PhysRevD.100.104030} {\bibfield  {journal} {\bibinfo
   {journal} {Physical Review D}\ }\textbf {\bibinfo {volume} {100}},\ \bibinfo
  {pages} {104030} (\bibinfo {year} {2019})}\BibitemShut {NoStop}%
\bibitem [{\citenamefont {Santos}\ and\ \citenamefont
  {Batista}(2020)}]{MPD_conservation_2020}%
  \BibitemOpen
  \bibfield  {author} {\bibinfo {author} {\bibfnamefont {E.~B.}\ \bibnamefont
  {Santos}}\ and\ \bibinfo {author} {\bibfnamefont {C.}~\bibnamefont
  {Batista}},\ }\bibfield  {title} {\bibinfo {title} {Conserved quantities for
  the free motion of particles with spin},\ }\href
  {https://doi.org/10.1103/PhysRevD.101.104049} {\bibfield  {journal} {\bibinfo
   {journal} {Physical Review D}\ }\textbf {\bibinfo {volume} {101}},\ \bibinfo
  {pages} {104049} (\bibinfo {year} {2020})}\BibitemShut {NoStop}%
\bibitem [{\citenamefont {Compère}\ and\ \citenamefont
  {Druart}(2022)}]{MPD_conservation_2021}%
  \BibitemOpen
  \bibfield  {author} {\bibinfo {author} {\bibfnamefont {G.}~\bibnamefont
  {Compère}}\ and\ \bibinfo {author} {\bibfnamefont {A.}~\bibnamefont
  {Druart}},\ }\bibfield  {title} {\bibinfo {title} {{Complete set of
  quasi-conserved quantities for spinning particles around Kerr}},\ }\href
  {https://doi.org/10.21468/SciPostPhys.12.1.012} {\bibfield  {journal}
  {\bibinfo  {journal} {SciPost Physics}\ }\textbf {\bibinfo {volume} {12}},\
  \bibinfo {pages} {12} (\bibinfo {year} {2022})}\BibitemShut {NoStop}%
\bibitem [{\citenamefont {Padgett}\ \emph {et~al.}(2015)\citenamefont
  {Padgett}, \citenamefont {Miatto}, \citenamefont {Lavery}, \citenamefont
  {Zeilinger},\ and\ \citenamefont {Boyd}}]{spotWidth}%
  \BibitemOpen
  \bibfield  {author} {\bibinfo {author} {\bibfnamefont {M.~J.}\ \bibnamefont
  {Padgett}}, \bibinfo {author} {\bibfnamefont {F.~M.}\ \bibnamefont {Miatto}},
  \bibinfo {author} {\bibfnamefont {M.~P.~J.}\ \bibnamefont {Lavery}}, \bibinfo
  {author} {\bibfnamefont {A.}~\bibnamefont {Zeilinger}},\ and\ \bibinfo
  {author} {\bibfnamefont {R.~W.}\ \bibnamefont {Boyd}},\ }\bibfield  {title}
  {\bibinfo {title} {Divergence of an orbital-angular-momentum-carrying beam
  upon propagation},\ }\href {https://doi.org/10.1088/1367-2630/17/2/023011}
  {\bibfield  {journal} {\bibinfo  {journal} {New Journal of Physics}\ }\textbf
  {\bibinfo {volume} {17}},\ \bibinfo {pages} {023011} (\bibinfo {year}
  {2015})}\BibitemShut {NoStop}%
\bibitem [{\citenamefont {Costa}\ \emph {et~al.}(2012)\citenamefont {Costa},
  \citenamefont {Herdeiro}, \citenamefont {Nat\'ario},\ and\ \citenamefont
  {Zilh\~ao}}]{Herdeiro2012}%
  \BibitemOpen
  \bibfield  {author} {\bibinfo {author} {\bibfnamefont {L.~F.}\ \bibnamefont
  {Costa}}, \bibinfo {author} {\bibfnamefont {C.}~\bibnamefont {Herdeiro}},
  \bibinfo {author} {\bibfnamefont {J.}~\bibnamefont {Nat\'ario}},\ and\
  \bibinfo {author} {\bibfnamefont {M.}~\bibnamefont {Zilh\~ao}},\ }\bibfield
  {title} {\bibinfo {title} {Mathisson's helical motions for a spinning
  particle: Are they unphysical?},\ }\href
  {https://doi.org/10.1103/PhysRevD.85.024001} {\bibfield  {journal} {\bibinfo
  {journal} {Physical Review D}\ }\textbf {\bibinfo {volume} {85}},\ \bibinfo
  {pages} {024001} (\bibinfo {year} {2012})}\BibitemShut {NoStop}%
\bibitem [{\citenamefont {Kontou}\ and\ \citenamefont
  {Sanders}(2020)}]{SandersEnergy}%
  \BibitemOpen
  \bibfield  {author} {\bibinfo {author} {\bibfnamefont {E.-A.}\ \bibnamefont
  {Kontou}}\ and\ \bibinfo {author} {\bibfnamefont {K.}~\bibnamefont
  {Sanders}},\ }\bibfield  {title} {\bibinfo {title} {Energy conditions in
  general relativity and quantum field theory},\ }\href
  {https://doi.org/10.1088/1361-6382/ab8fcf} {\bibfield  {journal} {\bibinfo
  {journal} {Classical and Quantum Gravity}\ }\textbf {\bibinfo {volume}
  {37}},\ \bibinfo {pages} {193001} (\bibinfo {year} {2020})}\BibitemShut
  {NoStop}%
\bibitem [{\citenamefont {Ehlers}\ and\ \citenamefont
  {Rudolph}(1977)}]{EhlersRudolph}%
  \BibitemOpen
  \bibfield  {author} {\bibinfo {author} {\bibfnamefont {J.}~\bibnamefont
  {Ehlers}}\ and\ \bibinfo {author} {\bibfnamefont {E.}~\bibnamefont
  {Rudolph}},\ }\bibfield  {title} {\bibinfo {title} {Dynamics of extended
  bodies in general relativity center-of-mass description and quasirigidity},\
  }\href {https://doi.org/https://doi.org/10.1007/BF00763547} {\bibfield
  {journal} {\bibinfo  {journal} {General Relativity and Gravitation}\ }\textbf
  {\bibinfo {volume} {8}},\ \bibinfo {pages} {197} (\bibinfo {year}
  {1977})}\BibitemShut {NoStop}%
\bibitem [{\citenamefont {Schattner}(1979)}]{SchattnerCM}%
  \BibitemOpen
  \bibfield  {author} {\bibinfo {author} {\bibfnamefont {R.}~\bibnamefont
  {Schattner}},\ }\bibfield  {title} {\bibinfo {title} {The center of mass in
  general relativity},\ }\href {https://doi.org/10.1007/bf00760221} {\bibfield
  {journal} {\bibinfo  {journal} {General Relativity and Gravitation}\ }\textbf
  {\bibinfo {volume} {10}},\ \bibinfo {pages} {377} (\bibinfo {year}
  {1979})}\BibitemShut {NoStop}%
\bibitem [{\citenamefont {Duval}\ and\ \citenamefont
  {Sch\"ucker}(2017)}]{Duval}%
  \BibitemOpen
  \bibfield  {author} {\bibinfo {author} {\bibfnamefont {C.}~\bibnamefont
  {Duval}}\ and\ \bibinfo {author} {\bibfnamefont {T.}~\bibnamefont
  {Sch\"ucker}},\ }\bibfield  {title} {\bibinfo {title} {{Gravitational
  birefringence of light in Robertson-Walker cosmologies}},\ }\href
  {https://doi.org/10.1103/PhysRevD.96.043517} {\bibfield  {journal} {\bibinfo
  {journal} {Physical Review D}\ }\textbf {\bibinfo {volume} {96}},\ \bibinfo
  {pages} {043517} (\bibinfo {year} {2017})}\BibitemShut {NoStop}%
\bibitem [{\citenamefont {Duval}\ \emph {et~al.}(2018)\citenamefont {Duval},
  \citenamefont {Pasquet}, \citenamefont {Sch\"ucker},\ and\ \citenamefont
  {Tilquin}}]{Duval2018}%
  \BibitemOpen
  \bibfield  {author} {\bibinfo {author} {\bibfnamefont {C.}~\bibnamefont
  {Duval}}, \bibinfo {author} {\bibfnamefont {J.}~\bibnamefont {Pasquet}},
  \bibinfo {author} {\bibfnamefont {T.}~\bibnamefont {Sch\"ucker}},\ and\
  \bibinfo {author} {\bibfnamefont {A.}~\bibnamefont {Tilquin}},\ }\bibfield
  {title} {\bibinfo {title} {Gravitational birefringence and an exotic formula
  for redshifts},\ }\href {https://doi.org/10.1103/PhysRevD.97.123508}
  {\bibfield  {journal} {\bibinfo  {journal} {Physical Review D}\ }\textbf
  {\bibinfo {volume} {97}},\ \bibinfo {pages} {123508} (\bibinfo {year}
  {2018})}\BibitemShut {NoStop}%
\bibitem [{\citenamefont {Duval}\ \emph {et~al.}(2019)\citenamefont {Duval},
  \citenamefont {Marsot},\ and\ \citenamefont {Sch\"ucker}}]{Duval2019}%
  \BibitemOpen
  \bibfield  {author} {\bibinfo {author} {\bibfnamefont {C.}~\bibnamefont
  {Duval}}, \bibinfo {author} {\bibfnamefont {L.}~\bibnamefont {Marsot}},\ and\
  \bibinfo {author} {\bibfnamefont {T.}~\bibnamefont {Sch\"ucker}},\ }\bibfield
   {title} {\bibinfo {title} {{Gravitational birefringence of light in
  Schwarzschild spacetime}},\ }\href
  {https://doi.org/10.1103/PhysRevD.99.124037} {\bibfield  {journal} {\bibinfo
  {journal} {Physical Review D}\ }\textbf {\bibinfo {volume} {99}},\ \bibinfo
  {pages} {124037} (\bibinfo {year} {2019})}\BibitemShut {NoStop}%
\bibitem [{\citenamefont {M\o{}ller}(1972)}]{Moller}%
  \BibitemOpen
  \bibfield  {author} {\bibinfo {author} {\bibfnamefont {C.}~\bibnamefont
  {M\o{}ller}},\ }\href@noop {} {\emph {\bibinfo {title} {The Theory of
  Relativity}}},\ \bibinfo {edition} {2nd}\ ed.\ (\bibinfo  {publisher}
  {Oxford, Clarendon Press},\ \bibinfo {year} {1972})\BibitemShut {NoStop}%
\bibitem [{\citenamefont {Vines}\ \emph {et~al.}(2016)\citenamefont {Vines},
  \citenamefont {Kunst}, \citenamefont {Steinhoff},\ and\ \citenamefont
  {Hinderer}}]{vines2016canonical}%
  \BibitemOpen
  \bibfield  {author} {\bibinfo {author} {\bibfnamefont {J.}~\bibnamefont
  {Vines}}, \bibinfo {author} {\bibfnamefont {D.}~\bibnamefont {Kunst}},
  \bibinfo {author} {\bibfnamefont {J.}~\bibnamefont {Steinhoff}},\ and\
  \bibinfo {author} {\bibfnamefont {T.}~\bibnamefont {Hinderer}},\ }\bibfield
  {title} {\bibinfo {title} {{Canonical Hamiltonian for an extended test body
  in curved spacetime: To quadratic order in spin}},\ }\href
  {https://doi.org/https://doi.org/10.1103/PhysRevD.93.103008} {\bibfield
  {journal} {\bibinfo  {journal} {Physical Review D}\ }\textbf {\bibinfo
  {volume} {93}},\ \bibinfo {pages} {103008} (\bibinfo {year}
  {2016})}\BibitemShut {NoStop}%
\bibitem [{\citenamefont {Allen}\ \emph {et~al.}(2003)\citenamefont {Allen},
  \citenamefont {Barnett},\ and\ \citenamefont {Padgett}}]{AM_Light2}%
  \BibitemOpen
  \bibfield  {author} {\bibinfo {author} {\bibfnamefont {L.}~\bibnamefont
  {Allen}}, \bibinfo {author} {\bibfnamefont {S.~M.}\ \bibnamefont {Barnett}},\
  and\ \bibinfo {author} {\bibfnamefont {M.~J.}\ \bibnamefont {Padgett}},\
  }\href {https://doi.org/https://doi.org/10.1201/9781482269017} {\emph
  {\bibinfo {title} {Optical angular momentum}}}\ (\bibinfo  {publisher} {CRC
  Press},\ \bibinfo {year} {2003})\BibitemShut {NoStop}%
\bibitem [{\citenamefont {Andrews}\ and\ \citenamefont
  {Babiker}(2012)}]{AM_Light}%
  \BibitemOpen
  \bibfield  {author} {\bibinfo {author} {\bibfnamefont {D.~L.}\ \bibnamefont
  {Andrews}}\ and\ \bibinfo {author} {\bibfnamefont {M.}~\bibnamefont
  {Babiker}},\ }\href {https://doi.org/10.1017/CBO9780511795213} {\emph
  {\bibinfo {title} {{The Angular Momentum of Light}}}}\ (\bibinfo  {publisher}
  {Cambridge University Press},\ \bibinfo {year} {2012})\BibitemShut {NoStop}%
\bibitem [{\citenamefont {Bliokh}(2006)}]{Bliokh2006}%
  \BibitemOpen
  \bibfield  {author} {\bibinfo {author} {\bibfnamefont {K.~Y.}\ \bibnamefont
  {Bliokh}},\ }\bibfield  {title} {\bibinfo {title} {{Geometrical optics of
  beams with vortices: Berry phase and orbital angular momentum Hall effect}},\
  }\href {https://doi.org/10.1103/PhysRevLett.97.043901} {\bibfield  {journal}
  {\bibinfo  {journal} {Physical Review Letters}\ }\textbf {\bibinfo {volume}
  {97}},\ \bibinfo {pages} {043901} (\bibinfo {year} {2006})}\BibitemShut
  {NoStop}%
\bibitem [{\citenamefont {Duval}\ \emph {et~al.}(2007)\citenamefont {Duval},
  \citenamefont {Horv\'ath},\ and\ \citenamefont {Horv\'athy}}]{Duval2007}%
  \BibitemOpen
  \bibfield  {author} {\bibinfo {author} {\bibfnamefont {C.}~\bibnamefont
  {Duval}}, \bibinfo {author} {\bibfnamefont {Z.}~\bibnamefont {Horv\'ath}},\
  and\ \bibinfo {author} {\bibfnamefont {P.~A.}\ \bibnamefont {Horv\'athy}},\
  }\bibfield  {title} {\bibinfo {title} {{Geometrical spinoptics and the
  optical Hall effect}},\ }\href
  {https://doi.org/https://doi.org/10.1016/j.geomphys.2006.07.003} {\bibfield
  {journal} {\bibinfo  {journal} {Journal of Geometry and Physics}\ }\textbf
  {\bibinfo {volume} {57}},\ \bibinfo {pages} {925} (\bibinfo {year}
  {2007})}\BibitemShut {NoStop}%
\bibitem [{\citenamefont {Ruiz}\ and\ \citenamefont {Dodin}(2015)}]{Ruiz2015}%
  \BibitemOpen
  \bibfield  {author} {\bibinfo {author} {\bibfnamefont {D.~E.}\ \bibnamefont
  {Ruiz}}\ and\ \bibinfo {author} {\bibfnamefont {I.~Y.}\ \bibnamefont
  {Dodin}},\ }\bibfield  {title} {\bibinfo {title} {First-principles
  variational formulation of polarization effects in geometrical optics},\
  }\href {https://doi.org/10.1103/PhysRevA.92.043805} {\bibfield  {journal}
  {\bibinfo  {journal} {Physical Review A}\ }\textbf {\bibinfo {volume} {92}},\
  \bibinfo {pages} {043805} (\bibinfo {year} {2015})}\BibitemShut {NoStop}%
\bibitem [{\citenamefont {Eddington}(1987)}]{Eddington}%
  \BibitemOpen
  \bibfield  {author} {\bibinfo {author} {\bibfnamefont {A.~S.}\ \bibnamefont
  {Eddington}},\ }\href@noop {} {\emph {\bibinfo {title} {{Space, time and
  gravitation: An outline of the general relativity theory}}}}\ (\bibinfo
  {publisher} {Cambridge University Press},\ \bibinfo {year}
  {1987})\BibitemShut {NoStop}%
\bibitem [{\citenamefont {Gordon}(1923)}]{Gordon}%
  \BibitemOpen
  \bibfield  {author} {\bibinfo {author} {\bibfnamefont {W.}~\bibnamefont
  {Gordon}},\ }\bibfield  {title} {\bibinfo {title} {Zur lichtfortpflanzung
  nach der relativit{\"a}tstheorie},\ }\href
  {https://doi.org/10.1002/andp.19233772202} {\bibfield  {journal} {\bibinfo
  {journal} {Annalen der Physik}\ }\textbf {\bibinfo {volume} {377}},\ \bibinfo
  {pages} {421} (\bibinfo {year} {1923})}\BibitemShut {NoStop}%
\bibitem [{\citenamefont {Plebanski}(1960)}]{Plebansky-Maxwell}%
  \BibitemOpen
  \bibfield  {author} {\bibinfo {author} {\bibfnamefont {J.}~\bibnamefont
  {Plebanski}},\ }\bibfield  {title} {\bibinfo {title} {Electromagnetic waves
  in gravitational fields},\ }\href {https://doi.org/10.1103/PhysRev.118.1396}
  {\bibfield  {journal} {\bibinfo  {journal} {Physical Review}\ }\textbf
  {\bibinfo {volume} {118}},\ \bibinfo {pages} {1396} (\bibinfo {year}
  {1960})}\BibitemShut {NoStop}%
\bibitem [{\citenamefont {Bialynicki-Birula}(1994)}]{Birula_wavefunction1}%
  \BibitemOpen
  \bibfield  {author} {\bibinfo {author} {\bibfnamefont {I.}~\bibnamefont
  {Bialynicki-Birula}},\ }\bibfield  {title} {{\selectlanguage
  {English}\bibinfo {title} {{On the wave function of the photon}}},\ }\href
  {https://doi.org/10.12693/APhysPolA.86.97} {\bibfield  {journal} {\bibinfo
  {journal} {Acta Physica Polonica A}\ }\textbf {\bibinfo {volume} {86}},\
  \bibinfo {pages} {97} (\bibinfo {year} {1994})}\BibitemShut {NoStop}%
\bibitem [{\citenamefont {Bialynicki-Birula}(1996)}]{Birula_wavefunction2}%
  \BibitemOpen
  \bibfield  {author} {\bibinfo {author} {\bibfnamefont {I.}~\bibnamefont
  {Bialynicki-Birula}},\ }\bibfield  {title} {\bibinfo {title} {{V Photon wave
  function}},\ }\href
  {https://doi.org/https://doi.org/10.1016/S0079-6638(08)70316-0} {\bibfield
  {journal} {\bibinfo  {journal} {Progress in Optics}\ }\textbf {\bibinfo
  {volume} {36}},\ \bibinfo {pages} {245} (\bibinfo {year} {1996})}\BibitemShut
  {NoStop}%
\bibitem [{\citenamefont {Fathi}\ and\ \citenamefont
  {Thompson}(2016)}]{cartographic_analog}%
  \BibitemOpen
  \bibfield  {author} {\bibinfo {author} {\bibfnamefont {M.}~\bibnamefont
  {Fathi}}\ and\ \bibinfo {author} {\bibfnamefont {R.~T.}\ \bibnamefont
  {Thompson}},\ }\bibfield  {title} {\bibinfo {title} {Cartographic distortions
  make dielectric spacetime analog models imperfect mimickers},\ }\href
  {https://doi.org/10.1103/PhysRevD.93.124026} {\bibfield  {journal} {\bibinfo
  {journal} {Physical Review D}\ }\textbf {\bibinfo {volume} {93}},\ \bibinfo
  {pages} {124026} (\bibinfo {year} {2016})}\BibitemShut {NoStop}%
\bibitem [{\citenamefont {Thompson}(2018)}]{covariant_dielectric}%
  \BibitemOpen
  \bibfield  {author} {\bibinfo {author} {\bibfnamefont {R.~T.}\ \bibnamefont
  {Thompson}},\ }\bibfield  {title} {\bibinfo {title} {Covariant
  electrodynamics in linear media: Optical metric},\ }\href
  {https://doi.org/10.1103/PhysRevD.97.065001} {\bibfield  {journal} {\bibinfo
  {journal} {Physical Review D}\ }\textbf {\bibinfo {volume} {97}},\ \bibinfo
  {pages} {065001} (\bibinfo {year} {2018})}\BibitemShut {NoStop}%
\bibitem [{\citenamefont {Onoda}\ \emph {et~al.}(2006)\citenamefont {Onoda},
  \citenamefont {Murakami},\ and\ \citenamefont {Nagaosa}}]{Onoda2006}%
  \BibitemOpen
  \bibfield  {author} {\bibinfo {author} {\bibfnamefont {M.}~\bibnamefont
  {Onoda}}, \bibinfo {author} {\bibfnamefont {S.}~\bibnamefont {Murakami}},\
  and\ \bibinfo {author} {\bibfnamefont {N.}~\bibnamefont {Nagaosa}},\
  }\bibfield  {title} {\bibinfo {title} {Geometrical aspects in optical
  wave-packet dynamics},\ }\href {https://doi.org/10.1103/PhysRevE.74.066610}
  {\bibfield  {journal} {\bibinfo  {journal} {Physical Review E}\ }\textbf
  {\bibinfo {volume} {74}},\ \bibinfo {pages} {066610} (\bibinfo {year}
  {2006})}\BibitemShut {NoStop}%
\bibitem [{\citenamefont {Bisnovatyi-Kogan}\ and\ \citenamefont
  {Tsupko}(2010)}]{plasma_lensing1}%
  \BibitemOpen
  \bibfield  {author} {\bibinfo {author} {\bibfnamefont {G.~S.}\ \bibnamefont
  {Bisnovatyi-Kogan}}\ and\ \bibinfo {author} {\bibfnamefont {O.~Y.}\
  \bibnamefont {Tsupko}},\ }\bibfield  {title} {\bibinfo {title}
  {{Gravitational lensing in a non-uniform plasma}},\ }\href
  {https://doi.org/10.1111/j.1365-2966.2010.16290.x} {\bibfield  {journal}
  {\bibinfo  {journal} {Monthly Notices of the Royal Astronomical Society}\
  }\textbf {\bibinfo {volume} {404}},\ \bibinfo {pages} {1790} (\bibinfo {year}
  {2010})}\BibitemShut {NoStop}%
\bibitem [{\citenamefont {Rogers}(2015)}]{plasma_lensing2}%
  \BibitemOpen
  \bibfield  {author} {\bibinfo {author} {\bibfnamefont {A.}~\bibnamefont
  {Rogers}},\ }\bibfield  {title} {\bibinfo {title} {{Frequency-dependent
  effects of gravitational lensing within plasma}},\ }\href
  {https://doi.org/10.1093/mnras/stv903} {\bibfield  {journal} {\bibinfo
  {journal} {Monthly Notices of the Royal Astronomical Society}\ }\textbf
  {\bibinfo {volume} {451}},\ \bibinfo {pages} {17} (\bibinfo {year}
  {2015})}\BibitemShut {NoStop}%
\bibitem [{\citenamefont {Bisnovatyi-Kogan}\ and\ \citenamefont
  {Tsupko}(2017)}]{plasma_lensing3}%
  \BibitemOpen
  \bibfield  {author} {\bibinfo {author} {\bibfnamefont {G.~S.}\ \bibnamefont
  {Bisnovatyi-Kogan}}\ and\ \bibinfo {author} {\bibfnamefont {O.~Y.}\
  \bibnamefont {Tsupko}},\ }\bibfield  {title} {\bibinfo {title} {Gravitational
  lensing in presence of plasma: Strong lens systems, black hole lensing and
  shadow},\ }\href {https://doi.org/10.3390/universe3030057} {\bibfield
  {journal} {\bibinfo  {journal} {Universe}\ }\textbf {\bibinfo {volume} {3}},\
  \bibinfo {pages} {57} (\bibinfo {year} {2017})}\BibitemShut {NoStop}%
\bibitem [{\citenamefont {Abdujabbarov}\ \emph {et~al.}(2017)\citenamefont
  {Abdujabbarov}, \citenamefont {Toshmatov}, \citenamefont {Schee},
  \citenamefont {Stuchlík},\ and\ \citenamefont {Ahmedov}}]{plasma_lensing4}%
  \BibitemOpen
  \bibfield  {author} {\bibinfo {author} {\bibfnamefont {A.}~\bibnamefont
  {Abdujabbarov}}, \bibinfo {author} {\bibfnamefont {B.}~\bibnamefont
  {Toshmatov}}, \bibinfo {author} {\bibfnamefont {J.}~\bibnamefont {Schee}},
  \bibinfo {author} {\bibfnamefont {Z.}~\bibnamefont {Stuchlík}},\ and\
  \bibinfo {author} {\bibfnamefont {B.}~\bibnamefont {Ahmedov}},\ }\bibfield
  {title} {\bibinfo {title} {Gravitational lensing by regular black holes
  surrounded by plasma},\ }\href {https://doi.org/10.1142/S0218271817410115}
  {\bibfield  {journal} {\bibinfo  {journal} {International Journal of Modern
  Physics D}\ }\textbf {\bibinfo {volume} {26}},\ \bibinfo {pages} {1741011}
  (\bibinfo {year} {2017})}\BibitemShut {NoStop}%
\bibitem [{\citenamefont {Tsupko}(2021)}]{plasma_lensing5}%
  \BibitemOpen
  \bibfield  {author} {\bibinfo {author} {\bibfnamefont {O.~Y.}\ \bibnamefont
  {Tsupko}},\ }\bibfield  {title} {\bibinfo {title} {Deflection of light rays
  by a spherically symmetric black hole in a dispersive medium},\ }\href
  {https://doi.org/10.1103/PhysRevD.103.104019} {\bibfield  {journal} {\bibinfo
   {journal} {Physical Review D}\ }\textbf {\bibinfo {volume} {103}},\ \bibinfo
  {pages} {104019} (\bibinfo {year} {2021})}\BibitemShut {NoStop}%
\bibitem [{\citenamefont {Newton}\ and\ \citenamefont
  {Wigner}(1949)}]{NewtonWigner}%
  \BibitemOpen
  \bibfield  {author} {\bibinfo {author} {\bibfnamefont {T.~D.}\ \bibnamefont
  {Newton}}\ and\ \bibinfo {author} {\bibfnamefont {E.~P.}\ \bibnamefont
  {Wigner}},\ }\bibfield  {title} {\bibinfo {title} {Localized states for
  elementary systems},\ }\href {https://doi.org/10.1103/RevModPhys.21.400}
  {\bibfield  {journal} {\bibinfo  {journal} {Reviews of Modern Physics}\
  }\textbf {\bibinfo {volume} {21}},\ \bibinfo {pages} {400} (\bibinfo {year}
  {1949})}\BibitemShut {NoStop}%
\bibitem [{\citenamefont {Jauch}\ and\ \citenamefont
  {Piron}(1967)}]{localizability1967}%
  \BibitemOpen
  \bibfield  {author} {\bibinfo {author} {\bibfnamefont {J.~M.}\ \bibnamefont
  {Jauch}}\ and\ \bibinfo {author} {\bibfnamefont {C.}~\bibnamefont {Piron}},\
  }\bibfield  {title} {\bibinfo {title} {Generalized localizability},\ }\href
  {https://doi.org/10.5169/seals-113783} {\bibfield  {journal} {\bibinfo
  {journal} {Helvetica Physica Acta}\ }\textbf {\bibinfo {volume} {40}},\
  \bibinfo {pages} {559} (\bibinfo {year} {1967})}\BibitemShut {NoStop}%
\bibitem [{\citenamefont {Amrein}(1969)}]{localizability1969}%
  \BibitemOpen
  \bibfield  {author} {\bibinfo {author} {\bibfnamefont {W.~O.}\ \bibnamefont
  {Amrein}},\ }\bibfield  {title} {\bibinfo {title} {Localizability for
  particles of mass zero},\ }\href {https://doi.org/10.5169/seals-114059}
  {\bibfield  {journal} {\bibinfo  {journal} {Helvetica Physica Acta}\ }\textbf
  {\bibinfo {volume} {42}},\ \bibinfo {pages} {149} (\bibinfo {year}
  {1969})}\BibitemShut {NoStop}%
\bibitem [{\citenamefont {Hegerfeldt}(1974)}]{Hegerfeldt1974}%
  \BibitemOpen
  \bibfield  {author} {\bibinfo {author} {\bibfnamefont {G.~C.}\ \bibnamefont
  {Hegerfeldt}},\ }\bibfield  {title} {\bibinfo {title} {Remark on causality
  and particle localization},\ }\href
  {https://doi.org/10.1103/PhysRevD.10.3320} {\bibfield  {journal} {\bibinfo
  {journal} {Physical Review D}\ }\textbf {\bibinfo {volume} {10}},\ \bibinfo
  {pages} {3320} (\bibinfo {year} {1974})}\BibitemShut {NoStop}%
\bibitem [{\citenamefont {Hegerfeldt}\ and\ \citenamefont
  {Ruijsenaars}(1980)}]{Hegerfeldt1980}%
  \BibitemOpen
  \bibfield  {author} {\bibinfo {author} {\bibfnamefont {G.~C.}\ \bibnamefont
  {Hegerfeldt}}\ and\ \bibinfo {author} {\bibfnamefont {S.~N.~M.}\ \bibnamefont
  {Ruijsenaars}},\ }\bibfield  {title} {\bibinfo {title} {Remarks on causality,
  localization, and spreading of wave packets},\ }\href
  {https://doi.org/10.1103/PhysRevD.22.377} {\bibfield  {journal} {\bibinfo
  {journal} {Physical Review D}\ }\textbf {\bibinfo {volume} {22}},\ \bibinfo
  {pages} {377} (\bibinfo {year} {1980})}\BibitemShut {NoStop}%
\bibitem [{\citenamefont {Bacry}(1988{\natexlab{a}})}]{bacry}%
  \BibitemOpen
  \bibfield  {author} {\bibinfo {author} {\bibfnamefont {H.}~\bibnamefont
  {Bacry}},\ }\href {https://link.springer.com/book/10.1007/BFb0019319} {\emph
  {\bibinfo {title} {{Localizability and space in Quantum Physics}}}},\ Lecture
  Notes in Physics 308\ (\bibinfo  {publisher} {Springer--Verlag},\ \bibinfo
  {year} {1988})\BibitemShut {NoStop}%
\bibitem [{\citenamefont {Bacry}(1988{\natexlab{b}})}]{bacry1988}%
  \BibitemOpen
  \bibfield  {author} {\bibinfo {author} {\bibfnamefont {H.}~\bibnamefont
  {Bacry}},\ }\bibfield  {title} {\bibinfo {title} {The position operator
  revisited},\ }\href {http://www.numdam.org/item/AIHPA_1988__49_2_245_0/}
  {\bibfield  {journal} {\bibinfo  {journal} {Annales de l'Institut Henri
  Poincar{\'e} Physique th\'eorique}\ }\textbf {\bibinfo {volume} {49}},\
  \bibinfo {pages} {245} (\bibinfo {year} {1988}{\natexlab{b}})}\BibitemShut
  {NoStop}%
\bibitem [{\citenamefont {Skagerstam}(1994)}]{SKAGERSTAM}%
  \BibitemOpen
  \bibfield  {author} {\bibinfo {author} {\bibfnamefont {B.-S.~K.}\
  \bibnamefont {Skagerstam}},\ }\bibinfo {title} {{Localization of massless
  spinning particles and the Berry phase}},\ in\ \href
  {https://doi.org/10.1142/9789814327060_0023} {\emph {\bibinfo {booktitle} {On
  Klauder's Path: A Field Trip}}}\ (\bibinfo  {publisher} {World Scientific},\
  \bibinfo {year} {1994})\ pp.\ \bibinfo {pages} {209--222}\BibitemShut
  {NoStop}%
\bibitem [{\citenamefont {B\'erard}\ and\ \citenamefont
  {Mohrbach}(2006)}]{SHE_QM2}%
  \BibitemOpen
  \bibfield  {author} {\bibinfo {author} {\bibfnamefont {A.}~\bibnamefont
  {B\'erard}}\ and\ \bibinfo {author} {\bibfnamefont {H.}~\bibnamefont
  {Mohrbach}},\ }\bibfield  {title} {\bibinfo {title} {{Spin Hall effect and
  Berry phase of spinning particles}},\ }\href
  {https://doi.org/https://doi.org/10.1016/j.physleta.2005.11.071} {\bibfield
  {journal} {\bibinfo  {journal} {Physics Letters A}\ }\textbf {\bibinfo
  {volume} {352}},\ \bibinfo {pages} {190} (\bibinfo {year}
  {2006})}\BibitemShut {NoStop}%
\bibitem [{\citenamefont {Kosiński}\ and\ \citenamefont
  {Maślanka}(2018)}]{KOSINSKI2018}%
  \BibitemOpen
  \bibfield  {author} {\bibinfo {author} {\bibfnamefont {P.}~\bibnamefont
  {Kosiński}}\ and\ \bibinfo {author} {\bibfnamefont {P.}~\bibnamefont
  {Maślanka}},\ }\bibfield  {title} {\bibinfo {title} {{Localizability, gauge
  symmetry and Newton–Wigner operator for massless particles}},\ }\href
  {https://doi.org/https://doi.org/10.1016/j.aop.2018.08.012} {\bibfield
  {journal} {\bibinfo  {journal} {Annals of Physics}\ }\textbf {\bibinfo
  {volume} {398}},\ \bibinfo {pages} {203} (\bibinfo {year}
  {2018})}\BibitemShut {NoStop}%
\bibitem [{\citenamefont {Finster}\ and\ \citenamefont
  {Paganini}(2020)}]{finster2020}%
  \BibitemOpen
  \bibfield  {author} {\bibinfo {author} {\bibfnamefont {F.}~\bibnamefont
  {Finster}}\ and\ \bibinfo {author} {\bibfnamefont {C.~F.}\ \bibnamefont
  {Paganini}},\ }\bibfield  {title} {\bibinfo {title} {{Incompatibility of
  frequency splitting and spatial localization: a quantitative analysis of
  Hegerfeldt's theorem}},\ }\href {https://arxiv.org/abs/2005.10120} {\bibfield
   {journal} {\bibinfo  {journal} {arXiv:2005.10120}\ } (\bibinfo {year}
  {2020})}\BibitemShut {NoStop}%
\end{thebibliography}%

\end{document}